\documentclass[11pt,a4paper]{article}
\usepackage{jheppub}
\usepackage{amsmath,amssymb}
\DeclareMathOperator{\tr}{tr}
\usepackage{orcidlink}
\usepackage{framed}
\usepackage{feynmp}
\usepackage{ytableau}
\allowdisplaybreaks
\newcommand{\ii}{\mathrm{i}}
\newcommand{\threefields}[3]{$\mathbf{#1}\otimes\mathbf{#2}\otimes\mathbf{#3}$}
\newcommand{\fourfields}[4]{$\mathbf{#1}\otimes\mathbf{#2}\otimes\mathbf{#3}\otimes\mathbf{#4}$}
\newcommand{\fivefields}[5]{%
  $\mathbf{#1}\otimes\mathbf{#2}\otimes\mathbf{#3}\otimes\mathbf{#4}\otimes\mathbf{#5}$}
\title{Birdtracks of Exotic $\mathrm{SU}(N)$~Color Structures}
\author{Thorsten Ohl \orcidlink{0000-0002-7526-2975}}
\emailAdd{ohl@physik.uni-wuerzburg.de}
\affiliation{%
  University of~W\"urzburg,
  Institute of Theoretical Physics and Astrophysics,
  Emil-Hilb-Weg 22,
  97074 W\"urzburg, Germany}
\abstract{%
  I introduce a systematic procedure for constructing complete and
  linearly independent sets of color structures for
  interactions of fields transforming under exotic
  representations of~$\mathrm{SU}(N)$, in particular
  the~$\mathrm{SU}(3)$ gauge group of QCD.
  It uncovers errors in previous results, starting with interactions of
  four fields including a single sextet.}
\arxivnumber{2403.04685}
\begin{document}
\maketitle
\begin{fmffile}{\jobname pics}
\fmfcmd{arrow_len := 3mm;}
\fmfcmd{curly_len := 2mm;}
\newcommand{\Ksixbar}[3]{[\overline{K}^{#1}_{\mathbf{6}}]_{#2#3}}
\newcommand{\Ksix}[3]{[K_{\mathbf{6}\,#1}]^{#2#3}}
\newcommand{\T}[3]{[T^{#1}]^{#2}_{\phantom{#2}#3}}
\newcommand{\Tsix}[3]{[T^{#1}_{\mathbf{6}}]^{#2}_{\phantom{#2}#3}}
\newcommand{\Teight}[3]{[T^{#1}_{\mathbf{8}}]^{#2}_{\phantom{#2}#3}}
\ytableausetup{centertableaux,smalltableaux}
\section{Introduction}
Physics Beyond the Standard Model~(BSM) can be searched for in
nonrenormalizable interactions of Standard Model~(SM) particles and in
the effects of new elementary particles that are not part of the~SM.
Due to the current absence of evidence for the latter, most efforts
have concentrated on the former, in particular on constraining the
parameter space of the dimension six and higher operators in the SM
Effective Field Theory~(SMEFT).  Nevertheless, UV complete embeddings
of these nonrenormalizable interactions typically need to introduce
new heavy particles.  These particles can transform under more
complicated irreducible representations~(irreps) of the color
group~$\mathrm{SU}(3)$ than the singlet, triplet and octet irreps that
exhaust the repertoire of the~SM.

The construction of a basis for the higher dimensional operators in
an~EFT involves the construction of a basis for the invariant tensors
of the symmetry groups.  In the case of flavor~$\mathrm{SU}(3)$, this
problem has been studied already a long time ago.  Classic references
for invariant tensors in products of triplet and octet representations
are~\cite{MacFarlane:1968vc,Dittner:1971fy,Dittner:1972}.  Most of these results
can be generalized to products of the fundamental and adjoint
representations of~$\mathrm{SU}(N)$~\cite{Dittner:1972,Haber:2019sgz}.  The major
technical difficulty for the construction of such bases lies in the
fact that these tensor algebras are not freely generated.  For
example, $\T{a}{i}{j}$ and~$f^{abc}$ are invariant tensors in
the products $\mathbf{8}\otimes\mathbf{3}\otimes\mathbf{\overline3}$
and $\mathbf{8}\otimes\mathbf{8}\otimes\mathbf{8}$, respectively,
but the sum
\begin{equation}
  \T{a}{i}{j}\T{b}{j}{k} - \T{b}{i}{j}\T{a}{j}{k}
    - \ii f^{abc} \T{c}{i}{k}
\end{equation}
of their products obviously vanishes
in~$\mathbf{8}\otimes\mathbf{8}\otimes\mathbf{3}\otimes\mathbf{\overline3}$.
There are many more non-trivial relations
among products of invariant
tensors~\cite{MacFarlane:1968vc,Dittner:1971fy,Dittner:1972,Haber:2019sgz} and a
naive approach risks producing overcomplete sets.  In the following,
independence will always refer to linear independence and not
algebraic independence.  Correspondingly, complete sets are to be
understood as spanning sets of a vector space of tensors and not as
generating sets of an algebra.

If we want to include new exotic fields that transform under
irreps other than the fundamental and adjoint,
these classic results are not sufficient, of course.  The
case of $\mathrm{SU}(3)$-sextets has been studied using methods
inspired by the investigations of dualities in supersymmetric field
theories~\cite{Almumin:2020yoq}.
The next step towards a complete classification of effective interactions
involving~$\mathrm{SU}(3)$ exotica has been done
in~\cite{Carpenter:2021rkl,Carpenter:2023giu}. Their approach is based on a
recursive decomposition of tensor products into irreps and a selection of
$\mathrm{SU}(3)$-singlets in the final product.  It corresponds to
integrating out heavy fields in different irreps of~$\mathrm{SU}(3)$
in a top-down construction of effective Lagrangians.
While such decompositions can be performed reliably with the aid of computers 
(e.g.~\cite{Horst:2010qj,Feger:2012bs,Feger:2019tvk,Fonseca:2020vke}),
one must verify that the tensors constructed this way are linearly
independent and that their set is
complete, as required for a systematic bottom-up
exploration of BSM physics, in  particular
if the renormalization of these interactions is taken into account
(see, e.g.,~\cite{Ohl:1992sr} for examples of nontrivial relations
among~$\mathrm{SU}(3)$ tensors in such calculations).


A different systematic approach is suggested by the fact that
all irreps of~$\mathrm{SU}(N)$ can be constructed as subspaces of the
tensor product of a suitable number of fundamental and conjugated
fundamental representations~$\mathbf{N}$ and~$\mathbf{\overline N}$ by
enforcing permutation symmetries in the factors.  Obviously, one can
represent an arbitrary tensor product of irreps in the same way.  It
has been known for a long time that every invariant tensor in a
product of fundamental and conjugated fundamental representations can
be expressed as a product of Kronecker
symbols~$\delta^{i}_{\phantom{i}j}$ and, in the case of unimodular
transformations, Levi-Civita symbols~$\epsilon^{i_1i_2\cdots i_N}$ or
$\overline\epsilon_{j_1j_2\cdots j_N}$~\cite{6bde026d-fcfc-356e-9eb5-74ec9ada1cff,%
  7f2f1d37-4403-36c7-9ff1-5a4a9dbfaa9e,Appleby_Duffy_Ogden_1987}.
Comprehensive proofs of this fact for~$\mathrm{GL}(N)$,
$\mathrm{SL}(N)$ and~$\mathrm{SO}(N)$ can be found
in~\cite{Appleby_Duffy_Ogden_1987}.  The proof for~$\mathrm{GL}(N)$ is
elementary und the one for~$\mathrm{SL}(N)$ is not much more
complicated.  However, in the case of proper subgroups
of~$\mathrm{GL}(N)$ and~$\mathrm{SL}(N)$, the conditions on invariant
tensors are weaker and care must be taken not to overlook additional
solutions.  The constraints on group elements from~$\mathrm{SO}(N)$
have been implemented in~\cite{Appleby_Duffy_Ogden_1987} by Lagrange
multipliers.  Fortunately, this proof translates directly to the cases
of~$\mathrm{SU}(N)$ by complex conjugating
the matrix elements of the adjoint transformations.

Unfortunately, this result also does not guarantee that the tensors
constructed in this way are linearly independent.
Indeed, as described in section~\ref{sec:epsilon}, tensors
containing~$\epsilon^{i_1i_2\cdots i_N}$
and~$\overline\epsilon_{j_1j_2\cdots j_N}$ simultaneously can be
expressed as a sum of products of Kronecker symbols.
Furthermore, there are many less obvious dependencies
among tensors.  This is complicated by the fact that some of them
are only valid for~$N$ smaller than some threshold.  Examples for
this will be presented in
sections~\ref{sec:spurious} and~\ref{sec:8866bar-revisited}.

If we want to systematically construct complete and
linearly independent sets of invariant tensors, we require a computational test
for the linear independence of tensors.
For this purpose, I define the natural sesquilinear form
\begin{equation}
\label{eq:mu}
  \mu_N\left(A,B\right)
    = \sum_{\substack{i_1\cdots\\ j_1\cdots}}
        \overline A^{i_1\cdots}_{j_1\cdots}
        B^{j_1\cdots}_{i_1\cdots}
    = \overline{\mu_N\left(B,A\right)}
\end{equation}
on the vector space of $\mathrm{SU}(N)$~tensors of a given rank.
Generalizing an observation for products of adjoint representations
of~$\mathrm{SU}(N)$~\cite{Dittner:1971fy},
all dependencies among tensors can then be found
by computing the radical of this bilinear form, i.e.~the
eigenvectors of a matrix representation of this
sesquilinear form with vanishing eigenvalue, as explained in
section~\ref{eq:eigenvectors}.
It turns out that the number
of vanishing eigenvalues can depend on~$N$.

In this paper, I propose a general algorithm for constructing
bases of invariant tensors describing
interactions involving particles transforming under
higher dimensional irreps of~$\mathrm{SU}(N)$.  This algorithm has been
implemented in the computer program \textsc{tangara}.\footnote{Note on the
name: \textit{Tangara} is a genus of spectacularly colored birds in
the family \textit{Thraupidae} found in South America.} using the
\textsc{O'Caml}~\cite{ocaml5/manual} birdtrack libraries developed
for~\textsc{O'Mega}~\cite{Moretti:2001zz,Ohl:2023bvv}.

Section~\ref{sec:colorflows} briefly introduces the colorflow
formalism in order to establish the notation and presents a non-trival
example that motivated the investigation presented here.
Section~\ref{sec:SU(N)-constraints} continues with a discussion of the
pecularities of~$\mathrm{SU}(N)$ colorflows that follow from the
tracelessness of the generators and the invariance of Levi-Civita
symbols.

Section~\ref{sec:algorithm} presents the novel algorithm for
identifying complete and linearly independent sets of invariant tensors.  I
apply it in section~\ref{sec:8866bar-revisited} to answer the
questions posed by the example studied in
section~\ref{sec:8866bar}.
Section~\ref{sec:results} presents a revised catalogue for the simplest
cases in detail and discusses in which cases it confirms
the results of~\cite{Carpenter:2021rkl} and in which cases these
results must be amended.

Appendix~\ref{sec:implementation} briefly describes
the program~\textsc{tangara}, which has been used to obtain the results
presented here.
I will also discuss how to make the results available to
Monte Carlo event generators and other tools for elementary particle
physics.

\section{Colorflows}
\label{sec:colorflows}
We are faced with the task of efficiently computing the
matrix elements of the inner product~$\mu_N$ defined in~\eqref{eq:mu}.
Fortunately, these are nothing but the ``color factors'' familiar
from squared QCD scattering amplitudes.  An efficient algorithm for their
computation that works directly in a product of fundamental and
conjugated fundamental representations has been advocated
in~\cite{Cvitanovic:1976am}: normalize the generators as
\begin{equation}
\label{eq:normalization}
  \tr(T^aT^b) = \delta^{ab}
\end{equation}
and replace all contractions of indices in
the adjoint representation of~$\mathrm{SU}(N)$ by
\begin{equation}
\label{eq:TT->dd-SU(N)}
  \T{a}{i}{j} \T{a}{k}{l}
     = \delta^{i}_{\phantom{i}l} \delta^{k}_{\phantom{k}j}
     - \frac{1}{N} \delta^{i}_{\phantom{i}j} \delta^{k}_{\phantom{k}l}\,.
\end{equation}
Then it only remains to keep track of the factors of~$1/N$ and count the
number of closed chains of Kronecker symbols
\begin{equation}
\label{eq:chains}
  \delta^{i_1}_{\phantom{i_1}i_2}
  \delta^{i_2}_{\phantom{i_2}i_3}
  \cdots
  \delta^{i_n}_{\phantom{i_n}i_1}
  = N\,,
\end{equation}
each contributing a factor of~$N$ to the color factor.  Representing
the Kronecker symbols by arrows leads to the colorflow representation
where each closed loop corresponds to one factor of~$N$ in the color factor.

This description has subsequently been
developed into the comprehensive ``birdtracks'' approach
to Lie algebras and their
representations~\cite{Cvitanovic:2008zz}.
It has also been used to
construct invariant tensors as building blocks for the color part of
scattering amplitudes of $\mathrm{SU}(3)$-triplets and
octets~\cite{Keppeler:2012ih,Keppeler:2013yla,Sjodahl:2013hra,Sjodahl:2018cca},
including implementations in computer
programs~\cite{Sjodahl:2012nk,Sjodahl:2014opa}.

The identity~\eqref{eq:TT->dd-SU(N)} must
be applied to two vertices in a Feynman diagram simultaneously
when evaluating~$\mathrm{SU}(N)$ color factors.
While this is not a problem for evaluating color sums for complete
Feynman diagrams~\cite{Maltoni:2002mq} or~\eqref{eq:mu},
it is an obstacle for the recursive algorithms that are the
state of the art in perturbative calculations (see~\cite{Ohl:2023bvv}
and references cited therein).  This can be avoided by reproducing
the subtraction term in~\eqref{eq:TT->dd-SU(N)} via
additional couplings to a fictitious particle, called a $\mathrm{U}(1)$-ghost,
whose sole purpose is to subtract the traces of the
generators~\cite{Kilian:2012pz}
\begin{equation}
\label{eq:U(1)-ghosts}
  \parbox{42\unitlength}{%
    \fmfframe(8,4)(8,4){%
      \begin{fmfgraph*}(26,20)
        \fmfstraight
        \fmfleft{j,i}
        \fmfright{k,l}
        \fmfv{label=$i$,label.angle=180}{i}
        \fmfv{label=$j$,label.angle=180}{j}
        \fmfv{label=$k$,label.angle=0}{k}
        \fmfv{label=$l$,label.angle=0}{l}
        \fmf{phantom}{j,tij,i}
        \fmf{phantom}{l,tkl,k}
        \fmf{phantom}{tij,tkl}
        \fmffreeze
        \fmf{fermion}{j,tij,i}
        \fmf{fermion}{l,tkl,k}
        \fmf{gluon,label=$\delta^{ab}$,label.dist=5thick,label.side=left}{tij,tkl}
        \fmfdot{tij,tkl}
        \fmfv{label=$\T{a}{i}{j}$,label.dist=6thick}{tij}
        \fmfv{label=$\T{b}{k}{l}$,label.dist=6thick}{tkl}
      \end{fmfgraph*}}} =
  \parbox{46\unitlength}{%
    \fmfframe(10,4)(10,4){%
      \begin{fmfgraph*}(26,20)
        \fmfstraight
        \fmfleft{j,i}
        \fmfright{k,l}
        \fmfv{label=$i$,label.angle=180}{i}
        \fmfv{label=$j$,label.angle=180}{j}
        \fmfv{label=$k$,label.angle=0}{k}
        \fmfv{label=$l$,label.angle=0}{l}
        \fmf{phantom}{j,tij,i}
        \fmf{phantom}{l,tkl,k}
        \fmf{phantom}{tij,tkl}
        \fmffreeze
        \fmfcmd{path il, kj;}
        \fmfcmd{il = vloc(__l){2down+left}
          .. (vloc(__tkl) shifted (thick*up)){left}
          .. {left}(vloc(__tij) shifted (thick*up))
          .. {2up+left}vloc(__i);}
        \fmfcmd{kj = vloc(__j){2up+right}
          .. (vloc(__tij) shifted (thick*down)){right}
          .. {right}(vloc(__tkl) shifted (thick*down))
          .. {2down+right}vloc(__k);}
        \fmfi{fermion}{subpath (0.00,0.25)*length(il) of il}
          \fmfi{plain}{subpath (0.25,0.34)*length(il) of il}
          \fmfi{plain}{subpath (0.34,0.50)*length(il) of il}
        \fmfi{fermion}{subpath (0.50,0.66)*length(il) of il}
          \fmfi{plain}{subpath (0.66,0.75)*length(il) of il}
        \fmfi{fermion}{subpath (0.75,1.00)*length(il) of il}
        \fmfi{fermion}{subpath (0.00,0.25)*length(kj) of kj}
        \fmfi{fermion}{subpath (0.25,0.34)*length(kj) of kj}
          \fmfi{plain}{subpath (0.34,0.50)*length(kj) of kj}
        \fmfi{fermion}{subpath (0.50,0.66)*length(kj) of kj}
          \fmfi{plain}{subpath (0.66,0.75)*length(kj) of kj}
        \fmfi{fermion}{subpath (0.75,1.00)*length(kj) of kj}
        \fmf{phantom,label=$\delta^{k'}_{\phantom{k}j'}\delta^{i'}_{\phantom{i}l'}$,
          label.dist=5thick,label.side=left}{tij,tkl}
        \fmfv{label=$\delta^{i}_{\phantom{i}i'}\delta^{j'}_{\phantom{j}j}$,label.dist=8thick}{tij}
        \fmfv{label=$\delta^{k}_{\phantom{k}k'}\delta^{l'}_{\phantom{l}l}$,label.dist=8thick}{tkl}
      \end{fmfgraph*}}} + 
  \parbox{34\unitlength}{%
    \fmfframe(4,4)(4,4){%
      \begin{fmfgraph*}(26,20)
        \fmfstraight
        \fmfleft{j,i}
        \fmfright{k,l}
        \fmfv{label=$i$,label.angle=180}{i}
        \fmfv{label=$j$,label.angle=180}{j}
        \fmfv{label=$k$,label.angle=0}{k}
        \fmfv{label=$l$,label.angle=0}{l}
        \fmf{phantom}{j,tij,i}
        \fmf{phantom}{l,tkl,k}
        \fmf{phantom}{tij,tkl}
        \fmffreeze
        \fmfcmd{path ij, kl;}
        \fmfcmd{ij = vloc(__j){2up+right}
          .. {up}(vloc(__tij) shifted (3thick*left))
          .. {2up+left}vloc(__i);}
        \fmfcmd{kl = vloc(__l){2down+left}
          .. {down}(vloc(__tkl) shifted (3thick*right))
          .. {2down+right}vloc(__k);}
        \fmfi{fermion}{subpath (0.0,0.5)*length(ij) of ij}
        \fmfi{fermion}{subpath (0.5,1.0)*length(ij) of ij}
        \fmfi{fermion}{subpath (0.0,0.5)*length(kl) of kl}
        \fmfi{fermion}{subpath (0.5,1.0)*length(kl) of kl}
        \fmfv{label=$\delta^{i}_{\phantom{i}j}$,label.dist=8thick}{tij}
        \fmfv{label=$\delta^{k}_{\phantom{k}l}$,label.dist=8thick}{tkl}
        \fmfdot{tij,tkl}
        \fmf{dots,label=$-\frac{1}{N}$,label.dist=5thick,label.side=left}{tij,tkl}
      \end{fmfgraph*}}}\,.
\end{equation}
These $\mathrm{U}(1)$-ghosts have to be included in the internal color
exchanges and in the sums over external colors, of course.  
The resulting colorflow Feynman rules
can automatically be derived from traditional Feynman rules as
specified, e.g., in UFO~\cite{Degrande:2011ua,Darme:2023jdn}.  This has
been implemented in the recursive matrix element generator
\textsc{O'Mega}~\cite{Moretti:2001zz,Ohl:2023bvv} that is used in the
general purpose Monte Carlo event generator
\textsc{Whizard}~\cite{Kilian:2007gr}.

In the colorflow Feynman rules, the couplings of the
$\mathrm{U}(1)$-ghost are fixed by a Ward identity to be the same as
the couplings to the $\mathrm{SU}(N)$-gluons~\cite{Kilian:2012pz}.
Therefore, they are not a new source of independent tensors and the formalism
can be used in arbitrary orders of
perturbation theory to construct complete
and independent sets of interactions in color space.

\subsection{A Note on Notation}
\label{sec:notation}
In the calculations, I will keep~$N\ge2$ general as long as possible.
This allows us to test the procedure by checking pecularities
of~$\mathrm{SU}(2)$ and to confirm simplifications in the
limit~$N\to\infty$.  Results involving the invariance of the
tensors~$\epsilon^{ijk}$ and $\overline\epsilon_{ijk}$ apply only
to~$\mathrm{SU}(3)$, of course.
Nevertheless, in applications I will only be interested
in~$\mathrm{SU}(3)$ and I shall engage in \textit{abus de langage}
throughout this paper when denoting the irreps
of~$\mathrm{SU}(N)$.  Instead of spelling out the
Young tableaux, I will often use the familiar dimensions of the
$\mathrm{SU}(3)$-irreps, as in formula~\eqref{eq:building-blocks} below.
With the exception of the~$\mathbf{15}$ and~$\mathbf{15'}$, this is
unambiguous for all small irreps and
allows me to take advantage of an abbreviated notation for which
much intuition as available among practitioners.

\subsection{Exotic Birdtracks}
\label{sec:building-blocks}
In the colorflow representation, states in the reducible product
of~$n$ fundamental representations are described by~$n$ parallel lines
with arrows pointing into the diagram.  The conjugated
representation has the direction of the arrows reversed.  As usual,
the reducible representations are decomposed into irreps by imposing
the permutation symmetries specified by the standard Young tableaux
consisting of~$n$ boxes~\cite{Cvitanovic:2008zz}.  For a given Young
tableau, one first antisymmetrizes the lines in each column and
subsequently symmetrizes the lines in each row.  The normalizations
are chosen such that the combined (anti)symmetrizations form a
projection.

Instead of repeating the comprehensive account given in section~9.5
of~\cite{Cvitanovic:2008zz} and
in~\cite{Keppeler:2012ih,Keppeler:2013yla,Sjodahl:2013hra}, I
only list the simplest building blocks\footnote{Note
that~\eqref{eq:building-blocks} depicts
the Young projectors described in~\cite{Cvitanovic:2008zz}, which are
readily available in the birdtracks library of
\textsc{O'Mega}~\cite{Moretti:2001zz,Ohl:2023bvv}.  We can replace
these projectors by the hermitian Young projectors advocated
in~\cite{Keppeler:2013yla,Alcock-Zeilinger:2016sxc}, without modifying the
other parts of~\textsc{tangara}.  In the general case, this will
change some matrix elements of the inner product~$\mu_N$~\eqref{eq:mu}, but
the number of vanishing eigenvalues will remain the same.
There will of course be no changes at all for totally symmetric or
antisymmetric irreps.}
in order to introduce the notation
\begin{subequations}
\label{eq:building-blocks}
  \begin{align}
    \mathbf{3}&&
    \ytableaushort{{j_1}}&&
      \parbox{25\unitlength}{%
        \fmfframe(4,4)(4,4){%
          \begin{fmfgraph*}(17,2)
            \fmfstraight
            \fmfleft{i1}
            \fmfright{o1}
            \fmfv{label=$\scriptstyle j_1$,label.angle=180}{i1}
            \fmfv{label=$\scriptstyle i_1$,label.angle=0}{o1}
            \fmf{fermion}{i1,o1}
        \end{fmfgraph*}}}\phantom{\,,} \\
    \mathbf{6}&&
    \ytableaushort{{j_1}{j_2}}&&
      \parbox{25\unitlength}{%
        \fmfframe(4,4)(4,4){%
          \begin{fmfgraph*}(17,3)
            \fmfstraight
            \fmfleft{i2,i1}
            \fmfright{o2,o1}
            \fmfv{label=$\scriptstyle j_1$,label.angle=180}{i1}
            \fmfv{label=$\scriptstyle j_2$,label.angle=180}{i2}
            \fmfv{label=$\scriptstyle i_1$,label.angle=0}{o1}
            \fmfv{label=$\scriptstyle i_2$,label.angle=0}{o2}
            \fmf{fermion}{i1,v1,o1}
            \fmf{fermion}{i2,v2,o2}
            \fmffreeze
            \fmfdraw
            \fmfcmd{path s;}
            \fmfcmd{s = unitsquare shifted (-1/2,-1/2) xscaled 2thick yscaled 1.6h
                           shifted .5[vloc(__v1),vloc(__v2)];}
            \fmfcmd{fill s withcolor white;}
            \fmfcmd{draw s withcolor black;}
        \end{fmfgraph*}}}\phantom{\,,} \\
    \mathbf{8}&&
    \ytableaushort{{j_2}{j_1},{j_3}}&&
      \parbox{25\unitlength}{%
        \fmfframe(4,4)(4,4){%
          \begin{fmfgraph*}(17,6)
            \fmfstraight
            \fmfleft{i3,i2,i1}
            \fmfright{o3,o2,o1}
            \fmfv{label=$\scriptstyle j_1$,label.angle=180}{i1}
            \fmfv{label=$\scriptstyle j_2$,label.angle=180}{i2}
            \fmfv{label=$\scriptstyle j_3$,label.angle=180}{i3}
            \fmfv{label=$\scriptstyle i_1$,label.angle=0}{o1}
            \fmfv{label=$\scriptstyle i_2$,label.angle=0}{o2}
            \fmfv{label=$\scriptstyle i_3$,label.angle=0}{o3}
            \fmf{phantom}{i1,vi1,vo1,o1}
            \fmf{fermion}{i2,vi2,vo2,o2}
            \fmf{phantom}{i3,vi3,vo3,o3}
            \fmffreeze
            \fmf{fermion}{i1,vo1,o1}
            \fmf{fermion}{i3,vi3,o3}
            \fmfdraw
            \fmfcmd{path a, s;}
            \fmfcmd{a = unitsquare shifted (-1/2,-1/2) xscaled 2thick yscaled .8h
                           shifted .5[vloc(__vi2),vloc(__vi3)];}
            \fmfcmd{s = unitsquare shifted (-1/2,-1/2) xscaled 2thick yscaled .8h
                           shifted .5[vloc(__vo1),vloc(__vo2)];}
            \fmfcmd{fill s withcolor white;}
            \fmfcmd{draw s withcolor black;}
            \fmfcmd{filldraw a withcolor black;}
          \end{fmfgraph*}}}\phantom{\,,} \\
    \mathbf{10}&&
    \ytableaushort{{j_1}{j_2}{j_3}}&&
      \parbox{25\unitlength}{%
        \fmfframe(4,4)(4,4){%
          \begin{fmfgraph*}(17,6)
            \fmfstraight
            \fmfleft{i3,i2,i1}
            \fmfright{o3,o2,o1}
            \fmfv{label=$\scriptstyle j_1$,label.angle=180}{i1}
            \fmfv{label=$\scriptstyle j_2$,label.angle=180}{i2}
            \fmfv{label=$\scriptstyle j_3$,label.angle=180}{i3}
            \fmfv{label=$\scriptstyle i_1$,label.angle=0}{o1}
            \fmfv{label=$\scriptstyle i_2$,label.angle=0}{o2}
            \fmfv{label=$\scriptstyle i_3$,label.angle=0}{o3}
            \fmf{fermion}{i1,v1,o1}
            \fmf{fermion}{i2,v2,o2}
            \fmf{fermion}{i3,v3,o3}
            \fmffreeze
            \fmfdraw
            \fmfcmd{path a;}
            \fmfcmd{a = unitsquare shifted (-1/2,-1/2) xscaled 2thick yscaled 1.4h shifted vloc(__v2);}
            \fmfcmd{fill a withcolor white;}
            \fmfcmd{draw a withcolor black;}
        \end{fmfgraph*}}}\phantom{\,,} \\
    \mathbf{15}&&
    \ytableaushort{{j_3}{j_1}{j_2},{j_4}}&&
      \parbox{25\unitlength}{%
        \fmfframe(4,4)(4,4){%
          \begin{fmfgraph*}(17,8)
            \fmfstraight
            \fmfleft{i4,i3,i2,i1}
            \fmfright{o4,o3,o2,o1}
            \fmfv{label=$\scriptstyle j_1$,label.angle=180}{i1}
            \fmfv{label=$\scriptstyle j_2$,label.angle=180}{i2}
            \fmfv{label=$\scriptstyle j_3$,label.angle=180}{i3}
            \fmfv{label=$\scriptstyle j_4$,label.angle=180}{i4}
            \fmfv{label=$\scriptstyle i_1$,label.angle=0}{o1}
            \fmfv{label=$\scriptstyle i_2$,label.angle=0}{o2}
            \fmfv{label=$\scriptstyle i_3$,label.angle=0}{o3}
            \fmfv{label=$\scriptstyle i_4$,label.angle=0}{o4}
            \fmf{phantom}{i1,vi1,vo1,o1}
            \fmf{phantom}{i2,vi2,vo2,o2}
            \fmf{fermion}{i3,vi3,vo3,o3}
            \fmf{phantom}{i4,vi4,vo4,o4}
            \fmffreeze
            \fmf{fermion}{i1,vo1,o1}
            \fmf{fermion}{i2,vo2,o2}
            \fmf{fermion}{i4,vi4,o4}
            \fmfdraw
            \fmfcmd{path a, s;}
            \fmfcmd{a = unitsquare shifted (-1/2,-1/2) xscaled 2thick yscaled .6h
                           shifted .5[vloc(__vi3),vloc(__vi4)];}
            \fmfcmd{s = unitsquare shifted (-1/2,-1/2) xscaled 2thick yscaled .9h
                           shifted .5[vloc(__vo1),vloc(__vo3)];}
            \fmfcmd{fill s withcolor white;}
            \fmfcmd{draw s withcolor black;}
            \fmfcmd{filldraw a withcolor black;}
          \end{fmfgraph*}}}\phantom{\,,} \\
    \mathbf{15'}&&
    \ytableaushort{{j_1}{j_2}{j_3}{j_4}}&&
      \parbox{25\unitlength}{%
        \fmfframe(4,4)(4,4){%
          \begin{fmfgraph*}(17,8)
            \fmfstraight
            \fmfleft{i4,i3,i2,i1}
            \fmfright{o4,o3,o2,o1}
            \fmfv{label=$\scriptstyle j_1$,label.angle=180}{i1}
            \fmfv{label=$\scriptstyle j_2$,label.angle=180}{i2}
            \fmfv{label=$\scriptstyle j_3$,label.angle=180}{i3}
            \fmfv{label=$\scriptstyle j_4$,label.angle=180}{i4}
            \fmfv{label=$\scriptstyle i_1$,label.angle=0}{o1}
            \fmfv{label=$\scriptstyle i_2$,label.angle=0}{o2}
            \fmfv{label=$\scriptstyle i_3$,label.angle=0}{o3}
            \fmfv{label=$\scriptstyle i_4$,label.angle=0}{o4}
            \fmf{fermion}{i1,v1,o1}
            \fmf{fermion}{i2,v2,o2}
            \fmf{fermion}{i3,v3,o3}
            \fmf{fermion}{i4,v4,o4}
            \fmffreeze
            \fmfdraw
            \fmfcmd{path a;}
            \fmfcmd{a = unitsquare shifted (-1/2,-1/2) xscaled 2thick yscaled 1.3h
                           shifted .5[vloc(__v2),vloc(__v3)];}
            \fmfcmd{fill a withcolor white;}
            \fmfcmd{draw a withcolor black;}
        \end{fmfgraph*}}}\phantom{\,,} \\
    \mathbf{21}&&
    \ytableaushort{{j_1}{j_2}{j_3}{j_4}{j_5}}&&
      \parbox{25\unitlength}{%
        \fmfframe(4,4)(4,4){%
          \begin{fmfgraph*}(17,10)
            \fmfstraight
            \fmfleft{i5,i4,i3,i2,i1}
            \fmfright{o5,o4,o3,o2,o1}
            \fmfv{label=$\scriptstyle j_1$,label.angle=180}{i1}
            \fmfv{label=$\scriptstyle j_2$,label.angle=180}{i2}
            \fmfv{label=$\scriptstyle j_3$,label.angle=180}{i3}
            \fmfv{label=$\scriptstyle j_4$,label.angle=180}{i4}
            \fmfv{label=$\scriptstyle j_5$,label.angle=180}{i5}
            \fmfv{label=$\scriptstyle i_1$,label.angle=0}{o1}
            \fmfv{label=$\scriptstyle i_2$,label.angle=0}{o2}
            \fmfv{label=$\scriptstyle i_3$,label.angle=0}{o3}
            \fmfv{label=$\scriptstyle i_4$,label.angle=0}{o4}
            \fmfv{label=$\scriptstyle i_5$,label.angle=0}{o5}
            \fmf{fermion}{i1,v1,o1}
            \fmf{fermion}{i2,v2,o2}
            \fmf{fermion}{i3,v3,o3}
            \fmf{fermion}{i4,v4,o4}
            \fmf{fermion}{i5,v5,o5}
            \fmffreeze
            \fmfdraw
            \fmfcmd{path a;}
            \fmfcmd{a = unitsquare shifted (-1/2,-1/2) xscaled 2thick yscaled 1.3h
                           shifted .5[vloc(__v1),vloc(__v5)];}
            \fmfcmd{fill a withcolor white;}
            \fmfcmd{draw a withcolor black;}
        \end{fmfgraph*}}}\phantom{\,,} \\
    \mathbf{24}&&
    \ytableaushort{{j_4}{j_1}{j_2}{j_3},{j_5}}&&
      \parbox{25\unitlength}{%
        \fmfframe(4,4)(4,4){%
          \begin{fmfgraph*}(17,10)
            \fmfstraight
            \fmfleft{i5,i4,i3,i2,i1}
            \fmfright{o5,o4,o3,o2,o1}
            \fmfv{label=$\scriptstyle j_1$,label.angle=180}{i1}
            \fmfv{label=$\scriptstyle j_2$,label.angle=180}{i2}
            \fmfv{label=$\scriptstyle j_3$,label.angle=180}{i3}
            \fmfv{label=$\scriptstyle j_4$,label.angle=180}{i4}
            \fmfv{label=$\scriptstyle j_5$,label.angle=180}{i5}
            \fmfv{label=$\scriptstyle i_1$,label.angle=0}{o1}
            \fmfv{label=$\scriptstyle i_2$,label.angle=0}{o2}
            \fmfv{label=$\scriptstyle i_3$,label.angle=0}{o3}
            \fmfv{label=$\scriptstyle i_4$,label.angle=0}{o4}
            \fmfv{label=$\scriptstyle i_5$,label.angle=0}{o5}
            \fmf{phantom}{i1,vi1,vo1,o1}
            \fmf{phantom}{i2,vi2,vo2,o2}
            \fmf{phantom}{i3,vi3,vo3,o3}
            \fmf{fermion}{i4,vi4,vo4,o4}
            \fmf{phantom}{i5,vi5,vo5,o5}
            \fmffreeze
            \fmf{fermion}{i1,vo1,o1}
            \fmf{fermion}{i2,vo2,o2}
            \fmf{fermion}{i3,vo3,o3}
            \fmf{fermion}{i5,vi5,o5}
            \fmfdraw
            \fmfcmd{path a, s;}
            \fmfcmd{a = unitsquare shifted (-1/2,-1/2) xscaled 2thick yscaled .45h
                           shifted .5[vloc(__vi4),vloc(__vi5)];}
            \fmfcmd{s = unitsquare shifted (-1/2,-1/2) xscaled 2thick yscaled .95h
                           shifted .5[vloc(__vo1),vloc(__vo4)];}
            \fmfcmd{fill s withcolor white;}
            \fmfcmd{draw s withcolor black;}
            \fmfcmd{filldraw a withcolor black;}
          \end{fmfgraph*}}}\phantom{\,,} \\
    \label{eq:27}
    \mathbf{27}&&
    \ytableaushort{{j_2}{j_5}{j_3}{j_4},{j_1}{j_6}}&&
      \parbox{25\unitlength}{%
        \fmfframe(4,4)(4,4){%
          \begin{fmfgraph*}(17,14)
            \fmfstraight
            \fmfleft{i6,i5,i4,i3,i2,i1}
            \fmfright{o6,o5,o4,o3,o2,o1}
            \fmfv{label=$\scriptstyle j_1$,label.angle=180}{i1}
            \fmfv{label=$\scriptstyle j_2$,label.angle=180}{i2}
            \fmfv{label=$\scriptstyle j_3$,label.angle=180}{i3}
            \fmfv{label=$\scriptstyle j_4$,label.angle=180}{i4}
            \fmfv{label=$\scriptstyle j_5$,label.angle=180}{i5}
            \fmfv{label=$\scriptstyle j_6$,label.angle=180}{i6}
            \fmfv{label=$\scriptstyle i_1$,label.angle=0}{o1}
            \fmfv{label=$\scriptstyle i_2$,label.angle=0}{o2}
            \fmfv{label=$\scriptstyle i_3$,label.angle=0}{o3}
            \fmfv{label=$\scriptstyle i_4$,label.angle=0}{o4}
            \fmfv{label=$\scriptstyle i_5$,label.angle=0}{o5}
            \fmfv{label=$\scriptstyle i_6$,label.angle=0}{o6}
            \fmf{fermion}{i1,vi1,vo1,o1}
            \fmf{fermion}{i2,vi2,vo2,o2}
            \fmf{phantom}{i3,vi3,vo3,o3}
            \fmf{phantom}{i4,vi4,vo4,o4}
            \fmf{fermion}{i5,vi5,vo5,o5}
            \fmf{fermion}{i6,vi6,vo6,o6}
            \fmffreeze
            \fmf{fermion}{i3,vo3,o3}
            \fmf{fermion}{i4,vo4,o4}
            \fmfdraw
            \fmfcmd{path a[], s, sw[];}
            \fmfcmd{a12 = unitsquare shifted (-1/2,-1/2) xscaled 2thick yscaled .35h
                           shifted .5[vloc(__vi1),vloc(__vi2)];}
            \fmfcmd{a56 = unitsquare shifted (-1/2,-1/2) xscaled 2thick yscaled .35h
                           shifted .5[vloc(__vi5),vloc(__vi6)];}
            \fmfcmd{s = unitsquare shifted (-1/2,-1/2) xscaled 2thick yscaled .75h
                           shifted .5[vloc(__vo2),vloc(__vo5)];}
            \fmfcmd{fill s withcolor white;}
            \fmfcmd{draw s withcolor black;}
            \fmfcmd{filldraw a12 withcolor black;}
            \fmfcmd{filldraw a56 withcolor black;}
            \fmfcmd{sw1 = unitsquare shifted (-1/2,-1/2) xscaled 2thick yscaled .18h
                             shifted vloc(__vo1) shifted (.03h*up);}
            \fmfcmd{sw6 = unitsquare shifted (-1/2,-1/2) xscaled 2thick yscaled .18h
                             shifted vloc(__vo6) shifted (.03h*down);}
            \fmfcmd{fill sw1 withcolor white;}
            \fmfcmd{fill sw6 withcolor white;}
            \fmfcmd{draw subpath (0,2) of sw1 withcolor black;}
            \fmfcmd{draw subpath (3,4) of sw1 withcolor black;}
            \fmfcmd{draw subpath (1,4) of sw6 withcolor black;}
          \end{fmfgraph*}}}\,,
  \end{align}
\end{subequations}
where the white boxes denote symmetrization and the black boxes
antisymmetrization
\begin{subequations}
  \begin{align}
      \parbox{18\unitlength}{%
        \fmfframe(4,4)(4,4){%
          \begin{fmfgraph*}(10,8)
            \fmfstraight
            \fmfleft{in,id,i2,i1}
            \fmfright{on,od,o2,o1}
            \fmfv{label=$\scriptstyle 1$,label.angle=180}{i1}
            \fmfv{label=$\scriptstyle 2$,label.angle=180}{i2}
            \fmfv{label=$\scriptstyle n$,label.angle=180}{in}
            \fmfv{label=$\scriptstyle 1$,label.angle=0}{o1}
            \fmfv{label=$\scriptstyle 2$,label.angle=0}{o2}
            \fmfv{label=$\scriptstyle n$,label.angle=0}{on}
            \fmf{fermion}{i1,v1,o1}
            \fmf{fermion}{i2,v2,o2}
            \fmf{fermion,foreground=.8white}{id,vd,od}
            \fmf{fermion}{in,vn,on}
            \fmffreeze
            \fmfdraw
            \fmfcmd{path s[];}
            \fmfcmd{s123 = unitsquare shifted (-1/2,-1/2) xscaled 2thick yscaled 1.4h
                             shifted .5[vloc(__v1),vloc(__vn)];}
            \fmfcmd{fill s123 withcolor white;}
            \fmfcmd{draw s123 withcolor black;}
          \end{fmfgraph*}}} &=
      \sum_{\sigma\in S_n} \frac{1}{n!}\cdot
      \parbox{17\unitlength}{%
        \fmfframe(4,4)(9,4){%
          \begin{fmfgraph*}(4,8)
            \fmfstraight
            \fmfleft{in,id,i2,i1}
            \fmfright{on,od,o2,o1}
            \fmfv{label=$\scriptstyle 1$,label.angle=180}{i1}
            \fmfv{label=$\scriptstyle 2$,label.angle=180}{i2}
            \fmfv{label=$\scriptstyle n$,label.angle=180}{in}
            \fmfv{label=$\scriptstyle \sigma(1)$,label.angle=0}{o1}
            \fmfv{label=$\scriptstyle \sigma(2)$,label.angle=0}{o2}
            \fmfv{label=$\scriptstyle \sigma(n)$,label.angle=0}{on}
            \fmf{fermion}{i1,o1}
            \fmf{fermion}{i2,o2}
            \fmf{fermion,foreground=.8white}{id,od}
            \fmf{fermion}{in,on}
          \end{fmfgraph*}}}\\
      \parbox{18\unitlength}{%
        \fmfframe(4,4)(4,4){%
          \begin{fmfgraph*}(10,8)
            \fmfstraight
            \fmfleft{in,id,i2,i1}
            \fmfright{on,od,o2,o1}
            \fmfv{label=$\scriptstyle 1$,label.angle=180}{i1}
            \fmfv{label=$\scriptstyle 2$,label.angle=180}{i2}
            \fmfv{label=$\scriptstyle n$,label.angle=180}{in}
            \fmfv{label=$\scriptstyle 1$,label.angle=0}{o1}
            \fmfv{label=$\scriptstyle 2$,label.angle=0}{o2}
            \fmfv{label=$\scriptstyle n$,label.angle=0}{on}
            \fmf{fermion}{i1,v1,o1}
            \fmf{fermion}{i2,v2,o2}
            \fmf{fermion,foreground=.8white}{id,vd,od}
            \fmf{fermion}{in,vn,on}
            \fmffreeze
            \fmfdraw
            \fmfcmd{path a[];}
            \fmfcmd{a123 = unitsquare shifted (-1/2,-1/2) xscaled 2thick yscaled 1.4h
                             shifted .5[vloc(__v1),vloc(__vn)];}
            \fmfcmd{filldraw a123 withcolor black;}
          \end{fmfgraph*}}} &=
      \sum_{\sigma\in S_n} \frac{(-)^{\epsilon(\sigma)}}{n!} \cdot
      \parbox{17\unitlength}{%
        \fmfframe(4,4)(9,4){%
          \begin{fmfgraph*}(4,8)
            \fmfstraight
            \fmfleft{in,id,i2,i1}
            \fmfright{on,od,o2,o1}
            \fmfv{label=$\scriptstyle 1$,label.angle=180}{i1}
            \fmfv{label=$\scriptstyle 2$,label.angle=180}{i2}
            \fmfv{label=$\scriptstyle n$,label.angle=180}{in}
            \fmfv{label=$\scriptstyle \sigma(1)$,label.angle=0}{o1}
            \fmfv{label=$\scriptstyle \sigma(2)$,label.angle=0}{o2}
            \fmfv{label=$\scriptstyle \sigma(n)$,label.angle=0}{on}
            \fmf{fermion}{i1,o1}
            \fmf{fermion}{i2,o2}
            \fmf{fermion,foreground=.8white}{id,od}
            \fmf{fermion}{in,on}
          \end{fmfgraph*}}}
  \end{align}
\end{subequations}
and the two parts of the symmetrizer for~$i_1$
and~$i_6$ in~\eqref{eq:27} are to be understood as glued together at the open boundary.

Denoting the combination of all (anti)symmetrizations and the
normalization factor corresponding to a Young tableau by a grey box,
the projection property can be verified by connecting the arrows
\begin{equation}
      \parbox{30\unitlength}{%
        \fmfframe(4,4)(4,4){%
          \begin{fmfgraph*}(22,8)
            \fmfstraight
            \fmfleft{in,id,i2,i1}
            \fmfright{on,od,o2,o1}
            \fmftop{a}
            \fmfv{label=$\scriptstyle 1$,label.angle=180}{i1}
            \fmfv{label=$\scriptstyle 2$,label.angle=180}{i2}
            \fmfv{label=$\scriptstyle n$,label.angle=180}{in}
            \fmfv{label=$\scriptstyle 1$,label.angle=0}{o1}
            \fmfv{label=$\scriptstyle 2$,label.angle=0}{o2}
            \fmfv{label=$\scriptstyle n$,label.angle=0}{on}
            \fmf{fermion}{i1,vi1,vd1,vo1,o1}
            \fmf{fermion}{i2,vi2,vd2,vo2,o2}
            \fmf{fermion}{in,vin,vdn,von,on}
            \fmf{fermion,foreground=.8white}{id,vid,vdd,vod,od}
            \fmffreeze
            \fmfdraw
            \fmfcmd{path s[];}
            \fmfcmd{s1 = unitsquare shifted (-1/2,-1/2) xscaled 2thick yscaled 1.4h
                             shifted .5[vloc(__vi1),vloc(__vin)];}
            \fmfcmd{s2 = unitsquare shifted (-1/2,-1/2) xscaled 2thick yscaled 1.4h
                             shifted .5[vloc(__vo1),vloc(__von)];}
            \fmfcmd{fill s1 withcolor .7white;}
            \fmfcmd{draw s1 withcolor black;}
            \fmfcmd{fill s2 withcolor .7white;}
            \fmfcmd{draw s2 withcolor black;}
          \end{fmfgraph*}}} =
      \parbox{18\unitlength}{%
        \fmfframe(4,4)(4,4){%
          \begin{fmfgraph*}(10,8)
            \fmfstraight
            \fmfleft{in,id,i2,i1}
            \fmfright{on,od,o2,o1}
            \fmftop{a}
            \fmfv{label=$\scriptstyle 1$,label.angle=180}{i1}
            \fmfv{label=$\scriptstyle 2$,label.angle=180}{i2}
            \fmfv{label=$\scriptstyle n$,label.angle=180}{in}
            \fmfv{label=$\scriptstyle 1$,label.angle=0}{o1}
            \fmfv{label=$\scriptstyle 2$,label.angle=0}{o2}
            \fmfv{label=$\scriptstyle n$,label.angle=0}{on}
            \fmf{fermion}{i1,v1,o1}
            \fmf{fermion}{i2,v2,o2}
            \fmf{fermion}{in,vn,on}
            \fmf{fermion,foreground=.8white}{id,vd,od}
            \fmffreeze
            \fmfdraw
            \fmfcmd{path s;}
            \fmfcmd{s = unitsquare shifted (-1/2,-1/2) xscaled 2thick yscaled 1.4h
                             shifted .5[vloc(__v1),vloc(__vn)];}
            \fmfcmd{fill s withcolor .7white;}
            \fmfcmd{draw s withcolor black;}
          \end{fmfgraph*}}}\,.
\end{equation}
When computing scattering amplitudes for a $\mathrm{SU}(N)$ gauge
theory, we use $\mathrm{U}(1)$-ghosts \cite{Kilian:2012pz}
both in internal propagators and in
in external states when computing color sums
\begin{subequations}
  \begin{align}
  \label{eq:U(N)-propagator}
      \parbox{24\unitlength}{%
        \fmfframe(4,2)(4,2){%
          \begin{fmfgraph*}(16,2)
            \fmfstraight
            \fmfleft{i}
            \fmfright{o}
            \fmfv{label=$a$}{i}
            \fmfv{label=$b$}{o}
            \fmffreeze
            \fmfcmd{path l[];}
            \fmfcmd{l0 = vloc(__i) -- vloc(__o);}
            \fmfcmd{l1 = l0 shifted (thick*down);}
            \fmfcmd{l2 = reverse l0 shifted (thick*up);}
            \fmfi{plain}{subpath (0,.3length(l1)) of l1}
            \fmfi{fermion}{subpath (.3length(l1),length(l1)) of l1}
            \fmfi{plain}{subpath (0,.3length(l2)) of l2}
            \fmfi{fermion}{subpath (.3length(l2),length(l2)) of l2}
          \end{fmfgraph*}}}
      &= \delta^{a_2}_{\phantom{a_2}b_1} \delta^{b_2}_{\phantom{b_2}a_1} \\
  \label{eq:U(1)-ghost-propagator}
      \parbox{24\unitlength}{%
        \fmfframe(4,2)(4,2){%
          \begin{fmfgraph*}(16,2)
            \fmfstraight
            \fmfleft{i}
            \fmfright{o}
            \fmfv{label=$a$}{i}
            \fmfv{label=$b$}{o}
            \fmf{dots}{i,o}
          \end{fmfgraph*}}}
      &= - \frac{1}{N}
  \end{align}
\end{subequations}
as in~\eqref{eq:U(1)-ghosts}.
Indices in the adjoint representation will be written as
single letters from the beginning of the latin alphabet, but they will
appear in calculations as pairs of indices from the fundamental and
conjugate representation~\eqref{eq:U(N)-propagator}.  In the case of
the singlet ghosts, the indices are only written for
illustration~\eqref{eq:U(1)-ghost-propagator}.  When constructing
colorflows representing invariant tensors, we do not have to keep
track of the $\mathrm{U}(1)$-ghosts, because they can be added at the
very end, as described in section~\ref{sec:U(1)-ghosts}.  One could
even ignore the ghosts altogether and use~\eqref{eq:TT->dd-SU(N)} for
the evaluation of color summed scattering amplitudes.

The representations of the generators are invariant tensors in the
product of the representation, its conjugate and the adjoint
representation generators~$\T{a}{i_1i_2\cdots i_n}{j_1j_2\cdots j_n}$.
They are written as
\begin{equation}
\label{eq:Ta}
  \left\{
  \sum_{i=1}^n
      \parbox{24\unitlength}{%
        \fmfframe(4,4)(4,4){%
          \begin{fmfgraph*}(16,16)
            \fmfstraight
            \fmfleft{in,id1,ii,id2,i1,id3,id4,id5}
            \fmfright{on,od1,oi,od2,o1,od3,od4,od5}
            \fmftop{a}
            \fmfv{label=$\scriptstyle 1$,label.angle=180}{i1}
            \fmfv{label=$\scriptstyle i$,label.angle=180}{ii}
            \fmfv{label=$\scriptstyle n$,label.angle=180}{in}
            \fmfv{label=$\scriptstyle 1$,label.angle=0}{o1}
            \fmfv{label=$\scriptstyle i$,label.angle=0}{oi}
            \fmfv{label=$\scriptstyle n$,label.angle=0}{on}
            \fmfv{label=$a$}{a}
            \fmf{fermion}{in,on}
            \fmf{fermion,foreground=.8white}{id1,od1}
            \fmffreeze
            \fmfdraw
            \fmfi{fermion}{vloc(__ii){right}
                    .. tension .9 .. {up}(vloc(__a) shifted (thick*left))}
            \fmfi{fermion}{(vloc(__a) shifted (thick*right)){down}
                    .. tension .9 .. {right}(vloc(__oi))}
            \fmfcmd{path l[], s[];}
            \fmfcmd{l1 = vloc(__i1) -- vloc(__o1);}
            \fmfcmd{l2 = vloc(__id2) -- vloc(__od2);}
            \fmfi{plain}{subpath (0,.36length(l1)) of l1}
            \fmfi{fermion}{subpath (.44length(l1),.56length(l1)) of l1}
            \fmfi{plain}{subpath (.64length(l1),length(l1)) of l1}
            \fmfi{plain,foreground=.8white}{subpath (0,.29length(l2)) of l2}
            \fmfi{fermion,foreground=.8white}{subpath (.39length(l2),.61length(l2)) of l2}
            \fmfi{plain,foreground=.8white}{subpath (.71length(l2),length(l2)) of l2}
            \fmfcmd{s1 = unitsquare shifted (-1/2,-1/2) xscaled 2thick yscaled .8h
                             shifted vloc(__ii) shifted (.2w*right);}
            \fmfcmd{s2 = unitsquare shifted (-1/2,-1/2) xscaled 2thick yscaled .8h
                             shifted vloc(__ii) shifted (.8w*right);}
            \fmfcmd{fill s1 withcolor .7white;}
            \fmfcmd{draw s1 withcolor black;}
            \fmfcmd{fill s2 withcolor .7white;}
            \fmfcmd{draw s2 withcolor black;}
          \end{fmfgraph*}}}\,,\; n\cdot
      \parbox{18\unitlength}{%
        \fmfframe(4,4)(4,4){%
          \begin{fmfgraph*}(10,16)
            \fmfstraight
            \fmfleft{in,id1,ii,id2,i1,id3,id4,id5}
            \fmfright{on,od1,oi,od2,o1,od3,od4,od5}
            \fmftop{a}
            \fmfv{label=$\scriptstyle 1$,label.angle=180}{i1}
            \fmfv{label=$\scriptstyle i$,label.angle=180}{ii}
            \fmfv{label=$\scriptstyle n$,label.angle=180}{in}
            \fmfv{label=$\scriptstyle 1$,label.angle=0}{o1}
            \fmfv{label=$\scriptstyle i$,label.angle=0}{oi}
            \fmfv{label=$\scriptstyle n$,label.angle=0}{on}
            \fmfv{label=$a$}{a}
            \fmf{fermion}{i1,v1,o1}
            \fmf{fermion}{ii,vi,oi}
            \fmf{fermion}{in,vn,on}
            \fmf{fermion,foreground=.8white}{id1,vd1,od1}
            \fmf{fermion,foreground=.8white}{id2,vd2,od2}
            \fmffreeze
            \fmf{dots}{a,va}
            \fmf{phantom}{va,v1}
            \fmfdot{va}
            \fmffreeze
            \fmfdraw
            \fmfcmd{path s;}
            \fmfcmd{s = unitsquare shifted (-1/2,-1/2) xscaled 2thick yscaled .8h
                             shifted vloc(__vi);}
            \fmfcmd{fill s withcolor .7white;}
            \fmfcmd{draw s withcolor black;}
        \end{fmfgraph*}}}\right\}\,.
\end{equation}
The commutator relation
\begin{equation}
\label{eq:commutator}
  \T{a}{i_1\cdots i_n}{j_1\cdots j_n}
  \T{b}{j_1\cdots j_n}{k_1\cdots k_n}
  -
  \T{b}{i_1\cdots i_n}{j_1\cdots j_n}
  \T{a}{j_1\cdots j_n}{k_1\cdots k_n}
  = \ii f^{abc} \T{c}{i_1\cdots i_n}{k_1\cdots k_n}
\end{equation}
can be checked explicitely for any
irrep considered here.  The coupling to
the $\mathrm{U}(1)$-ghost generalizes~\eqref{eq:U(1)-ghosts}. It
drops out of~\eqref{eq:commutator},
but is is required to make the generator traceless and the
coefficient~$n$ is determined by
\begin{equation}
  \T{a}{i_1\cdots i_n}{i_1\cdots i_n}
  \T{a}{j_1\cdots j_n}{j_1\cdots j_n} = 0\,.
\end{equation}
Note that, in the special cases of totally symmetric or antisymmetric
states, the sum over~$i$ in~\eqref{eq:Ta}
is equivalent to diverting only a single line
to the adjoint index and multiplying the result by~$n$, but this
shortcut is not available for mixed symmetries.

The totally antisymmetric rank-$n$ tensors~$\epsilon^{i_1i_2\cdots i_n}$
and~$\overline\epsilon_{j_1j_2\cdots j_n}$ are invariant
in~$\mathrm{SU}(N)$ iff~$n=N$.  We can represent them as
\begin{subequations}
  \begin{align}
      \epsilon^{i_1i_2\cdots i_N} &=
      \parbox{14\unitlength}{%
        \fmfframe(4,4)(4,4){%
          \begin{fmfgraph*}(6,8)
            \fmfstraight
            \fmfleft{in,id,i2,i1}
            \fmfright{on,od,o2,o1}
            \fmfv{label=$\scriptstyle 1$,label.angle=0}{o1}
            \fmfv{label=$\scriptstyle 2$,label.angle=0}{o2}
            \fmfv{label=$\scriptstyle N$,label.angle=0}{on}
            \fmf{fermion}{i1,o1}
            \fmf{fermion}{i2,o2}
            \fmf{fermion,foreground=.8white}{id,od}
            \fmf{fermion}{in,on}
            \fmffreeze
            \fmfdraw
            \fmfcmd{path s[];}
            \fmfcmd{s123 = unitsquare shifted (-1/2,-1/2) xscaled 2thick yscaled 1.4h
                             shifted .5[vloc(__i1),vloc(__in)];}
            \fmfcmd{filldraw s123 withcolor black;}
          \end{fmfgraph*}}} \\
      \overline\epsilon_{i_1i_2\cdots i_N} &=
      \parbox{14\unitlength}{%
        \fmfframe(4,4)(4,4){%
          \begin{fmfgraph*}(6,8)
            \fmfstraight
            \fmfleft{in,id,i2,i1}
            \fmfright{on,od,o2,o1}
            \fmfv{label=$\scriptstyle 1$,label.angle=180}{i1}
            \fmfv{label=$\scriptstyle 2$,label.angle=180}{i2}
            \fmfv{label=$\scriptstyle N$,label.angle=180}{in}
            \fmf{fermion}{i1,o1}
            \fmf{fermion}{i2,o2}
            \fmf{fermion,foreground=.8white}{id,od}
            \fmf{fermion}{in,on}
            \fmffreeze
            \fmfdraw
            \fmfcmd{path s[];}
            \fmfcmd{s123 = unitsquare shifted (-1/2,-1/2) xscaled 2thick yscaled 1.4h
                             shifted .5[vloc(__o1),vloc(__on)];}
            \fmfcmd{filldraw s123 withcolor black;}
          \end{fmfgraph*}}}\,.
  \end{align}
\end{subequations}
This has the consequence that the number of outgoing lines need not be the same
as the number of incoming lines
\begin{equation}
  \#_{\text{outgoing}} = \#_{\text{incoming}} \mod N\,.
\end{equation}
In the following I will only consider the case of~$\epsilon^{ijk}$
and~$\overline\epsilon_{ijk}$ in $\mathrm{SU}(3)$.

For a systematic approach to the construction of a basis of operators
involving exotic colorflows in the colorflow representation, we can
start with double lines only for the adjoint representation. In a second
step, we add systematically
$\mathrm{U}(1)$-ghosts~\cite{Kilian:2012pz} as in~\eqref{eq:Ta}
to obtain the~$\mathrm{SU}(N)$ colorflows with traceless generators
(see section~\ref{sec:U(1)-ghosts} for non-trivial examples).

\subsection{An Example Involving Sextets and Octets}
\label{sec:8866bar}
The example that has motivated the present paper is the search for
invariant tensors in the tensor
product~$\mathbf{8}\otimes\mathbf{8}\otimes\mathbf{6}\otimes\mathbf{\overline6}$
of~$\mathrm{SU}(3)$ irreps.  In order to make contact to the notation
used in~\cite{Carpenter:2021rkl}, I will
use the correspondence
\begin{equation}
  W^{ab\;s}_{\phantom{ab\;s}t} = W^{a_1b_1s_1s_2}_{a_2b_2t_1t_2}
\end{equation}
where~$a,b=1,\ldots,N^2-1$, $s,t=1,\ldots,N(N+1)/2$,
$a_i,b_i,s_i,t_i=1,\ldots,N$ and~$W$ is symmetric under the
separate exchanges~$s_1\leftrightarrow s_2$ and~$t_1\leftrightarrow t_2$
for the tensors in this product.
For example
\begin{equation}
  \delta^{s}_{\phantom{s}t} =
    \frac{1}{2!}
     \left( \delta^{s_1}_{t_1}\delta^{s_2}_{t_2} +
              +  \delta^{s_1}_{t_2}\delta^{s_2}_{t_1} \right)
\end{equation}
and analogously for the generators as tensors
in~$\mathbf{8}\otimes\mathbf{6}\otimes\mathbf{\overline6}$.

There are only four inequivalent ways
to connect $\mathbf{8}$, $\mathbf{8}$, $\mathbf{6}$ and
$\mathbf{\overline6}$ of~$\mathrm{SU}(3)$, two of which are related
be exchanging the factors of~$\mathbf{8}$. Starting from the
$\mathbf{6}$ we have the possibility to
\begin{enumerate}
  \item connect both lines to the $\mathbf{\overline6}$
    \begin{equation}
    \label{eq:cf-Xab}
      \parbox{32\unitlength}{%
        \fmfframe(3,4)(3,4){%
          \begin{fmfgraph*}(26,12)
            \fmfleft{k}
            \fmfright{kbar}
            \fmftop{ta}
            \fmfbottom{tb}
            \fmflabel{$\mathbf{8}$}{ta}
            \fmflabel{$\mathbf{8}$}{tb}
            \fmflabel{$\mathbf{6}$}{k}
            \fmflabel{$\mathbf{\overline 6}$}{kbar}
            \fmf{phantom}{k,vb,vd,va,kbar}
            \fmffreeze
            \fmfcmd{path kkbar[], tbta[];}
            \fmfcmd{kkbar1 = (vloc(__k) shifted (thick*up))
                  -- (vloc(__kbar) shifted (thick*up));}
            \fmfcmd{kkbar2 = (vloc(__k) shifted (thick*down))
                  -- (vloc(__kbar) shifted (thick*down));}
            \fmfcmd{tbta1 = (vloc(__tb) shifted (thick*left)) -- (vloc(__ta) shifted (thick*left));}
            \fmfcmd{tbta2 = (vloc(__ta) shifted (thick*right)) -- (vloc(__tb) shifted (thick*right));}
            \fmfi{fermion}{subpath (0,.85length(kkbar1)/2) of kkbar1}
            \fmfi{fermion}{subpath (1.15length(kkbar1)/2,length(kkbar1)) of kkbar1}
            \fmfi{fermion}{subpath (0,.85length(kkbar2)/2) of kkbar2}
            \fmfi{fermion}{subpath (1.15length(kkbar2)/2,length(kkbar2)) of kkbar2}
            \fmfi{fermion}{subpath (0,.7length(tbta1)/2) of tbta1}
            \fmfi{fermion}{subpath (.7length(tbta1)/2,length(tbta1)) of tbta1}
            \fmfi{fermion}{subpath (0,.79length(tbta2)/2) of tbta2}
            \fmfi{fermion}{subpath (.7length(tbta2)/2,length(tbta2)) of tbta2}
            \fmfdraw
            \fmfcmd{path s[];}
            \fmfcmd{s1 = unitsquare shifted (-1/2,-1/2) xscaled 2thick yscaled 4thick
                             shifted vloc(__k) shifted (2thick*right);}
            \fmfcmd{s2 = unitsquare shifted (-1/2,-1/2) xscaled 2thick yscaled 4thick
                             shifted vloc(__kbar) shifted (2thick*left);}
            \fmfcmd{fill s1 withcolor white;}
            \fmfcmd{draw s1 withcolor black;}
            \fmfcmd{fill s2 withcolor white;}
            \fmfcmd{draw s2 withcolor black;}
          \end{fmfgraph*}}}
    \end{equation}
    producing the tensor
    \begin{equation}
    \label{eq:Xab}
      [X^{ab}]^{s}_{\phantom{s}t} = \delta^{ab}\delta^{s}_{\phantom{s}t}
    \end{equation}
    after symmetrizing in the~$\mathbf{6}$ and~$\mathbf{\overline6}$ indices,
  \item connect one line to one~$\mathbf{8}$ and one to the other~$\mathbf{8}$
    \begin{equation}
    \label{eq:cf-Yab}
      \parbox{32\unitlength}{%
        \fmfframe(3,4)(3,4){%
          \begin{fmfgraph*}(26,12)
            \fmfleft{k}
            \fmfright{kbar}
            \fmftop{da1,ta,da2}
            \fmfbottom{db1,tb,db2}
            \fmflabel{$\mathbf{8}$}{ta}
            \fmflabel{$\mathbf{8}$}{tb}
            \fmflabel{$\mathbf{6}$}{k}
            \fmflabel{$\mathbf{\overline6}$}{kbar}
            \fmffreeze
            \fmfi{fermion}{(vloc(__k) shifted (thick*up)){right}
                              .. tension 3/4 .. {up}(vloc(__ta) shifted (thick*left))}
            \fmfi{fermion}{(vloc(__k) shifted (thick*down)){right}
                              .. tension 3/4 .. {down}(vloc(__tb) shifted (thick*left))}
            \fmfi{fermion}{(vloc(__ta) shifted (thick*right)){down}
                              .. tension 3/4 .. {right}(vloc(__kbar) shifted (thick*up))}
            \fmfi{fermion}{(vloc(__tb) shifted (thick*right)){up}
                              .. tension 3/4 .. {right}(vloc(__kbar) shifted (thick*down))}
            \fmfdraw
            \fmfcmd{path s[];}
            \fmfcmd{s1 = unitsquare shifted (-1/2,-1/2) xscaled 2thick yscaled 4thick
                             shifted vloc(__k) shifted (2thick*right);}
            \fmfcmd{s2 = unitsquare shifted (-1/2,-1/2) xscaled 2thick yscaled 4thick
                             shifted vloc(__kbar) shifted (2thick*left);}
            \fmfcmd{fill s1 withcolor white;}
            \fmfcmd{draw s1 withcolor black;}
            \fmfcmd{fill s2 withcolor white;}
            \fmfcmd{draw s2 withcolor black;}
          \end{fmfgraph*}}}
    \end{equation}
    producing the tensor\footnote{In the notation
    of~\cite{Carpenter:2021rkl}
    \begin{equation*}
      H^{n\;ab} = \T{a}{i}{k} \T{b}{j}{l},\;
      F^{n\;s}_{\phantom{n\;s}t} = \Ksixbar{s}{k}{l} \Ksix{t}{i}{j}\,,
    \end{equation*}
    where $n$ combines the indices~$i$, $j$, $k$, $l$, as in
    $\mathbf{27}\subset\mathbf{\overline3}\otimes\mathbf{\overline3}\otimes\mathbf{3}\otimes\mathbf{3}$.}
    \begin{equation}
      [Y^{ab}]^{s}_{\phantom{s}t} =
        \Ksixbar{s}{i}{j} \T{a}{i}{k} \T{b}{j}{l} \Ksix{t}{k}{l}
    \end{equation}
    which is symmetric in the exchange $a\leftrightarrow b$ since the
    tensors~$\Ksixbar{s}{i}{j}$ and~$\Ksix{t}{k}{l}$ are symmetric in
    both index pairs~$(i,j)$ and~$(k,l)$, or

  \item connect one line to the~$\mathbf{\overline6}$ and the other to
    one of the~$\mathbf{8}$s
    \begin{equation}
    \label{eq:cf-Zab}
      \parbox{32\unitlength}{%
        \fmfframe(3,4)(3,4){%
          \begin{fmfgraph*}(26,12)
            \fmfleft{k}
            \fmfright{kbar}
            \fmftop{ta}
            \fmfbottom{tb}
            \fmflabel{$\mathbf{8}$}{ta}
            \fmflabel{$\mathbf{8}$}{tb}
            \fmflabel{$\mathbf{6}$}{k}
            \fmflabel{$\mathbf{\overline 6}$}{kbar}
            \fmf{phantom}{k,vb,vd,va,kbar}
            \fmffreeze
            \fmfcmd{path kkbar, tbta;}
            \fmfcmd{kkbar = (vloc(__k) shifted (thick*up))
                  -- (vloc(__vb) shifted (thick*up)){right}
                  .. {right}(vloc(__va) shifted (thick*down))
                  -- (vloc(__kbar) shifted (thick*down));}
            \fmfcmd{tbta = (vloc(__tb) shifted (thick*right)){up}
                  .. tension 3/4 .. {up}(vloc(__ta) shifted (thick*left));}
            \fmfi{fermion}{subpath (0,length(kkbar)/2) of kkbar}
            \fmfi{fermion}{subpath (length(kkbar)/2,length(kkbar)) of kkbar}
            \fmfi{fermion}{(vloc(__ta) shifted (thick*right)){down}
                  .. tension 3/4 .. {right}(vloc(__kbar) shifted (thick*up))}
            \fmfi{fermion}{(vloc(__k) shifted (thick*down)){right}
                  .. tension 3/4 .. {down}(vloc(__tb) shifted (thick*left))}
            \fmfi{fermion}{subpath (0,0.9length(tbta)/2) of tbta}
            \fmfi{fermion}{subpath (1.1length(tbta)/2,length(tbta)) of tbta}
            \fmfdraw
            \fmfcmd{path s[];}
            \fmfcmd{s1 = unitsquare shifted (-1/2,-1/2) xscaled 2thick yscaled 4thick
                             shifted vloc(__k) shifted (2thick*right);}
            \fmfcmd{s2 = unitsquare shifted (-1/2,-1/2) xscaled 2thick yscaled 4thick
                             shifted vloc(__kbar) shifted (2thick*left);}
            \fmfcmd{fill s1 withcolor white;}
            \fmfcmd{draw s1 withcolor black;}
            \fmfcmd{fill s2 withcolor white;}
            \fmfcmd{draw s2 withcolor black;}
          \end{fmfgraph*}}}
    \end{equation}
    producing the tensor
    \begin{equation}
    \label{eq:Zab}
      [Z^{ab}]^{s}_{\phantom{s}t}
         = \Tsix{a}{s}{u} \Tsix{b}{u}{t} - 2 [Y^{ab}]^{s}_{\phantom{s}t}
    \end{equation}
    from which we can form two
    combinations,\footnote{Here~\cite{Carpenter:2021rkl}
    lists only the antisymmetric commutator
    \begin{equation*}
      \left[T_{\mathbf{6}}^a,T_{\mathbf{6}}^b\right]=\ii f^{abc}T_{\mathbf{6}}^c\,,
    \end{equation*}
    using $f^{abc}=\ii\Teight{a}{b}{c}$, and not the symmetric
    anticommutator.} symmetric and
    antisymmetric in the exchange $a\leftrightarrow b$
    after symmetrizing in the~$\mathbf{6}$ and~$\mathbf{\overline6}$ indices.
    Note that the line connecting the two~$\mathbf{8}$s is produced by the
    symmetrization between the factors in the products
    $T_{\mathbf{6}}^aT_{\mathbf{6}}^b$.
\end{enumerate}
All other connections are obtained from even permutations inside
the $\mathbf{6}$ and $\mathbf{\overline6}$.

Thus, there is a single colorflow~$Z_{\mathrm{A}}^{ab}=Z^{ab}-Z^{ba}$ that is
antisymmetric in the two $\mathbf{8}$s and there are three colorflows $X^{ab}$,
$Y^{ab}$ and~$Z_{\mathrm{S}}^{ab}=Z^{ab}+Z^{ba}$ that are symmetric in
the two $\mathbf{8}$s.
In section~\ref{sec:8866bar-SU(N)}, we will see that
one linear combination of the
symmetric flows vanishes in the special case of~$\mathrm{SU}(2)$ and
that they remain independent for~$\mathrm{SU}(N)$
with~$N\ge3$.

\section{Relations Among Colorflows}
\label{sec:SU(N)-constraints}

\subsection{$\mathrm{U}(1)$-Ghosts}
\label{sec:U(1)-ghosts}
In the approach of~\cite{Kilian:2012pz}, the
identity~\eqref{eq:TT->dd-SU(N)} is replaced by the introduction of
$\mathrm{U}(1)$-ghosts, as in~\eqref{eq:U(1)-ghosts}.
This corresponds to including colorflows in which all possible
subsets of the double lines representing an index in the adjoint
representation have been replaced by insertions of $\mathrm{U}(1)$-ghosts
\begin{equation}
\label{eq:ghost-subtraction}
  \parbox{22\unitlength}{%
    \fmfframe(3,4)(5,4){%
      \begin{fmfgraph*}(14,15)
        \fmfleft{b}
        \fmfright{a}
        \fmflabel{$a$}{a}
        \fmf{phantom}{a,v}
        \fmf{phantom,tension=6}{b,v}
        \fmffreeze
        \fmfcmd{path adj[];}
        \fmfcmd{adj1 = (vloc(__v) shifted (.3h*up)){up+right}
              .. tension 3/4 .. {right}(vloc(__a) shifted (thick*up));}
        \fmfcmd{adj2 = (vloc(__a) shifted (thick*down)){left}
              .. tension 3/4 .. {up+left}(vloc(__v) shifted (.3h*down));}
        \fmfi{fermion}{subpath (.1length(adj1),length(adj1)) of adj1}
        \fmfi{fermion}{subpath (0,.9length(adj2)) of adj2}
        \fmfcmd{fill fullcircle xscaled .4w yscaled h shifted vloc(__v) withcolor (.8,.8,.8);}
      \end{fmfgraph*}}} \longrightarrow\left\{
  \parbox{22\unitlength}{%
    \fmfframe(3,4)(5,4){%
      \begin{fmfgraph*}(14,15)
        \fmfleft{b}
        \fmfright{a}
        \fmflabel{$a$}{a}
        \fmf{phantom}{a,v}
        \fmf{phantom,tension=6}{b,v}
        \fmffreeze
        \fmfcmd{path adj[];}
        \fmfcmd{adj1 = (vloc(__v) shifted (.3h*up)){up+right}
              .. tension 3/4 .. {right}(vloc(__a) shifted (thick*up));}
        \fmfcmd{adj2 = (vloc(__a) shifted (thick*down)){left}
              .. tension 3/4 .. {up+left}(vloc(__v) shifted (.3h*down));}
        \fmfi{fermion}{subpath (.1length(adj1),length(adj1)) of adj1}
        \fmfi{fermion}{subpath (0,.9length(adj2)) of adj2}
        \fmfcmd{fill fullcircle xscaled .4w yscaled h shifted vloc(__v) withcolor (.8,.8,.8);}
      \end{fmfgraph*}}}\,,\;
  \parbox{22\unitlength}{%
    \fmfframe(3,4)(5,4){%
      \begin{fmfgraph*}(14,15)
        \fmfleft{b}
        \fmfright{a}
        \fmflabel{$a$}{a}
        \fmf{phantom}{a,v}
        \fmf{phantom,tension=8}{b,v}
        \fmffixed{.35w*right}{v,v'}
        \fmffixed{3thick*right}{v',v''}
        \fmf{dots}{a,v''}
        \fmfdot{v''}
        \fmffreeze
        \fmfcmd{path adj[];}
        \fmfcmd{adj1 = (vloc(__v) shifted (.3h*up)){up+right}
              ... vloc(__v')
              ... {up+left}(vloc(__v) shifted (.3h*down));}
        \fmfi{fermion}{adj1}
        \fmfcmd{fill fullcircle xscaled .4w yscaled h shifted vloc(__v) withcolor (.8,.8,.8);}
      \end{fmfgraph*}}}\right\}
\end{equation}
where I have represented the ghost by a dotted line and the rest of
the diagram by a grey blob.  A priori, this will replace each
colorflow containing $n$~external double lines by $2^n$~colorflows,
as in~\eqref{eq:V_abc} below.  Typically,
some of these will cancel after antisymmetrization, but
remain after symmetrization (see, e.g., \eqref{eq:f_abc}
and~\eqref{eq:d_abc}, below).
Note that, for the purpose of constructing inequivalent colorflows, the
substitution~\eqref{eq:ghost-subtraction} can be ignored until these
colorflows are used in the computation of matrix elements or of the
inner product~$\mu_N$~\eqref{eq:mu} using the diagrammatical
rule~\eqref{eq:U(1)-ghosts} instead of~\eqref{eq:TT->dd-SU(N)}.

As a non-trivial example which has already been discussed
in~\cite{Kilian:2012pz} in the context of the~$H\to ggg$ coupling,
consider the colorflow
\begin{equation}
  V_{abc} =
  \parbox{15\unitlength}{%
    \fmfframe(0,4)(5,4){%
      \begin{fmfgraph*}(10,12)
        \fmfleft{a,b}
        \fmfright{c}
        \fmflabel{$a$}{a}
        \fmflabel{$b$}{b}
        \fmflabel{$c$}{c}
        \fmf{phantom}{a,v}
        \fmf{phantom}{b,v}
        \fmf{phantom}{c,v}
        \fmffreeze
        \fmfcmd{path ab, bc, ca;}
        \fmfcmd{ab = (vloc(__a) shifted (thick*up)){up+right}
              .. tension 3/4 .. {up+left}(vloc(__b) shifted (thick*down));}
        \fmfcmd{bc = (vloc(__b) shifted (thick*right)){down+right}
              .. tension 3/4 .. {right}(vloc(__c) shifted (thick*up));}
        \fmfcmd{ca = (vloc(__c) shifted (thick*down)){left}
              .. tension 3/4 .. {down+left}(vloc(__a) shifted (thick*right));}
        \fmfi{plain}{subpath (0,length(ab)/2) of ab}
        \fmfi{fermion}{subpath (length(ab)/2,length(ab)) of ab}
        \fmfi{plain}{subpath (0,length(bc)/2) of bc}
        \fmfi{fermion}{subpath (length(bc)/2,length(bc)) of bc}
        \fmfi{plain}{subpath (0,length(ca)/2) of ca}
        \fmfi{fermion}{subpath (length(ca)/2,length(ca)) of ca}
      \end{fmfgraph*}}}
\end{equation}
coupling three adjoint representations.
Performing the substitutions~\eqref{eq:ghost-subtraction} for the three external states
results in $2^3=8$ colorflows
\begin{equation}
\label{eq:V_abc}
  V^{\mathrm{SU}}_{abc} =
  \left\{
  \parbox{19\unitlength}{%
    \fmfframe(4,4)(5,4){%
      \begin{fmfgraph*}(10,12)
        \fmfleft{a,b}
        \fmfright{c}
        \fmflabel{$a$}{a}
        \fmflabel{$b$}{b}
        \fmflabel{$c$}{c}
        \fmf{phantom}{a,v}
        \fmf{phantom}{b,v}
        \fmf{phantom}{c,v}
        \fmffreeze
        \fmfcmd{path ab, bc, ca;}
        \fmfcmd{ab = (vloc(__a) shifted (thick*up)){up+right}
              .. tension 3/4 .. {up+left}(vloc(__b) shifted (thick*down));}
        \fmfcmd{bc = (vloc(__b) shifted (thick*right)){down+right}
              .. tension 3/4 .. {right}(vloc(__c) shifted (thick*up));}
        \fmfcmd{ca = (vloc(__c) shifted (thick*down)){left}
              .. tension 3/4 .. {down+left}(vloc(__a) shifted (thick*right));}
        \fmfi{plain}{subpath (0,length(ab)/2) of ab}
        \fmfi{fermion}{subpath (length(ab)/2,length(ab)) of ab}
        \fmfi{plain}{subpath (0,length(bc)/2) of bc}
        \fmfi{fermion}{subpath (length(bc)/2,length(bc)) of bc}
        \fmfi{plain}{subpath (0,length(ca)/2) of ca}
        \fmfi{fermion}{subpath (length(ca)/2,length(ca)) of ca}
      \end{fmfgraph*}}}\,,\;
  \parbox{15\unitlength}{%
    \fmfframe(0,4)(5,4){%
      \begin{fmfgraph*}(10,12)
        \fmfleft{a,b}
        \fmfright{c}
        \fmflabel{$a$}{a}
        \fmflabel{$b$}{b}
        \fmflabel{$c$}{c}
        \fmf{phantom}{a,v}
        \fmf{phantom}{b,v}
        \fmf{phantom}{c,v}
        \fmffixed{3thick*right}{v,v'}
        \fmf{dots}{c,v'}
        \fmfdot{v'}
        \fmffreeze
        \fmfcmd{path ab, ba, vc;}
        \fmfcmd{ab = (vloc(__a) shifted (thick*up)){up+right}
              .. tension 3/4 .. {up+left}(vloc(__b) shifted (thick*down));}
        \fmfcmd{ba = (vloc(__b) shifted (thick*right)){down+right}
              .. tension 3/4 .. {down+left}(vloc(__a) shifted (thick*right));}
        \fmfcmd{vc = (vloc(__v) shifted (5thick*right)) -- vloc(__c);}
        \fmfi{plain}{subpath (0,length(ab)/2) of ab}
        \fmfi{fermion}{subpath (length(ab)/2,length(ab)) of ab}
        \fmfi{plain}{subpath (0,length(ba)/2) of ba}
        \fmfi{fermion}{subpath (length(ba)/2,length(ba)) of ba}
      \end{fmfgraph*}}}\,\text{and cyclic}\,,\;
  \parbox{15\unitlength}{%
    \fmfframe(0,4)(5,4){%
      \begin{fmfgraph*}(10,12)
        \fmfleft{a,b}
        \fmfright{c}
        \fmflabel{$a$}{a}
        \fmflabel{$b$}{b}
        \fmflabel{$c$}{c}
        \fmf{dots}{a,v}
        \fmf{dots}{b,v}
        \fmf{phantom}{c,v}
        \fmfdot{v}
        \fmffreeze
        \fmfcmd{path cc;}
        \fmfcmd{cc = (vloc(__c) shifted (thick*down))
              -- (vloc(__v) shifted (thick*(down+4right))){left}
              .. {right}(vloc(__v) shifted (thick*(up+4right)))
              -- (vloc(__c) shifted (thick*up));}
        \fmfi{plain}{subpath (0,2length(cc)/3) of cc}
        \fmfi{fermion}{subpath (2length(cc)/3,length(cc)) of cc}
      \end{fmfgraph*}}}\,\text{and cyclic}\,,\; N\cdot
  \parbox{16\unitlength}{%
    \fmfframe(-2,4)(5,4){%
      \begin{fmfgraph*}(13,15)
        \fmfleft{a,b}
        \fmfright{c}
        \fmflabel{$a$}{a}
        \fmflabel{$b$}{b}
        \fmflabel{$c$}{c}
        \fmf{dots}{a,v}
        \fmf{dots}{b,v}
        \fmf{dots}{c,v}
        \fmfdot{v}
      \end{fmfgraph*}}}\right\},
\end{equation}
where the cyclic permutations of~$(a,b,c)$ in the colorflows with one
or two ghosts have not been drawn separately.  The factor~$N$ in front
on the last colorflow arises from the closed loop remaining after
replacing the double line by the $\mathrm{U}(1)$-ghost on all external
states.  The antisymmetric combination of the~$V^{\mathrm{SU}}_{abc}$
corresponds to the structure constants of the $\mathrm{SU}(N)$~Lie algebra
\begin{equation}
\label{eq:f_abc}
  \ii f_{abc} = \tr \left(T_a\left[T_b,T_c\right]_-\right)
    = V^{\mathrm{SU}}_{abc} - V^{\mathrm{SU}}_{acb} \\ =
  \parbox{15\unitlength}{%
    \fmfframe(0,4)(5,4){%
      \begin{fmfgraph*}(10,12)
        \fmfleft{a,b}
        \fmfright{c}
        \fmflabel{$a$}{a}
        \fmflabel{$b$}{b}
        \fmflabel{$c$}{c}
        \fmf{phantom}{a,v}
        \fmf{phantom}{b,v}
        \fmf{phantom}{c,v}
        \fmffreeze
        \fmfcmd{path ab, bc, ca;}
        \fmfcmd{ab = (vloc(__a) shifted (thick*up)){up+right}
              .. tension 3/4 .. {up+left}(vloc(__b) shifted (thick*down));}
        \fmfcmd{bc = (vloc(__b) shifted (thick*right)){down+right}
              .. tension 3/4 .. {right}(vloc(__c) shifted (thick*up));}
        \fmfcmd{ca = (vloc(__c) shifted (thick*down)){left}
              .. tension 3/4 .. {down+left}(vloc(__a) shifted (thick*right));}
        \fmfi{plain}{subpath (0,length(ab)/2) of ab}
        \fmfi{fermion}{subpath (length(ab)/2,length(ab)) of ab}
        \fmfi{plain}{subpath (0,length(bc)/2) of bc}
        \fmfi{fermion}{subpath (length(bc)/2,length(bc)) of bc}
        \fmfi{plain}{subpath (0,length(ca)/2) of ca}
        \fmfi{fermion}{subpath (length(ca)/2,length(ca)) of ca}
      \end{fmfgraph*}}} -
  \parbox{15\unitlength}{%
    \fmfframe(0,4)(5,4){%
      \begin{fmfgraph*}(10,12)
        \fmfleft{a,b}
        \fmfright{c}
        \fmflabel{$a$}{a}
        \fmflabel{$b$}{b}
        \fmflabel{$c$}{c}
        \fmf{phantom}{a,v}
        \fmf{phantom}{b,v}
        \fmf{phantom}{c,v}
        \fmffreeze
        \fmfcmd{path ac, cb, ba;}
        \fmfcmd{ac = (vloc(__a) shifted (thick*up)){up+right}
              .. tension 3/4 .. {right}(vloc(__c) shifted (thick*up));}
        \fmfcmd{cb = (vloc(__c) shifted (thick*down)){left}
              .. tension 3/4 .. {left+up}(vloc(__b) shifted (thick*down));}
        \fmfcmd{ba = (vloc(__b) shifted (thick*right)){down+right}
              .. tension 3/4 .. {down+left}(vloc(__a) shifted (thick*right));}
        \fmfi{plain}{subpath (0,.45length(ac)/2) of ac}
        \fmfi{fermion}{subpath (1.1length(ac)/2,length(ac)) of ac}
        \fmfi{plain}{subpath (0,.3length(cb)/2) of cb}
        \fmfi{fermion}{subpath (.8length(cb)/2,length(cb)) of cb}
        \fmfi{plain}{subpath (0,.4length(ba)/2) of ba}
        \fmfi{fermion}{subpath (.9length(ba)/2,length(ba)) of ba}
      \end{fmfgraph*}}}\,,
\end{equation}
and all $\mathrm{U}(1)$-ghosts cancel, because they are symmetric.
In the symmetric combination, on the other hand, the
$\mathrm{U}(1)$-ghosts add up
\begin{multline}
\label{eq:d_abc}
  d_{abc} = \tr \left(T_a\left[T_b,T_c\right]_+\right)
    = V^{\mathrm{SU}}_{abc} + V^{\mathrm{SU}}_{acb} =
  \left\{
  \parbox{19\unitlength}{%
    \fmfframe(4,4)(5,4){%
      \begin{fmfgraph*}(10,12)
        \fmfleft{a,b}
        \fmfright{c}
        \fmflabel{$a$}{a}
        \fmflabel{$b$}{b}
        \fmflabel{$c$}{c}
        \fmf{phantom}{a,v}
        \fmf{phantom}{b,v}
        \fmf{phantom}{c,v}
        \fmffreeze
        \fmfcmd{path ab, bc, ca;}
        \fmfcmd{ab = (vloc(__a) shifted (thick*up)){up+right}
              .. tension 3/4 .. {up+left}(vloc(__b) shifted (thick*down));}
        \fmfcmd{bc = (vloc(__b) shifted (thick*right)){down+right}
              .. tension 3/4 .. {right}(vloc(__c) shifted (thick*up));}
        \fmfcmd{ca = (vloc(__c) shifted (thick*down)){left}
              .. tension 3/4 .. {down+left}(vloc(__a) shifted (thick*right));}
        \fmfi{plain}{subpath (0,length(ab)/2) of ab}
        \fmfi{fermion}{subpath (length(ab)/2,length(ab)) of ab}
        \fmfi{plain}{subpath (0,length(bc)/2) of bc}
        \fmfi{fermion}{subpath (length(bc)/2,length(bc)) of bc}
        \fmfi{plain}{subpath (0,length(ca)/2) of ca}
        \fmfi{fermion}{subpath (length(ca)/2,length(ca)) of ca}
      \end{fmfgraph*}}} +
  \parbox{15\unitlength}{%
    \fmfframe(0,4)(5,4){%
      \begin{fmfgraph*}(10,12)
        \fmfleft{a,b}
        \fmfright{c}
        \fmflabel{$a$}{a}
        \fmflabel{$b$}{b}
        \fmflabel{$c$}{c}
        \fmf{phantom}{a,v}
        \fmf{phantom}{b,v}
        \fmf{phantom}{c,v}
        \fmffreeze
        \fmfcmd{path ac, cb, ba;}
        \fmfcmd{ac = (vloc(__a) shifted (thick*up)){up+right}
              .. tension 3/4 .. {right}(vloc(__c) shifted (thick*up));}
        \fmfcmd{cb = (vloc(__c) shifted (thick*down)){left}
              .. tension 3/4 .. {left+up}(vloc(__b) shifted (thick*down));}
        \fmfcmd{ba = (vloc(__b) shifted (thick*right)){down+right}
              .. tension 3/4 .. {down+left}(vloc(__a) shifted (thick*right));}
        \fmfi{plain}{subpath (0,.45length(ac)/2) of ac}
        \fmfi{fermion}{subpath (1.1length(ac)/2,length(ac)) of ac}
        \fmfi{plain}{subpath (0,.3length(cb)/2) of cb}
        \fmfi{fermion}{subpath (.8length(cb)/2,length(cb)) of cb}
        \fmfi{plain}{subpath (0,.4length(ba)/2) of ba}
        \fmfi{fermion}{subpath (.9length(ba)/2,length(ba)) of ba}
      \end{fmfgraph*}}}\,,\right. \\ \left. 2\cdot
  \parbox{16\unitlength}{%
    \fmfframe(-2,4)(5,4){%
      \begin{fmfgraph*}(13,15)
        \fmfleft{a,b}
        \fmfright{c}
        \fmflabel{$a$}{a}
        \fmflabel{$b$}{b}
        \fmflabel{$c$}{c}
        \fmf{phantom}{a,v}
        \fmf{phantom}{b,v}
        \fmf{phantom}{c,v}
        \fmffixed{3thick*right}{v,v'}
        \fmf{dots}{c,v'}
        \fmfdot{v'}
        \fmffreeze
        \fmfcmd{path ab, ba, vc;}
        \fmfcmd{ab = (vloc(__a) shifted (thick*up)){up+right}
              .. tension 3/4 .. {up+left}(vloc(__b) shifted (thick*down));}
        \fmfcmd{ba = (vloc(__b) shifted (thick*right)){down+right}
              .. tension 3/4 .. {down+left}(vloc(__a) shifted (thick*right));}
        \fmfcmd{vc = (vloc(__v) shifted (5thick*right)) -- vloc(__c);}
        \fmfi{plain}{subpath (0,length(ab)/2) of ab}
        \fmfi{fermion}{subpath (length(ab)/2,length(ab)) of ab}
        \fmfi{plain}{subpath (0,length(ba)/2) of ba}
        \fmfi{fermion}{subpath (length(ba)/2,length(ba)) of ba}
      \end{fmfgraph*}}}\,\text{and cyclic}\,,\; 2\cdot
  \parbox{16\unitlength}{%
    \fmfframe(-2,4)(5,4){%
      \begin{fmfgraph*}(13,15)
        \fmfleft{a,b}
        \fmfright{c}
        \fmflabel{$a$}{a}
        \fmflabel{$b$}{b}
        \fmflabel{$c$}{c}
        \fmf{dots}{a,v}
        \fmf{dots}{b,v}
        \fmf{phantom}{c,v}
        \fmfdot{v}
        \fmffreeze
        \fmfcmd{path cc;}
        \fmfcmd{cc = (vloc(__c) shifted (thick*down))
              -- (vloc(__v) shifted (thick*(down+4right))){left}
              .. {right}(vloc(__v) shifted (thick*(up+4right)))
              -- (vloc(__c) shifted (thick*up));}
        \fmfi{plain}{subpath (0,2length(cc)/3) of cc}
        \fmfi{fermion}{subpath (2length(cc)/3,length(cc)) of cc}
      \end{fmfgraph*}}}\,\text{and cyclic}\,,\; 2N\cdot
  \parbox{16\unitlength}{%
    \fmfframe(-2,4)(5,4){%
      \begin{fmfgraph*}(13,15)
        \fmfleft{a,b}
        \fmfright{c}
        \fmflabel{$a$}{a}
        \fmflabel{$b$}{b}
        \fmflabel{$c$}{c}
        \fmf{dots}{a,v}
        \fmf{dots}{b,v}
        \fmf{dots}{c,v}
        \fmfdot{v}
      \end{fmfgraph*}}}\right\}\,.
\end{multline}

\subsection{Spurious Colorflows}
\label{sec:spurious}
The approach described in the previous subsection is straightforward for
the evaluation of color factors~\cite{Kilian:2012pz}, but in the
present application, special care must be taken to avoid counting
spurious colorflows.
Indeed, the expression~\eqref{eq:d_abc} does not appear to be correct
for the special case of~$\mathrm{SU}(2)$,
where~$d_{abc}=0$, as can be checked directly using the Pauli
matrices
\begin{equation}
 d^{\mathrm{SU}(2)}_{ijk}
   = \frac{1}{\sqrt{8}} \tr \left(\sigma_i\left[\sigma_j,\sigma_k\right]_+\right)
   = \frac{1}{\sqrt{2}} \tr \left(\sigma_i\right) \delta_{jk} = 0\,.
\end{equation}
Therefore, it appears that in the case of~$\mathrm{SU}(2)$, the
expression~\eqref{eq:d_abc} does \emph{not} represent an independent
invariant tensor, but is a complicated way of writing~$0$ instead. 

We can confirm this expectation by noticing that, up to
permutations of the indices~$a$, $b$ and~$c$,
$V_{abc}$ is the only possible colorflow for three
adjoint representations.  Thus, the symmetric and antisymmetric
combinations~$d_{abc}$ and~$f_{abc}$ form a complete set.
We can use the expressions~\eqref{eq:f_abc} and~\eqref{eq:d_abc} to
compute an inner product in the vector space spanned by~$d_{abc}$
and~$f_{abc}$ by computing color sums as in~\cite{Kilian:2012pz}
\begin{subequations}
  \begin{align}
    d^{\mathrm{SU}}_{abc} d^{\mathrm{SU}}_{abc} &= \frac{2(N^2-1)(N^2-4)}{N} \\
    d^{\mathrm{SU}}_{abc} f_{abc} &= 0 \\
    f_{abc} f_{abc} &= 2 N(N^2-1)\,.
  \end{align}
\end{subequations}
This result is consistent with~$d^{\mathrm{SU}(2)}_{ijk}=0$
and $f^{\mathrm{SU}(2)}_{ijk}=\sqrt{2}\,\epsilon^{ijk}$.

Therefore, we must be aware of the fact that a naive application of
the colorflow rules~\cite{Kilian:2012pz} for~$\mathrm{SU}(N)$ might
produce sums of colorflows that are, for special values of~$N$, just
a complicated way of writing~$0$ and don't enlarge the basis.  In
section~\ref{eq:eigenvectors}, I will describe a general algorithm
for finding such redundancies.

Of course, the same
results are obtained using~\eqref{eq:TT->dd-SU(N)} instead of
the $\mathrm{U}(1)$-ghosts~\eqref{eq:U(1)-ghosts}.

\subsection{Redundant $\epsilon$-Tensors}
\label{sec:epsilon}
In the case of matching dimension~$N=\delta_m^m$ and rank~$n$
of~$\epsilon$ and $\overline\epsilon$, the tensor algebra of
the $\delta_{i}^{j}$, $\epsilon^{i_1i_2\cdots i_n}$
and $\overline\epsilon_{j_1j_2\cdots j_n}$ is not freely generated.
Indeed, introducing the \emph{generalized Kronecker~$\delta$} symbol
\begin{multline}
\label{eq:generalized-delta}
   \delta^{i_1i_2\cdots i_n}_{j_1j_2\cdots j_n}
     = \sum_{\sigma\in S_n} (-1)^{\varepsilon(\sigma)}
         \delta^{i_1}_{\sigma(j_1)} 
         \delta^{i_2}_{\sigma(j_2)} 
         \cdots
         \delta^{i_n}_{\sigma(j_n)} \\
     = \sum_{\sigma\in S_n} (-1)^{\varepsilon(\sigma)}
         \delta^{\sigma(i_1)}_{j_1} 
         \delta^{\sigma(i_2)}_{j_2} 
         \cdots
         \delta^{\sigma(i_n)}_{j_n}
     = \det \begin{pmatrix}
         \delta^{i_1}_{j_1} & \delta^{i_1}_{j_2} & \cdots & \delta^{i_1}_{j_n} \\
         \delta^{i_2}_{j_1} & \delta^{i_2}_{j_2} & \cdots & \delta^{i_2}_{j_n} \\
         \vdots             & \vdots             & \ddots & \vdots             \\
         \delta^{i_n}_{j_1} & \delta^{i_n}_{j_2} & \cdots & \delta^{i_n}_{j_n}
       \end{pmatrix} \,,
\end{multline}
there is the relation~$\forall n=N\in\mathbf{N}$ with~$N\ge2$:
\begin{equation}
\label{eq:epsilon*epsilonbar-0}
   \epsilon^{i_1i_2\cdots i_n} \overline\epsilon_{j_1j_2\cdots j_n}
     = \delta^{i_1i_2\cdots i_n}_{j_1j_2\cdots j_n}\,,
\end{equation}
which follows from antisymmetry and the choice of normalization
$\epsilon^{12\cdots n} = 1 = \overline\epsilon_{12\cdots n}$ alone.
Contracting $k$ indices in the relation~\eqref{eq:epsilon*epsilonbar-0},
we find~$\forall k, n, N \in \mathbf{N}$ with $0 \le k \le n = N\ge2$:
\begin{equation}
\label{eq:epsilon*epsilonbar}
   \epsilon^{m_1\cdots m_ki_{k+1}\cdots i_n}
   \overline\epsilon_{m_1\cdots m_kj_{k+1}\cdots j_n}
     = k!\, \delta^{i_{k+1}i_{k+2}\cdots i_n}_{j_{k+1}j_{k+2}\cdots j_n}\,.
\end{equation}
Because the left hand side of~\eqref{eq:epsilon*epsilonbar-0} is the most
concise description of the $n!$~terms on the right hand side, it is
tempting to keep it in the basis.  On the other hand,
replacing the left hand side immediately by the right hand side is the
most symmetric evaluation rule possible and I will adopt it,
including the rules~\eqref{eq:epsilon*epsilonbar} obtained by
contracting pairs of indices.

\section{Enumerating Colorflows}
\label{sec:algorithm}

Having identified all the dependencies, I can now describe the
algorithm for constructing a basis for the invariant tensors in
products of irreps of~$\mathrm{SU}(N)$.

\subsection{Selection Rules}
Since all external states must be connected to the corresponding
number of incoming or outgoing colorflow lines,
not all products of irreps can contain invariant tensors.
We start by summing the number of boxes in the Young diagrams
corresponding to the irreps of particles and those of
antiparticles.  Each adjoint representation counts
as one box for a particle and one box for an antiparticle.
These sums correspond to the overall number of incoming and
outgoing lines, respectively.  They
can only differ by~$\nu N$ with~$\nu\in\mathbf{Z}$
for~$\mathrm{SU}(N)$.  Iff~$\nu<0$, the tensor contains exactly
$|\nu|$ factors of~$\epsilon^{i_1i_2\cdots i_N}$ and iff $\nu>0$,
there are $\nu$ factors of~$\overline\epsilon_{i_1i_2\cdots i_N}$.
According to the conventions described in section~\ref{sec:epsilon},
$\epsilon^{i_1i_2\cdots i_N}$ and~$\overline\epsilon_{i_1i_2\cdots i_N}$
must not appear together in the same tensor.

For the example from section~\ref{sec:8866bar}, we have $1+1+2$ incoming
lines from~$\mathbf{8}$, $\mathbf{8}$ and ~$\mathbf{6}$, the same
number of outgoing lines from~$\mathbf{8}$, $\mathbf{8}$
and~$\mathbf{\overline{6}}$.  Therefore, no~$\epsilon$
or~$\overline{\epsilon}$ appear in this example.

\subsection{Combinatorics}
Having established the number of~$\epsilon$s or~$\overline\epsilon$s
required, we can proceed by drawing \emph{all} combinations of arrows starting at
a particle or at an~$\epsilon$ and ending at an antiparticle
or at an~$\overline\epsilon$.
The lines starting at the same particle or at the same~$\epsilon$ obey
symmetrization and antisymmetrization conditions specified by the Young
tableau describing the irrep.  Therefore there will be equivalent
colorflows that should not be counted more than once.  The same
applies to lines ending at the same antiparticle or~$\overline\epsilon$.

In principle, the procedure described in
section~\ref{eq:eigenvectors} will weed out all double counting.
In the worst case, the size of the matrices to be diagonalized in that
step can grow with a factorial of the number of all arrows.
Thus it is worthwhile to reduce the size of these matrices by keeping
only one representative of obviously equivalent color flows.
Therefore, we proceed as follows:
\begin{enumerate}
  \item Create a list~$S$ of starting points of lines
    (adjoints, products of fundamental representations and $\epsilon$s).
    Adjoints and fundamental representations are represented by a
    single integer~$n$
    identifying the external state. The factors in products of fundamental
    representations are represented by the integer~$n$ denoting the external state
    combined with a second
    integer~$i$ identifying the factor, i.e.~$(n,i)$.
    Analogously for each~$\epsilon$, but we must
    treat them as indistinguishable.
    In the example of section~\ref{sec:8866bar}, we
    have~$S=\{1,2,(3,1),(3,2)\}$ if the four external states
    in~\fourfields{8}{8}{6}{\overline{6}} are enumerated from~$1$
    to~$4$. 
  \item Create a corresponding list~$E$ of endpoints of lines
    (adjoints, products of conjugate representations and $\overline\epsilon$s).
    In the example of section~\ref{sec:8866bar}, we
    have~$E=\{1,2,(4,1),(4,2)\}$.
  \item Generate all the permutations of~$E$, i.e.~all one-to-one
    maps~$S\to E$. In the example of section~\ref{sec:8866bar}, there
    are~$4!$ maps:
    \begin{equation}\small
      \begin{bmatrix}
        1     &\to& 2     \\
        2     &\to& 1     \\
        (3,1) &\to& (4,1) \\
        (3,2) &\to& (4,2)
      \end{bmatrix},\,
      \begin{bmatrix}
        1     &\to& (4,1) \\
        2     &\to& (4,2) \\
        (3,1) &\to& 1     \\
        (3,2) &\to& 2 
      \end{bmatrix},\,
      \begin{bmatrix}
        1     &\to& 2     \\
        2     &\to& (4,1) \\
        (3,1) &\to& 1     \\
        (3,2) &\to& (4,2)
      \end{bmatrix},\,
      \begin{bmatrix}
        1     &\to& (4,1) \\
        2     &\to& 1     \\
        (3,1) &\to& 2     \\
        (3,2) &\to& (4,2)
      \end{bmatrix},\,\ldots
    \end{equation}
    where I have spelled out four representatives for~\eqref{eq:cf-Xab},
    \eqref{eq:cf-Yab} and the two permutations of~\eqref{eq:cf-Zab}.
  \item Drop all maps~$S\to E$ with at least one line looping back to
    the same state, e.g.
    \begin{equation}\small
      \begin{bmatrix}
        1     &\to& 1     \\
        2     &\to& (4,1) \\
        (3,1) &\to& (4,2) \\
        (3,2) &\to& 2
      \end{bmatrix}\,,
    \end{equation}
    because they do not correspond to valid~$\mathrm{SU}(N)$
    colorflows. In the example there are 10 of those and 14 remain.
  \item Keep one representative of the equivalence classes
    under the permutations according to the Young tableaux describing
    the irreps, i.e.~according to permutations of the
    subsets~$\{(n,i)\}_i$ of~$S$ and~$E$. One of these equivalence
    classes in the example is
    \begin{equation}\small
    \label{eq:Zab-orbit}
      \begin{bmatrix}
        1     &\to& 2     \\
        2     &\to& (4,1) \\
        (3,1) &\to& 1     \\
        (3,2) &\to& (4,2)
      \end{bmatrix}\sim
      \begin{bmatrix}
        1     &\to& 2     \\
        2     &\to& (4,2) \\
        (3,1) &\to& 1     \\
        (3,2) &\to& (4,1)
      \end{bmatrix}\sim
      \begin{bmatrix}
        1     &\to& 2     \\
        2     &\to& (4,1) \\
        (3,2) &\to& 1     \\
        (3,1) &\to& (4,2)
      \end{bmatrix}\sim
      \begin{bmatrix}
        1     &\to& 2     \\
        2     &\to& (4,2) \\
        (3,2) &\to& 1     \\
        (3,1) &\to& (4,1)
      \end{bmatrix}\,.
    \end{equation}
    This can be done by computing the orbits of the
    permutations described by the Young tableau for the irrep of each
    external state. In the example,
    the orbits containing~\eqref{eq:cf-Xab}, \eqref{eq:cf-Yab}
    and one of~\eqref{eq:cf-Zab} consist of~2, 4 and 4~maps respectively,
    adding up to $2+4+2\cdot4=14$, as required. If necessary, this
    process can be sped up by restricting the permutations generated
    in step~3 to one representative of these orbits.
  \item Optionally symmetrize and antisymmetrize with respect to
    permutations of external states transforming under the same irrep
    of~$\mathrm{SU}(N)$.
  \item Apply the Young projection operators for all the factors.  In
    the example, only irreps are symmetric and the resulting
    colorflow~$Z^{ab}$~\eqref{eq:cf-Zab} is just
    the sum of the four maps in~\eqref{eq:Zab-orbit}.  This does not
    determine the overall normalization, which can be chosen to ensure that
    only integers appear as coefficients and to minimize the number of minus signs in
    the the case of antisymmetric and mixed irreps.
\end{enumerate}
Due to the subsequent test for redundancy described in
section~\ref{eq:eigenvectors}, it is less important to avoid
accidental double counting than it is to produce all colorflows.  In
particular, step~5 could be skipped without affecting the final
result.  It just speeds up the subsequent search for independent
tensors, because it keeps the matrices used in
section~\ref{eq:eigenvectors} substantially smaller.  This implies
that the implementation of any optimization in steps~3 and~5 can be checked for moderately sized
irreps by verifying that the constructed sets of independent invariant
tensors are the same with and without including the optimization.

\subsection{Finding Dependent Tensors}
\label{eq:eigenvectors}

Since all terms in the sum~\eqref{eq:mu} for~$\mu_N(A,A)$ are the
squared modulus of a component of the tensor~$A$, it is positive
by construction. Therefore the sesquilinear form~$\mu_N$ induces an inner
product and a norm on the vector space~$\mathcal{V}$ of invariant
tensors of a given rank
\begin{subequations}
  \begin{align}
    \|A\|_N = \sqrt{\mu_N(A,A)} &\ge 0 \\
    \|A\|_N = 0\;\; &\Leftrightarrow \;\; A = 0
  \end{align}
\end{subequations}
and it is not degenerate
\begin{equation}
\label{eq:non-degenerate}
  \forall B\in\mathcal{V}: \mu_N(B,A) = 0\;\;\Longleftrightarrow\;\; A = 0\,.
\end{equation}
The form~$\mu_N$ can be employed to generalize a
calculation~\cite{Dittner:1971fy} for small products of adjoint
representations of~$\mathrm{SU}(3)$:
given a complete, but not necessarily linearly independent, set
of~$n\ge\dim(\mathcal{V})$ tensors
\begin{equation}
  \mathcal{T}=\{T^i\}_{i=1,\ldots,n}\subseteq\mathcal{V}\,,
\end{equation}
we can expand every
tensor~$A\in\mathcal{V}$ as
\begin{equation}
  A = \sum_{i=1}^n a_i T^i\,,
\end{equation}
although this expansion will not be unique, in general.  The inner product
\begin{equation}
  \mu_N(A,B)
             = \left\langle a, M(N,\mathcal{T}) b\right\rangle
\end{equation}
can then be expressed by the natural sesquilinear form
\begin{equation}
  \left\langle a, b\right\rangle = \sum_{i=1}^n \overline{a_i} b_i
\end{equation}
on~$\mathbf{C}^n$ and the self-adjoint matrix~$M(N,\mathcal{T})$
\begin{equation}
\label{eq:colorfactor-matrix}
  M^{ij}(N,\mathcal{T}) = \mu_N(T^i,T^j) = \overline{M^{ji}(N,\mathcal{T})}
\end{equation}
which depends on the number of colors~$N$ and the set~$\mathcal{T}$.
It can be computed either by using the
identity~\eqref{eq:TT->dd-SU(N)} or by adding $\mathrm{U}(1)$~ghosts
as described in section~\ref{sec:U(1)-ghosts}.
The condition~\eqref{eq:non-degenerate} now reads
\begin{equation}
  \forall i: \sum_{j=1}^n M^{ij}(N,\mathcal{T})\,a_j = 0\;\;\Longleftrightarrow\;\;A = 0
\end{equation}
and we find that the linear relations among elements of~$\mathcal{T}$
are just the eigenvectors of the matrix~$M(N,\mathcal{T})$
corresponding to vanishing eigenvalues.
Conversely, the number of
independent invariant tensors is given by the rank~$r_N$ of the
matrix~$M(N,\mathcal{T})$.  The rank~$r_N$ is independent of the
set~$\mathcal{T}$ of invariant tensors used to
compute~$M(N,\mathcal{T})$, as long as it is complete.  The orthogonal
projector~$\mathcal{P}_N(\mathcal{T})$ on the subspace of~$\mathbf{C}^n$ spanned
by the eigenvectors corresponding to positive eigenvalues depends
on~$\mathcal{T}$, but the orthogonal projector~$P_N$ on the
corresponding subspace of~$\mathcal{V}$ does not.

Since~$M(N,\mathcal{T})$ is a finite and self-adjoint $n\times n$-matrix,
it is always possible to
compute~$r_N$ and~$P_N$ for any chosen value of $N$.  This
task is simplified by the observation that~$\mu_N(A,B)=0$, if~$A$
and~$B$ have different symmetries under
permutations of the factors in the tensor product.
Thus the matrix~$M(N,\mathcal{T})$
assumes a block diagonal form, if the elements of~$\mathcal{T}$ are
chosen to be symmetric or antisymmetric under permutations of the
factors. In the colorflow basis, the matrix elements
of~$M(N,\mathcal{T})$ will be polynomials in~$N$ with real
coefficients, possibly multiplied by a negative power of~$N$.
$M(N,\mathcal{T})$ will also be symmetric because
transposition corresponds to reversing all colorflow lines.

There is the option to
construct a basis of invariant tensors that are mutually orthogonal
with respect to~$\mu_N$.  Unfortunately, except for the simplest
cases, the real eigenvalues and eigenvectors can only be computed after
fixing a value for~$N$.  The resulting real numbers are then not
very illuminating.  Therefore, one should rather use~$P_N$
only to eliminate dependent tensors and to choose a linearly independent
set~$\{T^i\}_{i=1,\ldots,r_N}$ that is calculationally convenient, but
not necessarily orthonormal with respect to~$\mu_N$.

\subsubsection{Exceptional Values of~$N$}
\label{sec:N-dependence}

The identity~\eqref{eq:TT->dd-SU(N)} or the
rule~\eqref{eq:U(1)-ghosts} guarantee that all matrix
elements~$M^{ij}(N,\mathcal{T})$ are polynomials in~$N$, possibly
multiplied by~$N^{-k}$ with~$k$ a small natural number.  Thus the
characteristic polynomial has the form
\begin{equation}
      \det \left(M(N,\mathcal{T}) - \lambda \mathbf{1} \right)
    = N^{-k} \sum_{i=0}^{d} p_i(N)\, \lambda^i
    = N^{-k} \sum_{i=c_N+1}^{d} p_i(N)\, \lambda^i
\end{equation}
with polynomials~$\{p_i\}_{i=0,\ldots,d}$ in~$N$ as coefficients
and~$d$ the dimension of the matrix~$M(N,\mathcal{T})$.
The corank~$c_N$ of the
matrix~$M(N,\mathcal{T})$, i.e.~the number of eigenvectors with vanishing
eigenvalue, is the multiplicity of the root of the characteristic
polynomial at~$\lambda=0$.  For a given~$N$, this is the number of
consecutive~$p_i(N)$ starting from~$p_0(N)$ that vanish simultaneously
\begin{equation}
  c_N = \max_{i} \left\{\forall j\le i: p_j(N) = 0 \right\}\,.
\end{equation}
As a polynomial in~$N$, $p_i$~either vanishes for all~$N$ or has at
most~$\deg(p_i)$ positive real roots, where~$\deg(p)$ denotes the
degree of~$p$.  Thus there can be at most a finite number of
exceptional values of~$N$, where the rank and corank
of~$M(N,\mathcal{T})$ are not constant and additional relations among
invariant tensors appear. In particular
\begin{equation}
\label{eq:r-infty}
  \exists r_\infty, \hat N: \forall N>\hat N: r_{N}=r_\infty\,,
\end{equation}
i.e.~there is a maximum~$N$ above which the rank~$r_N$ no longer changes.

\section{Revisiting the Example}
\label{sec:8866bar-revisited}
We can now return to the example of the four-fold product
\fourfields{8}{8}{6}{\overline{6}} from section~\ref{sec:8866bar} and
compute the matrix~$M(N,\mathcal{T})$ for the colorflows
\begin{equation}
  \mathcal{T} = \left\{ X, Y, Z_{\mathrm{S}}, Z_{\mathrm{A}} \right\}\,.
\end{equation}

\subsection{$\mathrm{SU}(N)$}
\label{sec:8866bar-SU(N)}
We obtain for the
colorflows~$\mathcal{T}_{\mathrm{S}} = \{ X, Y, Z_{\mathrm{S}}\}$
that are symmetric in the two adjoint factors 
\begin{subequations}
  \begin{align}
    \mu_N(X,X) &= \frac{1}{2}N^2(N+1)C_{\mathrm{F}} \\
    \mu_N(Y,Y) &= \frac{1}{4}(N^3+2N^2-2)C_{\mathrm{F}} \\
    \mu_N(Z_{\mathrm{S}},Z_{\mathrm{S}}) &= \frac{1}{2}(N^3+2N^2-4)C_{\mathrm{F}} \\
    \mu_N(X,Y) = \mu_N(Y,X) &= \frac{1}{2}NC_{\mathrm{F}} \\
    \mu_N(X,Z_{\mathrm{S}}) = \mu_N(Z_{\mathrm{S}},X) &= N(N+1)C_{\mathrm{F}} \\
    \mu_N(Y,Z_{\mathrm{S}}) = \mu_N(Z_{\mathrm{S}},Y) &= \frac{1}{2}(N^2-2)C_{\mathrm{F}}
  \end{align}
\end{subequations}
and for the antisymmetric colorflow~$\mathcal{T}_{\mathrm{A}} =\{ Z_{\mathrm{A}} \}$
\begin{equation}
  \mu_N(Z_{\mathrm{A}},Z_{\mathrm{A}}) = \frac{1}{2}N^2(N+2)C_{\mathrm{F}}\,,
\end{equation}
where the quadratic Casimir operator
in the fundamental representation appears as a common factor.
It takes the value
\begin{equation}
\label{eq:C(F)}
  C_2(\mathrm{F}) = C_{\mathrm{F}} = \frac{N^2-1}{N}
\end{equation}
in the normalization~\eqref{eq:normalization}.

\begin{figure}
  \begin{center}
    \includegraphics[width=.8\columnwidth]{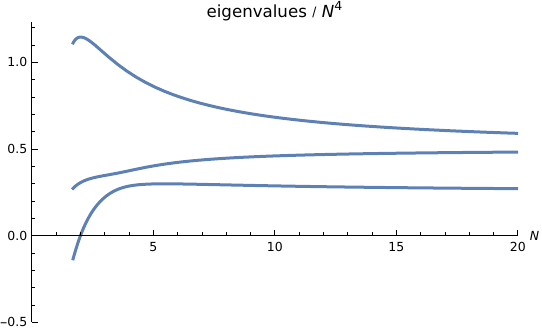}
  \end{center}
  \caption{\label{fig:EV(SU)}
    Eigenvalues of the matrix~$M(N,\mathcal{T})$ in~\eqref{eq:metric-8866bar}
    for~$\mathrm{SU}(N)$ as a function of~$N$, divided by the asymptotic scaling~$N^4$.}
\end{figure}

All products of the symmetric and antisymmetric
colorflows vanish, of course.  The eigenvalues of the real symmetric matrix
\begin{equation}
\label{eq:metric-8866bar}
  M(N,\mathcal{T}_{\mathrm{S}}) =  
  \begin{pmatrix}
    \mu_N(X,X) & \mu_N(X,Y) & \mu_N(X,Z_{\mathrm{S}}) \\
    \mu_N(Y,X) & \mu_N(Y,Y) & \mu_N(Y,Z_{\mathrm{S}}) \\
    \mu_N(Z_{\mathrm{S}},X) & \mu_N(Z_{\mathrm{S}},Y) & \mu_N(Z_{\mathrm{S}},Z_{\mathrm{S}})
  \end{pmatrix}
\end{equation}
can be computed numerically for arbitrary values of~$N$.  As illustrated in
figure~\ref{fig:EV(SU)}, they are all positive
for~$N>2$, but one eigenvalue vanishes for~$\mathrm{SU}(2)$.
It corresponds to the relation
\begin{equation}
  X = Z_{\mathrm{S}}\,.
\end{equation}
Thus we have found an invariant tensor that vanishes for~$\mathrm{SU}(2)$,
but is independent for~$\mathrm{SU}(N)$ with~$N\ge3$, similarly to
the~$d^{\mathrm{SU}}_{ijk}$ discussed in section~\ref{sec:spurious}.

\subsection{$\mathrm{U}(N)$}
For illustration, we can compute the elements of the
matrix~$M(N,\mathcal{T})$ also in the case of~$\mathrm{U}(N)$.  This
can be done by dropping all contributions of  $\mathrm{U}(1)$-ghosts
or by using~\eqref{eq:TT->dd-SU(N)} without the $1/N$-term.
Thus neither negative coefficients nor negative powers of~$N$
can appear in the results
for~$\mathrm{U}(N)$.  Indeed, we compute
\begin{subequations}
  \begin{align}
    \mu_N(X,X) &= \frac{1}{4}N^2 C \\
    \mu_N(Y,Y) &= \frac{1}{2} N(N+1) C \\
    \mu_N(Z_{\mathrm{S}},Z_{\mathrm{S}}) &= \frac{1}{2}(N^2+N+2)C \\
    \mu_N(X,Y) = \mu_N(Y,X) &= \frac{1}{2}C \\
    \mu_N(X,Z_{\mathrm{S}}) = \mu_N(Z_{\mathrm{S}},X) &= N C \\
    \mu_N(Y,Z_{\mathrm{S}}) = \mu_N(Z_{\mathrm{S}},Y) &= \frac{1}{2} (N+1) C\,,
  \end{align}
\end{subequations}
where the common factor is now~$C=N(N+1)$, and
\begin{equation}
  \mu_N(Z_{\mathrm{A}},Z_{\mathrm{A}}) = \frac{1}{2}N^2(N+2)C_{\mathrm{F}}
\end{equation}
for~$\mathcal{T}_{\mathrm{S}}$ and~$\mathcal{T}_{\mathrm{A}}$ respectively.
It is not surprising that the result for~$\mu_N(Z_{\mathrm{A}},Z_{\mathrm{A}})$ is the
same for~$\mathrm{U}(N)$ and~$\mathrm{SU}(N)$,
because the $\mathrm{U}(1)$-ghosts cancel in the antisymmetric
case, but not in the symmetric case.
\begin{figure}
  \begin{center}
    \includegraphics[width=.8\columnwidth]{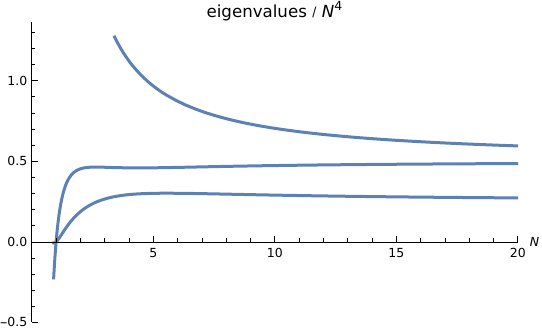}
  \end{center}
  \caption{\label{fig:EV(U)}
    Eigenvalues of the matrix~~$M(N,\mathcal{T})$ in~\eqref{eq:metric-8866bar}
    for~$\mathrm{U}(N)$ as a function of~$N$, divided by the asymptotic scaling~$N^4$.}
\end{figure}
As illustrated in figure~\ref{fig:EV(U)}, all eigenvalues are positive
for~$N\ge2$, but only one non-vanishing eigenvalue survives in the
abelian limit~$\mathrm{U}(1)$.  It can be written
\begin{equation}
  X + Y + 2Z_{\mathrm{S}}
\end{equation}
and the orthogonal combinations vanish.

\subsection{$N\to\infty$}
The coefficients of the leading powers of~$N$ agree
for~$\mathrm{U}(N)$ and~$\mathrm{SU}(N)$.  This was to be expected,
because the difference must contain powers of~$1/N$.

It is easy to see that, unless two colorflows~$A$ and~$B$ are related
by a permutation of the factors, their inner product~$\mu_N(A,B)$ contains
fewer closed chains~\eqref{eq:chains} than~$\mu_N(A,A)$
or~$\mu_N(B,B)$.  Therefore the off-diagonal elements of the
matrix~$M(N,\mathcal{T})$ will scale with a smaller power of~$N$
for~$N\to\infty$ and $M(N,\mathcal{T})$ will asymptotically
become diagonal.  This is indeed the case
\begin{equation}
  \lim_{N\to\infty}\left(\frac{M(N,\mathcal{T})}{N^4}\right) =  
  \frac{1}{4}
  \begin{pmatrix}
    2 & 0 & 0 \\
    0 & 1 & 0 \\
    0 & 0 & 2
  \end{pmatrix}
\end{equation}
and the two larger eigenvalues will approach~$N^4/2$, while the
smaller will approach~$N^4/4$.  In the case of the two smaller
eigenvalues, the asymptotic behaviour is already reached for small
values of~$N$, as illustrated in figures~\ref{fig:EV(SU)}
and~\ref{fig:EV(U)}.  This asymptotic behaviour is compatible with the
observations made in section~\ref{sec:N-dependence}, of course.

\section{A Catalogue of Exotic Birdtracks of~$\mathrm{SU}(N)$}
\label{sec:results}

The method described in section~\ref{sec:algorithm} can be used to
prepare a catalogue of bases of invariant tensors.  In this section, I
compare the results with the catalogue presented in~\cite{Carpenter:2021rkl}.

\subsection{Three Fields}
\label{sec:3-fields}
\begin{table}
  \begin{center}
  \begin{tabular}{l|ccc|l}
            & $n_{\epsilon}$ & $n_{\uparrow}$ & $r_3$ & remarks \\\hline
    \threefields{3}{3}{\overline{6}}  & 0 & 2 & 1 & CG: $\mathbf{6}\subset\mathbf{3}\otimes\mathbf{3}$\\
    \threefields{3}{\overline{3}}{8}  & 0 & 2 & 1 & $\T{a}{i}{j}$ \\
    \threefields{6}{\overline{6}}{8}  & 0 & 3 & 1 & $\Tsix{a}{s}{u}$ \\
    \threefields{8}{8}{8}             & 0 & 3 & 2 & $f_{abc}$, $d_{abc}$ (but $r_2=1$) \\
    \threefields{3}{6}{\overline{10}} & 0 & 3 & 1 & CG: $\mathbf{10}\subset\mathbf{3}\otimes\mathbf{6}$\\
    \threefields{6}{6}{\overline{15}} & 0 & 4 & 1
       & CG: $\mathbf{15}\subset\mathbf{6}\otimes\mathbf{6}$ \eqref{eq:cf-6-6-15}\\
    \threefields{6}{6}{\overline{15'}}& 0 & 4 & 1
       & CG: $\mathbf{15'}\subset\mathbf{6}\otimes\mathbf{6}$ \eqref{eq:cf-6-6-15'}
\\
    \threefields{3}{3}{3}             & 1 & 0 & 1 & totally antisymmetric \\
    \threefields{3}{6}{8}             & 1 & 1 & 1 & \eqref{eq:cf-3-6-8} \\
    \threefields{8}{8}{10}            & 1 & 2 & 1 & antisymmetric~\eqref{eq:cf-8-8-10} \\
    \threefields{3}{8}{\overline{15}} & 1 & 2 & 1 & \eqref{eq:cf-3-8-15} \\
    \threefields{3}{\overline{6}}{15} & 1 & 2 & 1 & \eqref{eq:cf-3-6-15} \\
    \threefields{6}{6}{6}             & 2 & 0 & 1 & totally symmetric \\
    \threefields{6}{8}{15}            & 2 & 1 & 1 & \eqref{eq:cf-6-8-15} \\
    \threefields{6}{\overline{6}}{27} & 2 & 2 & 1 & \eqref{eq:cf-6-6-27} \\
    \threefields{8}{8}{27}            & 2 & 2 & 1 & \eqref{eq:cf-8-8-27} 
  \end{tabular}
  \end{center}
  \caption{\label{tab:3-fields}%
    Invariant tensors in three-fold products of irreps
    of~$\mathrm{SU}(3)$, ordered in increasing
    numbers of epsilons~$n_{\epsilon}$, arrows~$n_{\uparrow}$ and
    rank~$r_3$, the number of independent colorflows for~$N=3$.
    This extends table~I of~\cite{Carpenter:2021rkl}.}
\end{table}

Table~\ref{tab:3-fields} lists the results for the three-fold
products presented in table~I of~\cite{Carpenter:2021rkl}.  They
confirm the latter.
To illustrate the colorflow formalism, I nevertheless display the
colorflows from table~\ref{tab:3-fields}
involving a mixed symmetry~$\mathbf{15}$ or one or
two~$\epsilon$s.  Some results will be used in
section~\ref{sec:4-fields}.
\begin{itemize}
\item \threefields{6}{6}{\overline{15}}: one line of both~$\mathbf{6}$ must
  be connected to the antisymmetrizer and the other to the symmetrizer
  of the~$\mathbf{15}$
\begin{equation}
\label{eq:cf-6-6-15}
  \parbox{34\unitlength}{%
    \fmfframe(6,4)(8,2){%
      \begin{fmfgraph*}(20,12)
        \fmfleft{r,s}
        \fmfright{q}
        \fmfv{label=$\mathbf{6}$,label.angle=180}{r}
        \fmfv{label=$\mathbf{6}$,label.angle=180}{s}
        \fmflabel{$\mathbf{\overline{15}}$}{q}
        \fmffreeze
        \fmfcmd{path a;}
        \fmfcmd{a = (vloc(__s) shifted (thick*down)){right}
             .. tension 3/4 .. {right}(vloc (__q) shifted (7thick*left+3thick*down))
             -- (vloc (__q) shifted (3thick*down));}
        \fmfi{fermion}{(vloc(__s) shifted (thick*up)){right}
             .. tension 3/4 .. {right}(vloc (__q) shifted (7thick*left+3thick*up))
             -- (vloc (__q) shifted (3thick*up))}
        \fmfi{fermion}{subpath (0,.32)*length a of a}
        \fmfi{plain}{subpath (.355,.37)*length a of a}
        \fmfi{plain}{subpath (.405,1)*length a of a}
        \fmfi{fermion}{(vloc(__r) shifted (thick*up)){right}
             .. tension 3/4 .. {right}(vloc (__q) shifted (9thick*left+thick*up))
             -- (vloc (__q) shifted (thick*up))}
        \fmfi{fermion}{(vloc(__r) shifted (thick*down)){right}
             .. tension 3/4 .. {right}(vloc (__q) shifted (9thick*left+thick*down))
             -- (vloc (__q) shifted (thick*down))}
        \fmfdraw
        \fmfcmd{path s[], sqa, sqs;}
        \fmfcmd{sqa = unitsquare shifted (-1/2,-1/2) xscaled 2thick yscaled 4thick
                         shifted vloc(__q) shifted (-5thick,-2thick);}
        \fmfcmd{filldraw sqa withcolor black;}
        \fmfcmd{sqs = unitsquare shifted (-1/2,-1/2) xscaled 2thick yscaled 6thick
                         shifted vloc(__q) shifted (-2thick,1thick);}
        \fmfcmd{fill sqs withcolor white;}
        \fmfcmd{draw sqs withcolor black;}
        \fmfcmd{s1 = unitsquare shifted (-1/2,-1/2) xscaled 2thick yscaled 4thick
                        shifted vloc(__r) shifted (2thick*right);}
        \fmfcmd{s2 = unitsquare shifted (-1/2,-1/2) xscaled 2thick yscaled 4thick
                        shifted vloc(__s) shifted (2thick*right);}
        \fmfcmd{fill s1 withcolor white;}
        \fmfcmd{draw s1 withcolor black;}
        \fmfcmd{fill s2 withcolor white;}
        \fmfcmd{draw s2 withcolor black;}
      \end{fmfgraph*}}}
    \end{equation}
    to obtain a non-vanishing colorflow.  It is antisymmetric
    under the exchange of the two factors of~$\mathbf{6}$.
  \item \threefields{3}{6}{8}: there is only one invariant tensor
\begin{equation}
\label{eq:cf-3-6-8}
  \parbox{32\unitlength}{%
    \fmfframe(6,4)(6,4){%
      \begin{fmfgraph*}(20,12)
        \fmfleft{a}
        \fmfright{q}
        \fmftop{i}
        \fmfbottom{e}
        \fmflabel{$\mathbf{3}$}{a}
        \fmflabel{$\mathbf{8}$}{i}
        \fmflabel{$\mathbf{6}$}{q}
        \fmflabel{$\overline{\epsilon}$}{e}
        \fmffreeze
        \fmfi{fermion}{(vloc(__i) shifted (thick*left)){down}
             .. tension 3/4 .. {down}(vloc (__e))}
        \fmfi{fermion}{(vloc (__a) shifted (thick*left)){right}
             .. tension 3/4 .. {down}(vloc(__e) shifted (2thick*left))}
        \fmfi{fermion}{(vloc (__q) shifted (thick*up)){left}
             .. tension 3/4 .. {up}(vloc(__i) shifted (thick*right))}
        \fmfi{fermion}{(vloc (__q) shifted (thick*down)){left}
             .. tension 3/4 .. {down}(vloc(__e) shifted (2thick*right))}
        \fmfdraw
        \fmfcmd{path s[];}
        \fmfcmd{s1 = unitsquare shifted (-1/2,-1/2) yscaled 2thick xscaled 6thick
                  shifted vloc(__e);}
        \fmfcmd{filldraw s1 withcolor black;}
        \fmfcmd{s2 = unitsquare shifted (-1/2,-1/2) xscaled 2thick yscaled 4thick
                  shifted vloc(__q) shifted (2thick*left);}
        \fmfcmd{fill s2 withcolor white;}
        \fmfcmd{draw s2 withcolor black;}
      \end{fmfgraph*}}}\,,
\end{equation}
because the antisymmetric~$\epsilon$ must not be connected twice to
the symmetric~$\mathbf{6}$.  This is the colorflow representation
of the formula~(A15) of~\cite{Carpenter:2021rkl}.

\item \threefields{8}{8}{10}: the only invariant tensor is antisymmetric
in the two factors~$\mathbf{8}$ due to the~$\overline{\epsilon}$
\begin{equation}
\label{eq:cf-8-8-10}
  \parbox{32\unitlength}{%
    \fmfframe(4,4)(4,4){%
      \begin{fmfgraph*}(24,12)
        \fmfleft{k}
        \fmfright{kbar}
        \fmftop{da1,ta,da2}
        \fmfbottom{db1,tb,db2}
        \fmflabel{$\mathbf{8}$}{ta}
        \fmflabel{$\mathbf{8}$}{tb}
        \fmflabel{$\mathbf{10}$}{k}
        \fmflabel{$\overline{\epsilon}$}{kbar}
        \fmf{fermion}{k,kbar}
        \fmffreeze
        \fmfi{fermion}{(vloc(__k) shifted (2thick*up)){right}
                          .. {up}(vloc(__ta) shifted (thick*left))}
        \fmfi{fermion}{(vloc(__k) shifted (2thick*down)){right}
                          .. {down}(vloc(__tb) shifted (thick*left))}
        \fmfi{fermion}{(vloc(__ta) shifted (thick*right)){down}
                          .. {right}(vloc(__kbar) shifted (2thick*up))}
        \fmfi{fermion}{(vloc(__tb) shifted (thick*right)){up}
                          .. {right}(vloc(__kbar) shifted (2thick*down))}
        \fmfdraw
        \fmfcmd{path s[];}
        \fmfcmd{s1 = unitsquare shifted (-1/2,-1/2) xscaled 2thick yscaled 6thick
                         shifted vloc(__k) shifted (2thick*right);}
        \fmfcmd{s2 = unitsquare shifted (-1/2,-1/2) xscaled 2thick yscaled 6thick
                         shifted vloc(__kbar);}
        \fmfcmd{fill s1 withcolor white;}
        \fmfcmd{draw s1 withcolor black;}
        \fmfcmd{filldraw s2 withcolor black;}
      \end{fmfgraph*}}}\,.
\end{equation}

\item \threefields{3}{8}{\overline{15}}: it is easier to see that there is indeed only
one inequivalent colorflow by looking at the
conjugate~\threefields{\overline{3}}{8}{15} instead:  there must be exactly
one~$\overline{\epsilon}$ to saturate all lines and one of the lines entering
this~$\overline{\epsilon}$ must be connected to the only line of
the~$\mathbf{15}$ that is not symmetrized
\begin{subequations}
\begin{equation}
\label{eq:cf-3-8-15}
  \parbox{32\unitlength}{%
    \fmfframe(6,4)(6,4){%
      \begin{fmfgraph*}(20,15)
        \fmfleft{a}
        \fmfright{q}
        \fmftop{i}
        \fmfbottom{e}
        \fmflabel{$\mathbf{8}$}{a}
        \fmflabel{$\mathbf{\overline{3}}$}{i}
        \fmflabel{$\mathbf{15}$}{q}
        \fmflabel{$\overline{\epsilon}$}{e}
        \fmffreeze
        \fmfi{fermion}{(vloc(__q) shifted (3thick*up)){left}
             .. tension 3/4 .. {up}(vloc (__i))}
        \fmfi{fermion}{(vloc(__q) shifted (thick*up)){left} .. {left}(vloc (__a) shifted (thick*up))}
        \fmfi{fermion}{(vloc (__a) shifted (thick*down)){right}
             .. tension 3/4 .. {down}(vloc(__e) shifted (2thick*left))}
        \fmfi{fermion}{(vloc (__q) shifted (1thick*down)){left}
             .. tension 3/4 .. {down}(vloc(__e))}
        \fmfi{fermion}{(vloc (__q) shifted (3thick*down)){left}
             .. {down}(vloc(__e) shifted (2thick*right))}
        \fmfdraw
        \fmfcmd{path s[];}
        \fmfcmd{s1 = unitsquare shifted (-1/2,-1/2) yscaled 2thick xscaled 7thick
                  shifted vloc(__e);}
        \fmfcmd{filldraw s1 withcolor black;}
        \fmfcmd{s2 = unitsquare shifted (-1/2,-1/2) xscaled 2thick yscaled 4thick
                  shifted vloc(__q) shifted (-2thick,-2thick);}
        \fmfcmd{filldraw s2 withcolor black;}
        \fmfcmd{s3 = unitsquare shifted (-1/2,-1/2) xscaled 2thick yscaled 6thick
                  shifted vloc(__q) shifted (-5thick,1thick);}
        \fmfcmd{fill s3 withcolor white;}
        \fmfcmd{draw s3 withcolor black;}
      \end{fmfgraph*}}}\,.
\end{equation}
All other contributions are then uniquely determined by symmetry.
With the opposite order of symmetrization and antisymmetrization in the
original~\threefields{3}{8}{\overline{15}}
\begin{equation}
\label{eq:cf-3-8-15a}
  \parbox{32\unitlength}{%
    \fmfframe(6,4)(6,4){%
      \begin{fmfgraph*}(20,15)
        \fmfleft{a}
        \fmfright{q}
        \fmftop{i}
        \fmfbottom{e}
        \fmflabel{$\mathbf{8}$}{a}
        \fmflabel{$\mathbf{3}$}{i}
        \fmflabel{$\mathbf{\overline{15}}$}{q}
        \fmflabel{$\epsilon$}{e}
        \fmffreeze
        \fmfi{fermion}{(vloc (__i)){down}
             .. tension 3/4 .. {right}(vloc(__q) shifted (3thick*up))}
        \fmfi{fermion}{(vloc (__a) shifted (thick*up)){right}
             .. tension 3/4 .. {right}(vloc(__q) shifted (thick*up))}
        \fmfi{fermion}{(vloc(__e) shifted (2thick*left)){up}
             .. tension 3/4 .. {left}(vloc (__a) shifted (thick*down))}
        \fmfi{fermion}{(vloc(__e)){up}
             .. tension 3/4 .. {right}(vloc (__q) shifted (1thick*down))}
        \fmfi{fermion}{(vloc(__e) shifted (2thick*right)){up}
             .. tension 3/4 .. {right}(vloc (__q) shifted (3thick*down))}
        \fmfdraw
        \fmfcmd{path s[];}
        \fmfcmd{s1 = unitsquare shifted (-1/2,-1/2) yscaled 2thick xscaled 6thick
                  shifted vloc(__e);}
        \fmfcmd{filldraw s1 withcolor black;}
        \fmfcmd{s2 = unitsquare shifted (-1/2,-1/2) xscaled 2thick yscaled 4thick
                  shifted vloc(__q) shifted (-5thick,-2thick);}
        \fmfcmd{filldraw s2 withcolor black;}
        \fmfcmd{s3 = unitsquare shifted (-1/2,-1/2) xscaled 2thick yscaled 6thick
                  shifted vloc(__q) shifted (-2thick,1thick);}
        \fmfcmd{fill s3 withcolor white;}
        \fmfcmd{draw s3 withcolor black;}
      \end{fmfgraph*}}}
\end{equation}
it is not immediately obvious that
\begin{equation}
\label{eq:cf-3-8-15b}
  \parbox{32\unitlength}{%
    \fmfframe(6,4)(6,4){%
      \begin{fmfgraph*}(20,15)
        \fmfleft{a}
        \fmfright{q}
        \fmftop{i}
        \fmfbottom{e}
        \fmflabel{$\mathbf{8}$}{a}
        \fmflabel{$\mathbf{3}$}{i}
        \fmflabel{$\mathbf{\overline{15}}$}{q}
        \fmflabel{$\epsilon$}{e}
        \fmffreeze
        \fmfcmd{path aq;}
        \fmfi{fermion}{(vloc (__i)){down}
                  .. tension 3/4 .. {right}(vloc(__q) shifted (3thick*up))}
        \fmfcmd{path aq;}
        \fmfcmd{aq = (vloc (__a) shifted (thick*up)){right}
                  .. (vloc(__a) shifted (.4w*right))
                  .. {right}(vloc(__q) shifted (9thick*left+3thick*down))
           .. {right}(vloc(__q) shifted (3thick*down));}
        \fmfi{fermion}{subpath (0,.45)*length(aq) of aq}
        \fmfi{plain}{subpath (.51,.55)*length(aq) of aq}
        \fmfi{plain}{subpath (.62,1)*length(aq) of aq}
        \fmfi{fermion}{(vloc(__e) shifted (2thick*left)){up}
             ... {left}(vloc (__a) shifted (thick*down))}
        \fmfi{fermion}{(vloc(__e)){up}
             .. {up}(vloc(__e) shifted (.2h*up))
             .. tension 3/4 .. {right}(vloc (__q) shifted (thick*up))}
        \fmfi{fermion}{(vloc(__e) shifted (2thick*right)){up}
             .. {up}(vloc(__e) shifted (.2h*up+2thick*right))
             .. tension 3/4 .. {right}(vloc (__q) shifted (thick*down))}
        \fmfdraw
        \fmfcmd{path s[];}
        \fmfcmd{s1 = unitsquare shifted (-1/2,-1/2) yscaled 2thick xscaled 6thick
                  shifted vloc(__e);}
        \fmfcmd{filldraw s1 withcolor black;}
        \fmfcmd{s2 = unitsquare shifted (-1/2,-1/2) xscaled 2thick yscaled 4thick
                  shifted vloc(__q) shifted (-5thick,-2thick);}
        \fmfcmd{filldraw s2 withcolor black;}
        \fmfcmd{s3 = unitsquare shifted (-1/2,-1/2) xscaled 2thick yscaled 6thick
                  shifted vloc(__q) shifted (-2thick,1thick);}
        \fmfcmd{fill s3 withcolor white;}
        \fmfcmd{draw s3 withcolor black;}
      \end{fmfgraph*}}}
\end{equation}
is equivalent after applying the Young projector\footnote{The
hermitian Young projectors advocated
in~\cite{Keppeler:2012ih,Keppeler:2013yla} make both variants equally
complicated.} for the~$\mathbf{\overline{15}}$.
However, the method described in section~\ref{eq:eigenvectors}
confirms that~$r_N=1$ in both cases.
\end{subequations}

\item \threefields{3}{\overline{6}}{15}: the single invariant tensor looks
very similar to~\eqref{eq:cf-3-8-15}
\begin{equation}
\label{eq:cf-3-6-15}
  \parbox{32\unitlength}{%
    \fmfframe(6,4)(6,4){%
      \begin{fmfgraph*}(20,15)
        \fmfleft{a}
        \fmfright{q}
        \fmftop{i}
        \fmfbottom{e}
        \fmflabel{$\mathbf{3}$}{a}
        \fmflabel{$\mathbf{\overline{6}}$}{i}
        \fmflabel{$\mathbf{15}$}{q}
        \fmflabel{$\overline{\epsilon}$}{e}
        \fmffreeze
        \fmfi{fermion}{(vloc(__q) shifted (3thick*up)){left}
              .. tension 3/4 .. {up}(vloc (__i) shifted (thick*right))}
        \fmfi{fermion}{(vloc(__q) shifted (thick*up)){left}
              .. tension 3/4 .. {up}(vloc (__i) shifted (thick*left))}
        \fmfi{fermion}{(vloc (__a) shifted (thick*down)){right}
             .. tension 3/4 .. {down}(vloc(__e) shifted (2thick*left))}
        \fmfi{fermion}{(vloc (__q) shifted (1thick*down)){left}
             .. tension 3/4 .. {down}(vloc(__e))}
        \fmfi{fermion}{(vloc (__q) shifted (3thick*down)){left}
             .. tension 3/4 .. {down}(vloc(__e) shifted (2thick*right))}
        \fmfdraw
        \fmfcmd{path s[];}
        \fmfcmd{s1 = unitsquare shifted (-1/2,-1/2) yscaled 2thick xscaled 6thick
                  shifted vloc(__e);}
        \fmfcmd{filldraw s1 withcolor black;}
        \fmfcmd{s2 = unitsquare shifted (-1/2,-1/2) xscaled 2thick yscaled 4thick
                  shifted vloc(__q) shifted (-2thick,-2thick);}
        \fmfcmd{filldraw s2 withcolor black;}
        \fmfcmd{s3 = unitsquare shifted (-1/2,-1/2) xscaled 2thick yscaled 6thick
                  shifted vloc(__q) shifted (-5thick,1thick);}
        \fmfcmd{fill s3 withcolor white;}
        \fmfcmd{draw s3 withcolor black;}
        \fmfcmd{s4 = unitsquare shifted (-1/2,-1/2) yscaled 2thick xscaled 4thick
                  shifted vloc(__i) shifted (2thick*down);}
        \fmfcmd{fill s4 withcolor white;}
        \fmfcmd{draw s4 withcolor black;}
      \end{fmfgraph*}}}\,.
\end{equation}

\item \threefields{6}{8}{15}: this needs two~$\overline{\epsilon}$ and
their lines must avoid the symmetrization of the~$\mathbf{15}$.
\begin{equation}
\label{eq:cf-6-8-15}
  \parbox{34\unitlength}{%
    \fmfframe(3,4)(6,4){%
      \begin{fmfgraph*}(25,15)
        \fmfleft{i}
        \fmfright{q}
        \fmfbottom{db1,e,db3,db4,db5}
        \fmftop{dt1,e',a,dt3,dt4}
        \fmflabel{$\mathbf{6}$}{i}
        \fmflabel{$\mathbf{15}$}{q}
        \fmfv{label=$\mathbf{8}$,label.angle=90}{a}
        \fmfv{label=$\overline{\epsilon}$,label.angle=90}{e'}
        \fmfv{label=$\overline{\epsilon}$,label.angle=-90}{e}
        \fmffreeze
        \fmfi{fermion}{(vloc(__q) shifted (3thick*up)){left}
             .. tension 3/4 .. {up}(vloc (__a) shifted (thick*right))}
        \fmfi{fermion}{(vloc(__q) shifted (thick*up)){left} .. {up}(vloc (__e'))}
        \fmfi{fermion}{(vloc(__a) shifted (thick*left)){down}
             .. tension 3/4 .. {up}(vloc (__e') shifted (2thick*right))}
        \fmfi{fermion}{(vloc(__i) shifted (thick*up)){right} .. {up}(vloc (__e') shifted (2thick*left))}
        \fmfi{fermion}{(vloc(__i) shifted (thick*down)){right} .. {down}(vloc (__e) shifted (2thick*left))}
        \fmfi{fermion}{(vloc (__q) shifted (1thick*down)){left}
             .. {down}(vloc(__e))}
        \fmfi{fermion}{(vloc (__q) shifted (3thick*down)){left}
             ... {down}(vloc(__e) shifted (2thick*right))}
        \fmfdraw
        \fmfcmd{path si, sqa, sqs, se, se';}
        \fmfcmd{se = unitsquare shifted (-1/2,-1/2) yscaled 2thick xscaled 6thick
                  shifted vloc(__e);}
        \fmfcmd{filldraw se withcolor black;}
        \fmfcmd{se' = unitsquare shifted (-1/2,-1/2) yscaled 2thick xscaled 6thick
                  shifted vloc(__e');}
        \fmfcmd{filldraw se' withcolor black;}
        \fmfcmd{sqa = unitsquare shifted (-1/2,-1/2) xscaled 2thick yscaled 4thick
                  shifted vloc(__q) shifted (-2thick,-2thick);}
        \fmfcmd{filldraw sqa withcolor black;}
        \fmfcmd{sqs = unitsquare shifted (-1/2,-1/2) xscaled 2thick yscaled 6thick
                  shifted vloc(__q) shifted (-5thick,thick);}
        \fmfcmd{fill sqs withcolor white;}
        \fmfcmd{draw sqs withcolor black;}
        \fmfcmd{si = unitsquare shifted (-1/2,-1/2) xscaled 2thick yscaled 4thick
                  shifted vloc(__i) shifted (2thick*right);}
        \fmfcmd{fill si withcolor white;}
        \fmfcmd{draw si withcolor black;}
      \end{fmfgraph*}}}\,.
\end{equation}
Note that all other ways of inserting the~$\mathbf{8}$ can be obtained
by exchanging the $\overline{\epsilon}$s and the lines ending at them.  Since
the combinatorics is already not completely obvious, the method described in
section~\ref{eq:eigenvectors} has been helpful for confirming the result~$r_N=1$.

\item \threefields{6}{\overline{6}}{27}: here is again only one way to
saturate all antisymmetric lines ending in the~$\overline{\epsilon}$s
\begin{equation}
\label{eq:cf-6-6-27}
  \parbox{34\unitlength}{%
    \fmfframe(3,4)(6,4){%
      \begin{fmfgraph*}(25,20)
        \fmfleft{da1,r,s,da2}
        \fmfright{q}
        \fmfbottom{db1,e1,db2,db3,db4}
        \fmftop{dt1,e2,dt2,dt3,dt4}
        \fmflabel{$\mathbf{6}$}{r}
        \fmflabel{$\mathbf{\overline{6}}$}{s}
        \fmflabel{$\mathbf{27}$}{q}
        \fmfv{label=$\overline{\epsilon}$,label.angle=-90}{e1}
        \fmfv{label=$\overline{\epsilon}$,label.angle=90}{e2}
        \fmffreeze
        \fmfdraw
        \fmfi{fermion}{(vloc(__q) shifted (5thick*up)){left}
             ... {up}(vloc (__e2) shifted (2thick*right))}
        \fmfi{fermion}{(vloc(__q) shifted (3thick*up)){left}
             ... {up}(vloc (__e2))}
        \fmfi{fermion}{(vloc(__q) shifted (1thick*up)){left}
             .. {left+up/3}(vloc (__s) shifted (1thick*up))}
        \fmfcmd{path re[];}
        \fmfcmd{re2 = (vloc(__r) shifted (thick*up)){right}
               .. {up}(vloc (__e2) shifted (2thick*left));}
        \fmfi{plain}{subpath (0,.46)*length(re2) of re2}
        \fmfi{plain}{subpath (.54,.56)*length(re2) of re2}
        \fmfi{fermion}{subpath (.64,1)*length(re2) of re2}
        \fmfi{fermion}{(vloc(__r) shifted (thick*down)){right}
             .. {down}(vloc (__e1) shifted (2thick*left))}
        \fmfi{fermion}{(vloc(__q) shifted (1thick*down)){left}
             .. {left+up/3}(vloc (__s) shifted (1thick*down))}
        \fmfi{fermion}{(vloc(__q) shifted (3thick*down)){left}
             ... {down}(vloc (__e1))}
        \fmfi{fermion}{(vloc(__q) shifted (5thick*down)){left}
             ... {down}(vloc (__e1) shifted (2thick*right))}
        \fmfcmd{path sqa[], sqs, se[], ss, sr, sw[];}
        \fmfcmd{se1 = unitsquare shifted (-1/2,-1/2) yscaled 2thick xscaled 6thick
                  shifted vloc(__e1);}
        \fmfcmd{filldraw se1 withcolor black;}
        \fmfcmd{se2 = unitsquare shifted (-1/2,-1/2) yscaled 2thick xscaled 6thick
                  shifted vloc(__e2);}
        \fmfcmd{filldraw se2 withcolor black;}
        \fmfcmd{sqa1 = unitsquare shifted (-1/2,-1/2) xscaled 2thick yscaled 4thick
                  shifted vloc(__q) shifted (-2thick,-4thick);}
        \fmfcmd{sqa2 = unitsquare shifted (-1/2,-1/2) xscaled 2thick yscaled 4thick
                  shifted vloc(__q) shifted (-2thick,4thick);}
        \fmfcmd{filldraw sqa1 withcolor black;}
        \fmfcmd{filldraw sqa2 withcolor black;}
        \fmfcmd{sqs = unitsquare shifted (-1/2,-1/2) xscaled 2thick yscaled 7.5thick
                  shifted vloc(__q) shifted (5thick*left);}
        \fmfcmd{fill sqs withcolor white;}
        \fmfcmd{draw sqs withcolor black;}
        \fmfcmd{ss = unitsquare shifted (-1/2,-1/2) xscaled 2thick yscaled 4thick
                  shifted vloc(__s) shifted (2thick*right+.5thick*down);}
        \fmfcmd{fill ss withcolor white;}
        \fmfcmd{draw ss withcolor black;}
        \fmfcmd{sr = unitsquare shifted (-1/2,-1/2) xscaled 2thick yscaled 4thick
                  shifted vloc(__r) shifted (2thick*right);}
        \fmfcmd{fill sr withcolor white;}
        \fmfcmd{draw sr withcolor black;}
        \fmfcmd{sw1 = unitsquare shifted (-1/2,-1/2) xscaled 2thick yscaled 2thick
                  shifted vloc(__q) shifted (5thick*left+5.5thick*up);}
        \fmfcmd{sw6 = unitsquare shifted (-1/2,-1/2) xscaled 2thick yscaled 2thick
                  shifted vloc(__q) shifted (5thick*left+5.5thick*down);}
        \fmfcmd{fill sw1 withcolor white;}
        \fmfcmd{fill sw6 withcolor white;}
        \fmfcmd{draw subpath (0,2) of sw1 withcolor black;}
        \fmfcmd{draw subpath (3,4) of sw1 withcolor black;}
        \fmfcmd{draw subpath (1,4) of sw6 withcolor black;}
      \end{fmfgraph*}}}\,,
\end{equation}
with the symmetrizer of the pair of outer lines wrapping around, as
in~\eqref{eq:27}.
En passant we note that this graphical representation makes it is obvious that there can be
no invariant tensor in the product~\threefields{3}{\overline{3}}{27}.

\item \threefields{8}{8}{27}: the symmetry in the two adjoint factors is
  obvious
\begin{equation}
\label{eq:cf-8-8-27}
  \parbox{34\unitlength}{%
    \fmfframe(3,4)(6,4){%
      \begin{fmfgraph*}(25,20)
        \fmfleft{da1,a1,a2,da2}
        \fmfright{q}
        \fmfbottom{db1,e1,db2,db3,db4,db5}
        \fmftop{dt1,e2,dt2,dt3,dt4,dt5}
        \fmflabel{$\mathbf{8}$}{a1}
        \fmflabel{$\mathbf{8}$}{a2}
        \fmflabel{$\mathbf{27}$}{q}
        \fmfv{label=$\overline{\epsilon}$,label.angle=-90}{e1}
        \fmfv{label=$\overline{\epsilon}$,label.angle=90}{e2}
        \fmffreeze
        \fmfdraw
        \fmfi{fermion}{(vloc(__q) shifted (5thick*up)){left}
             ... {up}(vloc (__e2) shifted (2thick*right))}
        \fmfi{fermion}{(vloc(__q) shifted (3thick*up)){left}
             ... {up}(vloc (__e2))}
        \fmfi{fermion}{(vloc(__q) shifted (1thick*up)){left}
             .. (vloc (__a2) shifted (1thick*down))}
        \fmfi{fermion}{(vloc(__a2) shifted (thick*up)){right}
             .. {up}(vloc (__e2) shifted (2thick*left))}
        \fmfi{fermion}{(vloc(__a1) shifted (thick*down)){right}
             .. {down}(vloc (__e1) shifted (2thick*left))}
        \fmfi{fermion}{(vloc(__q) shifted (1thick*down)){left}
             .. (vloc (__a1) shifted (1thick*up))}
        \fmfi{fermion}{(vloc(__q) shifted (3thick*down)){left}
             ... {down}(vloc (__e1))}
        \fmfi{fermion}{(vloc(__q) shifted (5thick*down)){left}
             ... {down}(vloc (__e1) shifted (2thick*right))}
        \fmfcmd{path sqa[], sqs, se[], sw[];}
        \fmfcmd{se1 = unitsquare shifted (-1/2,-1/2) yscaled 2thick xscaled 6thick
                  shifted vloc(__e1);}
        \fmfcmd{filldraw se1 withcolor black;}
        \fmfcmd{se2 = unitsquare shifted (-1/2,-1/2) yscaled 2thick xscaled 6thick
                  shifted vloc(__e2);}
        \fmfcmd{filldraw se2 withcolor black;}
        \fmfcmd{sqa1 = unitsquare shifted (-1/2,-1/2) xscaled 2thick yscaled 4thick
                  shifted vloc(__q) shifted (-2thick,-4thick);}
        \fmfcmd{sqa2 = unitsquare shifted (-1/2,-1/2) xscaled 2thick yscaled 4thick
                  shifted vloc(__q) shifted (-2thick,4thick);}
        \fmfcmd{filldraw sqa1 withcolor black;}
        \fmfcmd{filldraw sqa2 withcolor black;}
        \fmfcmd{sqs = unitsquare shifted (-1/2,-1/2) xscaled 2thick yscaled 7.5thick
                  shifted vloc(__q) shifted (5thick*left);}
        \fmfcmd{fill sqs withcolor white;}
        \fmfcmd{draw sqs withcolor black;}
        \fmfcmd{sw1 = unitsquare shifted (-1/2,-1/2) xscaled 2thick yscaled 2thick
                  shifted vloc(__q) shifted (5thick*left+5.5thick*up);}
        \fmfcmd{sw6 = unitsquare shifted (-1/2,-1/2) xscaled 2thick yscaled 2thick
                  shifted vloc(__q) shifted (5thick*left+5.5thick*down);}
        \fmfcmd{fill sw1 withcolor white;}
        \fmfcmd{fill sw6 withcolor white;}
        \fmfcmd{draw subpath (0,2) of sw1 withcolor black;}
        \fmfcmd{draw subpath (3,4) of sw1 withcolor black;}
        \fmfcmd{draw subpath (1,4) of sw6 withcolor black;}
      \end{fmfgraph*}}}\,,
\end{equation}
with the symmetrizer of the pair of outer lines wrapping around again.

\item \threefields{6}{6}{\overline{15'}}
The authors of~\cite{Carpenter:2021rkl} did not spell out the single
invariant tensor 
\begin{equation}
\label{eq:cf-6-6-15'}
  \parbox{32\unitlength}{%
    \fmfframe(6,4)(6,2){%
      \begin{fmfgraph*}(20,10)
        \fmfleft{r,s}
        \fmfright{q}
        \fmfv{label=$\mathbf{6}$,label.angle=180}{r}
        \fmfv{label=$\mathbf{6}$,label.angle=180}{s}
        \fmflabel{$\mathbf{\overline{15'}}$}{q}
        \fmffreeze
        \fmfi{fermion}{(vloc(__s) shifted (thick*up)){right}
             .. tension 3/4 .. {right}(vloc (__q) shifted (3thick*up))}
        \fmfi{fermion}{(vloc(__s) shifted (thick*down)){right}
             .. tension 3/4 .. {right}(vloc (__q) shifted (thick*up))}
        \fmfi{fermion}{(vloc(__r) shifted (thick*up)){right}
             .. tension 3/4 .. {right}(vloc (__q) shifted (thick*down))}
        \fmfi{fermion}{(vloc(__r) shifted (thick*down)){right}
             .. tension 3/4 .. {right}(vloc (__q) shifted (3thick*down))}
        \fmfdraw
        \fmfcmd{path s[], sqs;}
        \fmfcmd{sqs = unitsquare shifted (-1/2,-1/2) xscaled 2thick yscaled 8thick
                         shifted vloc(__q) shifted (2thick*left);}
        \fmfcmd{fill sqs withcolor white;}
        \fmfcmd{draw sqs withcolor black;}
        \fmfcmd{s1 = unitsquare shifted (-1/2,-1/2) xscaled 2thick yscaled 4thick
                        shifted vloc(__r) shifted (2thick*right);}
        \fmfcmd{s2 = unitsquare shifted (-1/2,-1/2) xscaled 2thick yscaled 4thick
                        shifted vloc(__s) shifted (2thick*right);}
        \fmfcmd{fill s1 withcolor white;}
        \fmfcmd{draw s1 withcolor black;}
        \fmfcmd{fill s2 withcolor white;}
        \fmfcmd{draw s2 withcolor black;}
      \end{fmfgraph*}}}
\end{equation}
in their catalogue, maybe because it is just a trivial symmetric Clebsch-Gordan
coefficient.
\end{itemize}

\subsection{Four Fields}
\label{sec:4-fields}
\begin{table}
  \begin{center}
  \begin{tabular}{l|ccc|l}
             & $n_{\epsilon}$ & $n_{\uparrow}$ & $r_3$ & remarks \\\hline
    \fourfields{3}{\overline{3}}{6}{\overline{6}}
                                       & 0 & 3 & 2 & \\
    \fourfields{3}{3}{\overline{6}}{8} & 0 & 3 & 2 & \eqref{eq:cf-3-3-6-8} \\
    \fourfields{6}{6}{\overline{6}}{\overline{6}}
                                       & 0 & 4 & 3 & 1 anti-, 2 symmetric \\
    \fourfields{6}{\overline{6}}{8}{8} & 0 & 4 & 4 & sections~\ref{sec:8866bar}, \ref{sec:8866bar-SU(N)} \\
    \fourfields{3}{3}{3}{\overline{10}}& 0 & 3 & 1 &
          CG: $\mathbf{10}\subset\mathbf{3}\otimes\mathbf{3}\otimes\mathbf{3}$\\
    \fourfields{3}{3}{6}{\overline{15}}& 0 & 4 & 2 &
          CG: $\mathbf{15}\subset\mathbf{3}\otimes\mathbf{3}\otimes\mathbf{6}$\\
    \fourfields{3}{\overline{3}}{\overline{3}}{\overline{6}}
                                       & 1 & 1 & 1 & \eqref{eq:cf-3-3-3-6} \\
    \fourfields{3}{3}{3}{8}            & 1 & 1 & 2 & \\
    \fourfields{3}{6}{6}{\overline{6}} & 1 & 2 & 1 & \eqref{eq:cf-3-6-6-6} \\
    \fourfields{3}{\overline{6}}{\overline{6}}{8}
                                       & 1 & 2 & 2 & 1 anti-, 1 symmetric \\
    \fourfields{3}{6}{8}{8}            & 1 & 2 & 3 & \eqref{eq:cf-3-6-8-8} \\
    \fourfields{3}{3}{6}{6}            & 2 & 0 & 1 & symmetric \\
    \fourfields{6}{6}{6}{8}            & 2 & 1 & 2 & \eqref{eq:cf-6-6-6-8}
  \end{tabular}
  \end{center}
  \caption{\label{tab:4-fields}%
    Invariant tensors in four-fold products of irreps
    of~$\mathrm{SU}(3)$, ordered in increasing
    numbers of epsilons~$n_{\epsilon}$, arrows~$n_{\uparrow}$ and
    rank~$r_3$, the number of independent colorflows for~$N=3$.
    This extends table~II of~\cite{Carpenter:2021rkl}.}
\end{table}

Table~\ref{tab:4-fields} lists the results for the four-fold
products presented in table~II of~\cite{Carpenter:2021rkl}.
In this case, we can not confirm them all:
\begin{enumerate}
  \item \fourfields{3}{3}{\overline{6}}{8}:
    here~\cite{Carpenter:2021rkl} reports two additional invariant tensors.
    However, there are only two ways to insert a gluon into the Clebsch-Gordan
    coefficient~\threefields{3}{3}{\overline{6}}. Thus there is only one
    invariant tensor
    in~$\mathbf{3}\otimes_{\mathrm{S}}\mathbf{3}\otimes\mathbf{\overline{6}}\otimes\mathbf{8}$
    and one in
    in~$\mathbf{3}\otimes_{\mathrm{A}}\mathbf{3}\otimes\mathbf{\overline{6}}\otimes\mathbf{8}$
    \begin{equation}
    \label{eq:cf-3-3-6-8}
    \parbox{25\unitlength}{%
        \fmfframe(3,4)(5,4){%
          \begin{fmfgraph*}(17,12)
            \fmfleft{k}
            \fmfright{kbar}
            \fmftop{da1,ta,da2}
            \fmfbottom{db1,tb,db2}
            \fmflabel{$\mathbf{8}$}{ta}
            \fmflabel{$\mathbf{3}$}{tb}
            \fmflabel{$\mathbf{3}$}{k}
            \fmflabel{$\mathbf{\overline6}$}{kbar}
            \fmffreeze
            \fmfi{fermion}{(vloc(__k)){right}
                              .. tension 3/4 .. {up}(vloc(__ta) shifted (thick*left))}
            \fmfi{fermion}{(vloc(__ta) shifted (thick*right)){down}
                              .. tension 3/4 .. {right}(vloc(__kbar) shifted (thick*up))}
            \fmfi{fermion}{(vloc(__tb)){up}
                              .. tension 3/4 .. {right}(vloc(__kbar) shifted (thick*down))}
            \fmfdraw
            \fmfcmd{path s[];}
            \fmfcmd{s2 = unitsquare shifted (-1/2,-1/2) xscaled 2thick yscaled 4thick
                             shifted vloc(__kbar) shifted (2thick*left);}
            \fmfcmd{fill s2 withcolor white;}
            \fmfcmd{draw s2 withcolor black;}
          \end{fmfgraph*}}} \pm
      \parbox{27\unitlength}{%
        \fmfframe(5,4)(5,4){%
          \begin{fmfgraph*}(17,12)
            \fmfleft{k}
            \fmfright{kbar}
            \fmftop{ta}
            \fmfbottom{tb}
            \fmflabel{$\mathbf{8}$}{ta}
            \fmflabel{$\mathbf{3}$}{tb}
            \fmflabel{$\mathbf{3}$}{k}
            \fmflabel{$\mathbf{\overline 6}$}{kbar}
            \fmf{phantom}{k,vb,vd,va,kbar}
            \fmffreeze
            \fmfcmd{path kkbar, tbta;}
            \fmfcmd{kkbar = (vloc(__k)){right}
                  -- (.5[vloc(__ta),vloc(__tb)]){right}
                  .. {right}(vloc(__kbar) shifted (thick*down));}
            \fmfcmd{tbta = (vloc(__tb)){up}
                  .. tension 3/4 .. {up}(vloc(__ta) shifted (thick*left));}
            \fmfi{fermion}{subpath (0,length(kkbar)/2) of kkbar}
            \fmfi{fermion}{subpath (length(kkbar)/2,length(kkbar)) of kkbar}
            \fmfi{fermion}{(vloc(__ta) shifted (thick*right)){down}
                  .. tension 3/4 .. {right}(vloc(__kbar) shifted (thick*up))}
            \fmfi{fermion}{subpath (0,0.9length(tbta)/2) of tbta}
            \fmfi{fermion}{subpath (1.1length(tbta)/2,length(tbta)) of tbta}
            \fmfdraw
            \fmfcmd{path s[];}
            \fmfcmd{s2 = unitsquare shifted (-1/2,-1/2) xscaled 2thick yscaled 4thick
                             shifted vloc(__kbar) shifted (2thick*left);}
            \fmfcmd{fill s2 withcolor white;}
            \fmfcmd{draw s2 withcolor black;}
          \end{fmfgraph*}}}\,.
    \end{equation}
    They can be expressed as combinations of~$K_{\mathbf{6}}T_{\mathbf{3}}$ and~$K_{\mathbf{6}}T_{\mathbf{6}}$.
    The other two tensors in table~II of~\cite{Carpenter:2021rkl},
    $L\overline{J}$ and~$QV$, both contain
    $\epsilon^{i_1i_2k} \overline\epsilon_{j_1j_2k}
         = \delta^{i_1}_{j_1}\delta^{i_2}_{j_2} - \delta^{i_1}_{j_2}\delta^{i_2}_{j_1}$
    and are therefore redundant, as described in section~\ref{sec:epsilon}.
  \item \fourfields{6}{\overline{6}}{8}{8}:
    one symmetric tensor is missing in~\cite{Carpenter:2021rkl}.  This has been
    discussed at length in sections~\ref{sec:8866bar}
    and~\ref{sec:8866bar-SU(N)}.
  \item \fourfields{3}{\overline{3}}{\overline{3}}{\overline{6}}:
    the only independent invariant tensor is antisymmetric
\begin{equation}
\label{eq:cf-3-3-3-6}
  \parbox{32\unitlength}{%
    \fmfframe(6,4)(6,4){%
      \begin{fmfgraph*}(20,15)
        \fmfleft{i}
        \fmfright{dr,j1,j2,dr2}
        \fmftop{s}
        \fmfbottom{e}
        \fmflabel{$\mathbf{3}$}{i}
        \fmflabel{$\mathbf{\overline{6}}$}{s}
        \fmfv{label=$\mathbf{\overline{3}}$,label.angle=0}{j1}
        \fmfv{label=$\mathbf{\overline{3}}$,label.angle=0}{j2}
        \fmflabel{$\epsilon$}{e}
        \fmffreeze
        \fmfdraw
        \fmfi{fermion}{(vloc (__i)){right}
          .. tension 3/4 .. {up}(vloc (__s) shifted (thick*left))}
        \fmfi{fermion}{(vloc (__e) shifted (2thick*left)){up}
          .. tension 3/4 .. {up}(vloc (__s) shifted (thick*right))}
        \fmfi{fermion}{(vloc (__e)){up}
          .. tension 3/4 .. {right}(vloc (__j2))}
        \fmfi{fermion}{(vloc (__e) shifted (2thick*right)){up}
          .. tension 3/4 .. {right}(vloc (__j1))}
        \fmfcmd{path ss, se;}
        \fmfcmd{se = unitsquare shifted (-1/2,-1/2) yscaled 2thick xscaled 6thick
                  shifted vloc(__e);}
        \fmfcmd{filldraw se withcolor black;}
        \fmfcmd{ss = unitsquare shifted (-1/2,-1/2) yscaled 2thick xscaled 4thick
                  shifted vloc(__s) shifted (2thick*down);}
        \fmfcmd{fill ss withcolor white;}
        \fmfcmd{draw ss withcolor black;}
      \end{fmfgraph*}}}
\end{equation}
  \item \fourfields{3}{6}{6}{\overline{6}}:
    since each leg of the~$\overline{\epsilon}$ must be connected to a
    different~$\mathbf{3}$ or~$\mathbf{6}$,
    there is again only one
    independent invariant tensor and it is antisymmetric
\begin{equation}
\label{eq:cf-3-6-6-6}
  \parbox{32\unitlength}{%
    \fmfframe(6,4)(6,4){%
      \begin{fmfgraph*}(20,15)
        \fmfleft{dl1,i,dl3,j1,dl4,dl5}
        \fmfright{dr1,dr2,dr3,j2,dr4,dr5}
        \fmftop{s}
        \fmfbottom{e}
        \fmflabel{$\mathbf{\overline{6}}$}{s}
        \fmfv{label=$\mathbf{3}$,label.angle=180}{i}
        \fmfv{label=$\mathbf{6}$,label.angle=180}{j1}
        \fmfv{label=$\mathbf{6}$,label.angle=0}{j2}
        \fmflabel{$\overline{\epsilon}$}{e}
        \fmffreeze
        \fmfdraw
        \fmfi{fermion}{(vloc (__i)){right}
          .. tension 3/4 .. {down}(vloc (__e) shifted (2thick*left))}
        \fmfi{fermion}{(vloc (__j1) shifted (thick*down)){right}
          .. tension 3/4 .. {down}(vloc (__e))}
        \fmfi{fermion}{(vloc (__j1) shifted (thick*up)){right}
          .. tension 3/4 .. {up}(vloc (__s) shifted (thick*left))}
        \fmfi{fermion}{(vloc (__j2) shifted (thick*up)){left}
          .. tension 3/4 .. {up}(vloc (__s) shifted (thick*right))}
        \fmfi{fermion}{(vloc (__j2) shifted (thick*down)){left}
          .. tension 3/4 .. {down}(vloc (__e) shifted (2thick*right))}
        \fmfcmd{path ss[], se;}
        \fmfcmd{se = unitsquare shifted (-1/2,-1/2) yscaled 2thick xscaled 6thick
                  shifted vloc(__e);}
        \fmfcmd{filldraw se withcolor black;}
        \fmfcmd{ss1 = unitsquare shifted (-1/2,-1/2) xscaled 2thick yscaled 4thick
                  shifted vloc(__j1) shifted (2thick*right);}
        \fmfcmd{fill ss1 withcolor white;}
        \fmfcmd{draw ss1 withcolor black;}
        \fmfcmd{ss2 = unitsquare shifted (-1/2,-1/2) xscaled 2thick yscaled 4thick
                  shifted vloc(__j2) shifted (2thick*left);}
        \fmfcmd{fill ss2 withcolor white;}
        \fmfcmd{draw ss2 withcolor black;}
        \fmfcmd{ss3 = unitsquare shifted (-1/2,-1/2) yscaled 2thick xscaled 4thick
                  shifted vloc(__s) shifted (2thick*down);}
        \fmfcmd{fill ss3 withcolor white;}
        \fmfcmd{draw ss3 withcolor black;}
      \end{fmfgraph*}}}\,.
\end{equation}
  \item \fourfields{3}{6}{8}{8}: there are two invariant tensors
    antisymmetric in the factors~$\mathbf{8}$ and one symmetric.
    \begin{subequations}
    \label{eq:cf-3-6-8-8}
    Up to permutations of the external~$\mathbf{8}$s, there are three
    different ways to connect the~$\overline{\epsilon}$ to the other external states:
    to both the~$\mathbf{3}$ and the~$\mathbf{6}$
    \begin{equation}
    \label{eq:cf-3-6-8-8a}
      \parbox{30\unitlength}{%
        \fmfframe(3,4)(5,4){%
          \begin{fmfgraph*}(22,12)
            \fmfleft{i}
            \fmfright{s}
            \fmftop{da1,a1,da2,a2,da3}
            \fmfbottom{e}
            \fmfv{label=$\mathbf{8}$,label.angle=90}{a1}
            \fmfv{label=$\mathbf{8}$,label.angle=90}{a2}
            \fmflabel{$\mathbf{3}$}{i}
            \fmflabel{$\mathbf{6}$}{s}
            \fmflabel{$\overline{\epsilon}$}{e}
            \fmffreeze
            \fmfdraw
            \fmfi{fermion}{(vloc(__i)){right}
                   .. {down}(vloc(__e) shifted (2thick*left))}
            \fmfi{fermion}{(vloc(__a1) shifted (thick*left)){down}
                   .. tension 3/4 .. {down}(vloc(__e))}
            \fmfi{fermion}{(vloc(__a2) shifted (thick*left)){down}
                   ... {up}(vloc(__a1) shifted (thick*right))}
            \fmfi{fermion}{(vloc(__s) shifted (thick*down)){left}
                   .. tension 3/4 .. {down}(vloc(__e) shifted (2thick*right))}
            \fmfi{fermion}{(vloc(__s) shifted (thick*up))
                   -- (vloc(__s) shifted (7thick*left+thick*up)){left}
                   .. tension 3/4 .. {up}(vloc(__a2) shifted (thick*right))}
            \fmfcmd{path se, ss;}
            \fmfcmd{ss = unitsquare shifted (-1/2,-1/2) xscaled 2thick yscaled 4thick
                             shifted vloc(__s) shifted (2thick*left);}
            \fmfcmd{fill ss withcolor white;}
            \fmfcmd{draw ss withcolor black;}
            \fmfcmd{se = unitsquare shifted (-1/2,-1/2) yscaled 2thick xscaled 6thick
                             shifted vloc(__e);}
            \fmfcmd{filldraw se withcolor black;}
          \end{fmfgraph*}}}\,,
    \end{equation}
    to the~$\mathbf{6}$ only 
    \begin{equation}
    \label{eq:cf-3-6-8-8b}
      \parbox{30\unitlength}{%
        \fmfframe(3,4)(5,4){%
          \begin{fmfgraph*}(22,12)
            \fmfleft{i}
            \fmfright{s}
            \fmftop{da1,a1,da2,a2,da3}
            \fmfbottom{e}
            \fmfv{label=$\mathbf{8}$,label.angle=90}{a1}
            \fmfv{label=$\mathbf{8}$,label.angle=90}{a2}
            \fmflabel{$\mathbf{3}$}{i}
            \fmflabel{$\mathbf{6}$}{s}
            \fmflabel{$\overline{\epsilon}$}{e}
            \fmffreeze
            \fmfdraw
            \fmfi{fermion}{(vloc(__i)){right}
                   .. tension 3/4 .. {up}(vloc(__a1) shifted (thick*left))}
            \fmfi{fermion}{(vloc(__a1) shifted (thick*right)){down}
                   .. tension 3/4 .. {down}(vloc(__e) shifted (2thick*left))}
            \fmfi{fermion}{(vloc(__a2) shifted (thick*left)){down}
                   .. tension 3/4 .. {down}(vloc(__e))}
            \fmfi{fermion}{(vloc(__s) shifted (thick*down)){left}
                   .. tension 3/4 .. {down}(vloc(__e) shifted (2thick*right))}
            \fmfi{fermion}{(vloc(__s) shifted (thick*up))
                   -- (vloc(__s) shifted (5thick*left+thick*up)){left}
                   .. tension 3/4 .. {up}(vloc(__a2) shifted (thick*right))}
            \fmfcmd{path se, ss;}
            \fmfcmd{ss = unitsquare shifted (-1/2,-1/2) xscaled 2thick yscaled 4thick
                             shifted vloc(__s) shifted (2thick*left);}
            \fmfcmd{fill ss withcolor white;}
            \fmfcmd{draw ss withcolor black;}
            \fmfcmd{se = unitsquare shifted (-1/2,-1/2) yscaled 2thick xscaled 6thick
                             shifted vloc(__e);}
            \fmfcmd{filldraw se withcolor black;}
          \end{fmfgraph*}}}
    \end{equation}
    and to the~$\mathbf{3}$ only
    \begin{equation}
    \label{eq:cf-3-6-8-8c}
      \parbox{30\unitlength}{%
        \fmfframe(3,4)(5,4){%
          \begin{fmfgraph*}(22,12)
            \fmfleft{i}
            \fmfright{s}
            \fmftop{da1,a1,da2,a2,da3}
            \fmfbottom{e}
            \fmfv{label=$\mathbf{8}$,label.angle=90}{a1}
            \fmfv{label=$\mathbf{8}$,label.angle=90}{a2}
            \fmflabel{$\mathbf{3}$}{i}
            \fmflabel{$\mathbf{6}$}{s}
            \fmflabel{$\overline{\epsilon}$}{e}
            \fmffreeze
            \fmfdraw
            \fmfi{fermion}{(vloc(__i)){right}
                   .. {down}(vloc(__e) shifted (2thick*left))}
            \fmfi{fermion}{(vloc(__a1) shifted (thick*left)){down}
                   .. tension 3/4 .. {down}(vloc(__e))}
            \fmfi{fermion}{(vloc(__a2) shifted (thick*left)){down}
                  ... {down}(vloc(__e) shifted (2thick*right))}
            \fmfcmd{path sa[];}
            \fmfcmd{sa1 = (vloc(__s) shifted (thick*down)){left}
                   .. {up}(vloc(__a1) shifted (thick*right));}
            \fmfi{plain}{subpath (0,.32)*length(sa1) of sa1}
            \fmfi{fermion}{subpath (.42,1)*length(sa1) of sa1}
            \fmfi{fermion}{(vloc(__s) shifted (thick*up))
                   -- (vloc(__s) shifted (6thick*left+thick*up)){left}
                   .. tension 3/4 .. {up}(vloc(__a2) shifted (thick*right))}
            \fmfcmd{path se, ss;}
            \fmfcmd{ss = unitsquare shifted (-1/2,-1/2) xscaled 2thick yscaled 4thick
                             shifted vloc(__s) shifted (2thick*left);}
            \fmfcmd{fill ss withcolor white;}
            \fmfcmd{draw ss withcolor black;}
            \fmfcmd{se = unitsquare shifted (-1/2,-1/2) yscaled 2thick xscaled 6thick
                             shifted vloc(__e);}
            \fmfcmd{filldraw se withcolor black;}
          \end{fmfgraph*}}}\,.
    \end{equation}
    \end{subequations}
    The colorflow~\eqref{eq:cf-3-6-8-8c} is antisymmetric in the two
    $\mathbf{8}$s, while~\eqref{eq:cf-3-6-8-8a}
    and~\eqref{eq:cf-3-6-8-8b} contain both symmetric and antisymmetric
    contributions.
    These three colorflows correspond to the invariant tensors
    \begin{subequations}
      \begin{align}
        A^{ab}_{is}&=
           \overline{\epsilon}_{ij_1k}\T{a}{k}{l}\T{b}{l}{j_2}\Ksix{s}{j_1}{j_2}\\
        B^{ab}_{is}&=
           \overline{\epsilon}_{j_1kl}\T{a}{k}{i}\T{b}{l}{j_2}\Ksix{s}{j_1}{j_2}\\
        C^{ab}_{is}&=
           \overline{\epsilon}_{ikl}\T{a}{k}{j_1}\T{b}{l}{j_2}\Ksix{s}{j_1}{j_2}
      \end{align}
    \end{subequations}
    Due to the presence of an~$\overline{\epsilon}$, the
    matrix~\eqref{eq:colorfactor-matrix} can only be computed
    for~$N=3$
    \begin{equation}
      M\left(3,\{A,B,C\}\right) =
      \begin{pmatrix}
         7 &  4 & -3 \\
         4 &  8 & 4 \\
         -3 &  4 & 7
      \end{pmatrix} \cdot 16
    \end{equation}
    and has the eigenvalues~$0$, $160$ and~$192$.  The eigenvector
    for the eigenvalue~$0$ is~$(1,-1,1)^T$ and corresponds to the relation
    \begin{equation}
    \label{eq:cf-3-6-8-8}
      A - B + C = 0\,,
    \end{equation}
    revealing that one symmetric and one antisymmetric tensor is redundant.
    It is easy to verify that the relation~\eqref{eq:cf-3-6-8-8} is
    just the invariance of the tensor~\eqref{eq:cf-3-6-8}
    \begin{equation}
      D^a_{is} = \overline{\epsilon}_{ij_1k}\T{a}{k}{j_2}\Ksix{s}{j_1}{j_2}
    \end{equation}
    in the product \threefields{3}{6}{8}.


    The corresponding row in table~II of~\cite{Carpenter:2021rkl} lists
    six invariant tensors: three of mixed symmetry, two antisymmetric
    and one symmetric. Therefore there are three
    non-trivial relation among them.


  \item \fourfields{6}{6}{6}{8}: the only two independent invariant
    tensors are combinations of permutations of
\begin{equation}
\label{eq:cf-6-6-6-8}
  \parbox{34\unitlength}{%
    \fmfframe(3,4)(6,4){%
      \begin{fmfgraph*}(25,20)
        \fmfleft{dl1,s2,s1,dl2}
        \fmfright{dr1,s3,a,dl2}
        \fmfright{q}
        \fmfbottom{e1}
        \fmftop{e2}
        \fmfv{label=$\mathbf{8}$,label.angle=0}{a}
        \fmfv{label=$\mathbf{6}$,label.angle=0}{s3}
        \fmfv{label=$\mathbf{6}$,label.angle=180}{s1}
        \fmfv{label=$\mathbf{6}$,label.angle=180}{s2}
        \fmflabel{$\overline{\epsilon}$}{e1}
        \fmflabel{$\overline{\epsilon}$}{e2}
        \fmffreeze
        \fmfdraw
        \fmfi{fermion}{(vloc (__s1) shifted (thick*down)){right}
                .. tension 3/4 .. {down}(vloc (__e1))}
        \fmfi{fermion}{(vloc (__s1) shifted (thick*up)){right}
                .. tension 3/4 .. {up}(vloc (__e2) shifted (2thick*left))}
        \fmfi{fermion}{(vloc (__s2) shifted (thick*down)){right}
                .. tension 3/4 .. {down}(vloc (__e1) shifted (2thick*left))}
        \fmfcmd{path sebelow;}
        \fmfcmd{sebelow = (vloc (__s2) shifted (thick*up)){right}
                .. tension 3/4 .. {up}(vloc (__e2));}
        \fmfi{fermion}{subpath (0,.42)*length(sebelow) of sebelow}
        \fmfi{plain}{subpath (0.55,1)*length(sebelow) of sebelow}
        \fmfi{fermion}{(vloc (__s3) shifted (thick*down)){left}
                .. tension 3/4 .. {down}(vloc (__e1) shifted (2thick*right))}
        \fmfi{fermion}{(vloc (__a) shifted (thick*up)){left}
                .. tension 3/4 .. {up}(vloc (__e2) shifted (2thick*right))}
        \fmfi{fermion}{(vloc (__s3) shifted (thick*up)){left}
                .. {up}(.6w,.5h)
                .. {right}(vloc (__a) shifted (thick*down))}
        \fmfcmd{path ss[], se[];}
        \fmfcmd{se1 = unitsquare shifted (-1/2,-1/2) yscaled 2thick xscaled 6thick
                  shifted vloc(__e1);}
        \fmfcmd{se2 = unitsquare shifted (-1/2,-1/2) yscaled 2thick xscaled 6thick
                  shifted vloc(__e2);}
        \fmfcmd{filldraw se1 withcolor black;}
        \fmfcmd{filldraw se2 withcolor black;}
        \fmfcmd{ss1 = unitsquare shifted (-1/2,-1/2) xscaled 2thick yscaled 4thick
                  shifted vloc(__s1) shifted (2thick*right);}
        \fmfcmd{ss2 = unitsquare shifted (-1/2,-1/2) xscaled 2thick yscaled 4thick
                  shifted vloc(__s2) shifted (2thick*right);}
        \fmfcmd{ss3 = unitsquare shifted (-1/2,-1/2) xscaled 2thick yscaled 4thick
                  shifted vloc(__s3) shifted (2thick*left);}
        \fmfcmd{fill ss1 withcolor white;}
        \fmfcmd{draw ss1 withcolor black;}
        \fmfcmd{fill ss2 withcolor white;}
        \fmfcmd{draw ss2 withcolor black;}
        \fmfcmd{fill ss3 withcolor white;}
        \fmfcmd{draw ss3 withcolor black;}
      \end{fmfgraph*}}}\,.
\end{equation}
    which can be viewed as insertions of a gluon into the only
    invariant tensor in~\threefields{6}{6}{6}.
    The corresponding invariant tensors are
    \begin{equation}
      A^{a}_{s_1s_2s_3}
        = \T{a}{k}{i_1} \overline{\epsilon}_{i_2i_3k} \overline{\epsilon}_{j_1j_2j_3}
          \Ksix{s_1}{i_1}{j_1}\Ksix{s_2}{i_2}{j_2}\Ksix{s_3}{i_3}{j_3}
    \end{equation}
    and its cyclic permutations in~$\{s_1,s_2,s_3\}$, while
    the non-cyclic permutations are trivially related by the
    antisymmetry of the~$\overline{\epsilon}$s.
    The eigenvector of the matrix~$M$ corresponding to the eigenvalue~$0$ turns out to
    be the sum of the cyclic permutations.  This can again be understood
    as the invariance of the invariant tensor in~\threefields{6}{6}{6}.
    Therefore only two combinations of the~$A$ are independent.

    The tensors~$A$ correspond to
    the~$ST_{\mathbf{6}}$ tensors in table~II of~\cite{Carpenter:2021rkl}.
    The~$WX$ tensors are linear combinations of these, as can be seen
    by gluing the conjugate of~\eqref{eq:cf-6-6-15} to~\eqref{eq:cf-6-8-15}
    at the~$\mathbf{15}$.
\end{enumerate}
There are three more products that have been left out of table~II
of~\cite{Carpenter:2021rkl}
\begin{enumerate}
  \item \fourfields{3}{3}{3}{\overline{10}}:
    there is only one colorflow and it is totally symmetric.
  \item \fourfields{3}{3}{6}{\overline{15}}: there is one symmetric
    and one antisymmetric colorflow, corresponding
    to~$\mathbf{15}\subset\mathbf{6}\otimes\mathbf{6}$
    and~$\mathbf{15}\subset\mathbf{\overline{3}}\otimes\mathbf{6}$.
  \item \fourfields{3}{3}{3}{8}: the two independent colorflows are
    just like~\eqref{eq:cf-6-6-6-8}, with one of the $\overline{\epsilon}$s removed
    and all~$\mathbf{6}$s replaced by~$\mathbf{3}$s.
  As in the case of~\fourfields{6}{6}{6}{8}, only two of the
  \begin{equation}
    A^{a}_{i_1i_2i_3} = \T{a}{k}{i_1} \overline{\epsilon}_{i_2i_3k}\,,
  \end{equation}
  are independent: the sum of the cyclic permutations
  vanishes because~$\overline{\epsilon}_{i_1i_2i_3}$ is an invariant
  invariant tensor in \threefields{3}{3}{3}.

\end{enumerate}
I can confirm the remaining results of~\cite{Carpenter:2021rkl} for the
four-fold products and only use this opportunity to clarify
permutation symmetries in the factors:
\begin{itemize}
  \item \fourfields{3}{\overline{3}}{6}{\overline{6}}: there are two
    independent colorflows and they are linear combinations of the
    invariant tensors listed in~\cite{Carpenter:2021rkl}.
  \item \fourfields{6}{6}{\overline{6}}{\overline{6}}: there are one
    antisymmetric and two symmetric invariant tensors. This agrees
    with~\cite{Carpenter:2021rkl},
    since~$\delta^{s_1}_{t_1}\delta^{s_2}_{t_2}$ contains both a
    symmetric and an antisymmetric contribution.
  \item \fourfields{3}{\overline{6}}{\overline{6}}{8}: there is one
    symmetric and one antisymmetric invariant tensor.  This is
    compatible with~\cite{Carpenter:2021rkl}, except for obvious typos
    in the indices of the~$K\overline{J}$ term.
  \item \fourfields{3}{3}{6}{6}: since each~$\overline{\epsilon}$ must be
    connected with both~$\mathbf{6}$s, the only colorflow is symmetric.
\end{itemize}

\subsection{Five and More Fields}
\begin{table}
  \begin{center}
  \begin{tabular}{l|ccc|l}
       & $n_{\epsilon}$ & $n_{\uparrow}$ & $r_3$ & remarks \\\hline
    \fivefields{3}{3}{6}{\overline{6}}{\overline{6}}
                               & 0 & 4 & 4 & \\
    \fivefields{3}{\overline{3}}{6}{\overline{6}}{8}
                               & 0 & 4 & 5 & but $r_2=4$ \\
    \fivefields{6}{6}{\overline{6}}{\overline{6}}{8}
                               & 0 & 5 & 8 & but $r_2=6$\\
    \fivefields{3}{3}{3}{\overline{3}}{\overline{6}}
                               & 1 & 0 & 3 & \\
    \fivefields{3}{3}{3}{3}{6} & 2 & 0 & 2 & \\
    \fivefields{3}{6}{\overline{6}}{\overline{6}}{\overline{6}}
                               & 2 & 0 & 3 & \\
    \fivefields{3}{\overline{3}}{6}{6}{6}
                               & 2 & 1 & 3 & \\
    \fivefields{6}{6}{6}{8}{8} & 2 & 2 & 10& 4 anti-, 6 symmetric\\
    \fivefields{3}{6}{6}{6}{6} & 3 & 0 & 3 & \\
    \fivefields{6}{\overline{6}}{\overline{6}}{\overline{6}}{\overline{6}}
                               & 3 & 0 & 6 &
  \end{tabular}
  \end{center}
  \caption{\label{tab:5-fields}%
    Invariant tensors in five-fold products of irreps
    of~$\mathrm{SU}(3)$, ordered in increasing
    numbers of epsilons~$n_{\epsilon}$, arrows~$n_{\uparrow}$ and
    rank~$r_3$, the number of independent colorflows for~$N=3$,
    cf.~table~III of~\cite{Carpenter:2021rkl}.}
\end{table}
Table~III of~\cite{Carpenter:2021rkl} sketches a catalogue of
invariant tensors in five-fold products of irreps of~$\mathrm{SU}(3)$.
Since a complete catalogue can easily be produced with the program
\textsc{tangara} together with
\textsc{Mathematica}~\cite{Mathematica}, I only count them
in table~\ref{tab:5-fields} and refrain from presenting a graphical
representation and a detailled discussion.

\begin{table}
  \begin{center}
  \begin{tabular}{c|rrrrrrrcr}
     n & $r_2$ & $r_3$ & $r_4$ & $r_5$ & $r_6$ & $r_7$& $r_8$ & $\cdots$ & $r_\infty$ \\\hline
     3 & 1     & 2     & 2     & 2     & 2     & 2    & 2    & $\cdots$ & 2 \\
     4 & 3     & 8     & 9     & 9     & 9     & 9    & 9    & $\cdots$ & 9 \\
     5 & 6     & 32    & 43    & 44    & 44    & 44   & 44   & $\cdots$ & 44 \\
     6 & 15    & 145   & 245   & 264   & 265   & 265  & 265  & $\cdots$ & 265 \\
     7 & 36    & 702   & 1557  & 1824  & 1853  & 1854 & 1854 & $\cdots$ & 1854
  \end{tabular}
  \end{center}
  \caption{\label{tab:adjoints}%
    The rank $r_N$ of the matrix $M(N,\mathcal{T})$ of
    color factors~\eqref{eq:colorfactor-matrix}, i.e.~the number of
    independent invariant tensors, in the product of $n$~adjoint
    representations of~$\mathrm{SU}(N)$ for~$2\le N\le 8$.}
\end{table}

There are again products involving adjoint representations
in table~\ref{tab:5-fields} for which
the number of independent invariant tensors changes when going
from~$\mathrm{SU}(2)$ to~$\mathrm{SU}(3)$.
As a curiosity, table~\ref{tab:adjoints} displays the number of
independent invariant tensors in products of $n$~adjoint representations
of~$\mathrm{SU}(N)$ for different values of~$N$. The
products in this table contain no exotic irreps and the
results for~$r_3$ can already be found in~\cite{Dittner:1971fy,Dittner:1972}.
The values of~$r_3$ and~$r_{\infty}$ for~$n=6$ are given in the
caption of table~6 in~\cite{Keppeler:2012ih}, where they have been
derived using purely combinatorial arguments.

An inspection of table~\ref{tab:adjoints} suggests a curious pattern
for the products of~$n\ge3$ adjoints of~$\mathrm{SU}(N)$
\begin{subequations}
\label{eq:adjoints}
  \begin{align}
    r_n &= r_{n-1} + 1 \\
    \forall N\ge n: r_N &= r_n\,.
  \end{align}
\end{subequations}
The considerations in section~\ref{sec:N-dependence} show that the
limit~$r_\infty$ in~\eqref{eq:r-infty} exists, but they are not
sufficient to show that~$r_\infty=r_n$.  
I don't know if there is a deeper reason for, a general proof or even a
practical application of~\eqref{eq:adjoints}.

\section{Conclusions}

I have presented a systematic construction of complete and linearly independent
sets of invariant tensors in products of irreducible representations
of~$\mathrm{SU}(N)$.  This construction is algorithmic and has been
implemented in the computer code~\textsc{tangara}.
There is no fundamental limit on the
size of the irreps and the number of factors.  However, there are
practical limits since the computational complexity of the most
straightforward unoptimized algorithms grows combinatorially.

In section~\ref{sec:results}, I have compared the results of the new
algorithm to a catalogue of invariant tensors published
previously~\cite{Carpenter:2021rkl}. There are several discrepancies
and I explain for each case in detail why the new result is the
correct one.

The study of invariant tensors in products of
representations larger than the~$\mathbf{10}$ appears at the moment to
be more of mathematical than phenomenological interest.  But
section~\ref{sec:4-fields} also lists six examples of colorflows involving
only four triplets, sextets or octets, where previous published
results are wrong.  Three of these contain only a single sextet and
another one a single
pair.  These are of immediate phenomenological interest for the study of
BSM models containing such particles.

Nothing precludes the application of the method to other Lie algebras
that appear in more exotic BSM models, such as~$\mathrm{SO}(N)$.  For
an implementation in~\textsc{tangara}, only the underlying birdtrack
library must be extended to support undirected lines.

\appendix
\section{Implementation and Interoperation}
\label{sec:implementation}

\subsection{\textsc{tangara}}
\label{sec:tangara}
\begin{figure}
\begin{quote}
\begin{framed}
\begin{small}
\begin{verbatim}
$ tangara_tool -s '8 *S 8 * 6 * ~6'

 0: [(1) * [1>2; 2>1; 3.0>4.0; 3.1>4.1]]
 1: [(1) * [1>2; 2>4.0; 3.0>1; 3.1>4.1];
     (1) * [1>4.0; 2>1; 3.0>2; 3.1>4.1]]
 2: [(1) * [1>4.0; 2>4.1; 3.0>1; 3.1>2]]

colorfactors[n_] :=
 { { (1/2)*n^4+(1/2)*n^3-(1/2)*n^2-(1/2)*n,  
     n^3+n^2-n-1                    ,        
     (1/2)*n^2-(1/2)                         },
   { n^3+n^2-n-1                          ,  
     (1/2)*n^4+n^3-(1/2)*n^2-3*n+2/n,        
     (1/2)*n^3-(3/2)*n+1/n                   },
   { (1/2)*n^2-(1/2)                      ,  
     (1/2)*n^3-(3/2)*n+1/n          ,        
     (1/4)*n^4+(1/2)*n^3-(1/4)*n^2-n+(1/2)/n } }
\end{verbatim}
\end{small}
\end{framed}
\end{quote}
\caption{\label{fig:tangara-8866S}
  \textsc{tangara} command line parameters and output: colorflows of
  invariant tensors in the symmetric tensor
  product~$\mathbf{8}\otimes_{\mathrm{S}}\mathbf{8}\otimes\mathbf{6}\otimes\mathbf{\overline6}$
  and their color factor matrix~$M(N,\mathcal{T})$.}
\end{figure}

\begin{figure}
\begin{quote}
\begin{framed}
\begin{small}
\begin{verbatim}
SU(2): rank = 2

  eval(1) = (3*(31 + Sqrt[321]))/8 = 18.3
  eval(2) = (3*(31 - Sqrt[321]))/8 = 4.9
  eval(3) = 0 = 0.

SU(3): rank = 3

  eval(1) = (10*(17 + Sqrt[73]))/3 = 85.2
  eval(2) = (10*(17 - Sqrt[73]))/3 = 28.2
  eval(3) = 18 = 18.
\end{verbatim}
\end{small}
\end{framed}
\end{quote}
\caption{\label{fig:Mathematica-8866S}
   \textsc{Mathematica}~\cite{Mathematica} output for the rank~$r_N$
   and the eigenvalues of the color factor matrix~$M(N,\mathcal{T})$ in
   figure~\ref{fig:tangara-8866S} for~$\mathrm{SU}(2)$ and~$\mathrm{SU}(3)$.}
\end{figure}

\begin{figure}
\begin{quote}
\begin{framed}
\begin{small}
\begin{verbatim}
$ tangara_tool -s '8 *A 8 * 6 * ~6'

0: [ ( 1)*[1>2; 2>4.0; 3.0>1; 3.1>4.1];
     (-1)*[1>4.0; 2>1; 3.0>2; 3.1>4.1] ]

colorfactors[n_] :=
 { { (1/2)*n^4+n^3-(1/2)*n^2-n } }
\end{verbatim}
\end{small}
\end{framed}
\end{quote}
\caption{\label{fig:tangara-8866A}
  \textsc{tangara} command line parameters and output: colorflows of
  invariant tensors in the antisymmetric tensor
  product~$\mathbf{8}\otimes_{\mathrm{A}}\mathbf{8}\otimes\mathbf{6}\otimes\mathbf{\overline6}$
  and their color factor matrix~$M(N,\mathcal{T})$.}
\end{figure}

The algorithm of section~\ref{sec:algorithm} has been implemented in
the computer program~\textsc{tangara}.  As illustrated in
figures~\ref{fig:tangara-8866S} and~\ref{fig:tangara-8866A}, the
program is given a tensor product, optionally with a representation
of the permutation
symmetry groups of identical factors, and computes a
list~$\mathcal{T}$ of candidates for a
complete and linearly independent set of invariant tensors together with the
matrix $M(N,\mathcal{T})$ of
color factors~\eqref{eq:colorfactor-matrix}.

Note that colorflows representing the invariant tensors
do \emph{not} contain the Young projectors nor the
ghosts, because these can be added trivially by other programs using
this output as input.  As can be seen in
figure~\ref{fig:tangara-8866S}, \textsc{tangara} lists the colorflows
from section~\ref{sec:8866bar} in the order~$\{X,Z_{\mathrm{S}},Y\}$
and normalizes them in such a way that the coefficients are integers
with the smallest modulus possible.  This normalization is fixed,
but the order of the colorflows is not guaranteed to be the same for
different versions of~\textsc{tangara}.  The chosen normalization is the most
convenient one, since it directly corresponds to the graphical
representation of colorflows, as in section~\ref{sec:results}.
Conceptually, the normalization~$\mu_N(T,T)=1$ might appear to be more
satisfactory, but it would require dividing by square roots of
polynomials in~$N$ and make the output much more complicated.

The matrix of color factors is accompanied by a short script that
computes the rank~$r_N$, the eigenvalues and the eigenvectors using
\textsc{Mathematica}~\cite{Mathematica}. The output is shown in
figure~\ref{fig:Mathematica-8866S} without the eigenvectors.
The computed eigenvectors can then be used to eliminate dependent
tensors from the set~$\mathcal{T}$. The script does not try to make a
recommendation for a canonical or ``best'' choice of linearly independent
invariant tensors, since mutually excluding goals are bound to enter into this
decision. Optionally, if the
colorflows contain no~$\epsilon$s, the $N$-dependence of the
eigenvalues can be plotted for illustration, as shown in figures~\ref{fig:EV(SU)}
and~\ref{fig:EV(U)}.

The complete source code of~\textsc{tangara} will be made publicly available in
the \textsc{O'Mega} subdirectory of a forthcoming
\textsc{Whizard}~\cite{Moretti:2001zz,Kilian:2007gr,Ohl:2023bvv} release.


\subsection{UFO}

Counting the number of linearly independent invariant tensors is an interesting
exercise, but the ultimate goal is their application in the study of
BSM physics.  For this purpose, the results must be made available to
other tools.  The UFO format~\cite{Degrande:2011ua,Darme:2023jdn} has
established itself as the \emph{lingua franca} for describing models
of BSM physics to automatic computation systems that compute
renormalization group running, decay rates and cross sections.  The
latter are subsequently used by Monte Carlo event generators to
simulate scattering processes at colliders.

The building blocks for color structures specified in the current UFO
format~\cite{Darme:2023jdn} are sufficient to express all
possible invariant tensors describing interactions of particles
transforming under the~$\mathbf{3}$, $\mathbf{\overline{3}}$,
$\mathbf{6}$, $\mathbf{\overline{6}}$,
and~$\mathbf{8}$ of~$\mathrm{SU}(3)$, including baryon number
violating terms containing~$\epsilon^{ijk}$
or~$\overline\epsilon_{ijk}$.
The $\mathbf{6}$ and $\mathbf{\overline{6}}$ irreps are described in
UFO by the Clebsch-Gordan coefficients~$\Ksix{s}{i}{j}$
and~$\Ksixbar{s}{i}{j}$ and generators~$\Tsix{a}{s}{t}$, where the
latter could have be expressed by the Clebsch-Gordan coefficients and
the generators in the fundamental representation
\begin{equation}
  \Tsix{a}{s}{t}
    = 2 \Ksixbar{s}{i}{k}\T{a}{i}{j}\Ksix{t}{j}{k}\,.
\end{equation}
All this can be translated automatically into a colorflow basis, as
has been demonstrated by the
implementation in~\textsc{O'Mega}~\cite{Moretti:2001zz,Ohl:2023bvv}
and~\textsc{Whizard}~\cite{Kilian:2007gr}.

Replacing the arrows by pairs of summation indices and symmetrizers by
Clebsch-Gordan coefficients, all colorflows connecting triplets,
sextets and octets can be translated to UFO directly, using only these
building blocks.  For example, the colorflow~\eqref{eq:cf-3-6-8}
in~\threefields{3}{6}{8} corresponds to
\begin{subequations}
\begin{equation}
\label{eq:cf-3-6-8-expression}
  C_{ir}^a = \overline{\epsilon}_{ijk} \T{a}{j}{l} \Ksix{r}{k}{l},.
\end{equation}
Using the UFO notation~\cite{Degrande:2011ua,Darme:2023jdn}, this is
written
\begin{equation}
\label{eq:cf-3-6-8-expression-UFO''}
  C_{ir}^{a} = \epsilon_{ijk} \T{a}{\bar\jmath}{l} \Ksix{r}{\bar k}{\bar l}
\end{equation}
and can be encoded as a UFO expression
\begin{equation}
\label{eq:cf-3-6-8-UFO}
  \text{\texttt{Epsilon(1,-1,-2) * T(3,-3,-1) * K6(2,-2,-3)}}\,,
\end{equation}
\end{subequations}
taking into account that \texttt{T(a,i,j)} is translated to~$\T{a}{\bar\jmath}{i}$.
The conjugate
\begin{subequations}
\begin{equation}
\label{eq:cf-3-6-8-expression-UFO'}
  C^{a\bar\imath\bar r}
    = \epsilon^{\bar\imath\bar\jmath\bar k} \T{a}{\bar l}{j} \Ksixbar{\,\bar r}{k}{l}
\end{equation}
is written
\begin{equation}
\label{eq:cf-3-6-8-UFO'}
  \text{\texttt{EpsilonBar(1,-1,-2) * T(3,-1,-3) * K6Bar(2,-2,-3)}}\,.
\end{equation}
\end{subequations}
Such translations can be performed directly by \textsc{tangara} and similar
programs.
In principle, this approach can be continued by adding dedicated
Clebsch-Gordan coefficients and generators for each higher
representation, where the catalogue~\eqref{eq:building-blocks} should
be more than enough for all practical purposes in the foreseeable future:
\texttt{K10}, \texttt{K15}, \texttt{K15prime}, \texttt{K21},
\texttt{K24}, \texttt{K27}.

An even more flexible solution would be to extend the syntax of UFO by
generic particle declarations, Clebsch-Gordan coefficients and
generators that accept a Young tableau as an additional argument
specifying the irrep: e.g.~\texttt{K[[1,2,3],[4]]} instead of
\texttt{K15}.  At the same time, one could add the option to encode
interactions more concisely in a colorflow basis using only Kronecker
and Levi-Civita symbols of external indices instead of the current
building blocks that force the user to introduce summation indices as
in~\eqref{eq:cf-3-6-8-UFO}.  In order to avoid a fragmentation of the
UFO format~\cite{Degrande:2011ua,Darme:2023jdn},
this should be decided as a
community effort for a future iteration of the format, after some experience has
been gained with example implementations of concrete syntax in
\textsc{Whizard}~\cite{Moretti:2001zz,Kilian:2007gr,Ohl:2023bvv} and
other programs.

\acknowledgments{%
I thank Manuel Kunkel for triggering this paper by asking me to count
the invariant tensors in the
$\mathbf{8}\otimes\mathbf{8}\otimes\mathbf{6}\otimes\mathbf{\overline6}$
representation of~$\mathrm{SU}(3)$.
I thank Wolfgang Kilian, J\"urgen Reuter for many productive
discussions on making \textsc{O'Mega} more colorful. I also thank JR
for useful comments on the manuscript.
I thank the referee Andreas Trautner (MPIK Heidelberg) for useful suggestions on the
presentation and for pointing out errors in the original manuscript.

This work is supported by the German Federal Ministry for Education and
Research (BMBF) under contract no.~05H21WWCAA.}


\end{fmffile}
\bibliography{tangara}
\end{document}